\documentclass[modern]{aastex701}

\newcommand{\teff}{T$_{eff}$}

\newcommand{\msun}{M$_\odot$}

\received{TBD}
\revised{\#2 \today}
\accepted{TBD}

\submitjournal{ApJ}

\shorttitle{Random Forest Analysis of VLM Spectral Binaries}
\shortauthors{Draxl Giannoni et al.}

\begin{document}

\title{Identifying and Characterizing Very Low Mass Spectral Blend Binaries with Machine Learning Methods}

\correspondingauthor{Adam J.\ Burgasser}

\author[0009-0004-4394-3890,sname=Draxl Giannoni]{Juan Diego Draxl Giannoni}
\affiliation{Department of Physics, UC San Diego, La Jolla, CA, USA}
\affiliation{{Department of Aerospace and Geodesy
TUM School of Engineering and Design, Technische Universit\"{a}t M\"{u}nchen, M\"{u}nchen, Germany}}
\email{jd.draxlgiannoni@gmail.com}

\author[0009-0003-4448-3681,sname=Desai]{Malina Desai}
\affiliation{Department of Physics, MIT, Cambridge, MA, USA}
\email{mmdesai@mit.edu}

\author[orcid=0000-0002-6523-9536,sname=Burgasser]{Adam J.\ Burgasser}
\affiliation{Department of Astronomy \& Astrophysics, UC San Diego, La Jolla, CA 92093, USA}
\email[show]{aburgasser@ucsd.edu}

\author[sname=Dunning]{A.\ Camille Dunning}
\affiliation{Halıcıoğlu Data Science Institute, UC San Diego, La Jolla, CA, USA}
\email{dunningcamille@gmail.com}

\author[0000-0003-2094-9128,sname=Aganze]{Christian Aganze}
\affiliation{Kavli Institute for Particle Astrophysics \& Cosmology, Stanford University, Stanford, CA 94305, USA}
\email{caganze@stanford.edu}

\author[sname=McDermott]{Luke McDermott}
\affiliation{Department of Computer Science and Engineering, UC San Diego, La Jolla, CA, USA}
\email{lcmcdermo@ucsd.edu}

\author[0000-0002-9807-5435,sname=Theissen]{Christopher A.\ Theissen}
\affiliation{Department of Astronomy \& Astrophysics, UC San Diego, La Jolla, CA, USA}
\email{ctheissen@ucsd.edu}

\author[0000-0001-8170-7072,sname=Bardalez Gagliuffi]{Daniella C.\ Bardalez Gagliuffi}
\affiliation{Department of Physics \& Astronomy, Amherst College, Amherst, MA, USA}
\email{dbardalezgagliuffi@amherst.edu}

\begin{abstract}
We present an approach to identifying and characterizing unresolved, very low mass spectral blend binaries composed of late-M, L, and T dwarfs using machine learning methodologies. 
We generated and evaluated a series of hierarchical random forest models to distinguish spectral blends from single very low-mass dwarfs, and to classify their primary and secondary components.
Models were trained on a sample of single and synthesized binary templates generated from empirical spectra.
We explored various aspects of the design of our models, 
and find that models trained on {a} full range of single and binary combinations
{have} the best performance for identification and component classification. 
These models achieve binary identification recall and precision of $\gtrsim$85\%, 
median component classification errors of $\lesssim$0.1~subtypes,
and systematic classification uncertainties of $\lesssim$1~subtype,
outperforming 
index-based methods in terms of fidelity, range, and speed.
Optimal performance is achieved for binaries composed of L and T dwarf primaries and late-L and T dwarf secondaries.
When applied to the spectra of previously confirmed very low-mass binaries, model performance is degraded due to the prevalence of systems with similar component types, but remains high in the optimal performance range.
We propose potential improvements to these models, which can be used to explore binary populations among the thousands to millions of very low-mass stars and brown dwarfs anticipated with large-scale spectral surveys such as {SPHEREx} and Euclid.
\end{abstract}

\keywords{
Binary stars (154) ---
Brown dwarfs (185) --- 
L dwarfs (894) --- 
M dwarfs (982) --- 
T dwarfs (1679) --- 
Random forests (1935)
}

\section{Introduction} \label{sec:intro}

Binary systems containing very low mass (VLM; M $\leq$ 0.1~M$_{\odot}$) stars and brown dwarfs are important benchmarks for testing theories of star and brown dwarf formation and evolution, measuring fundamental parameters such as mass and radius, and calibrating empirical relations connecting observables to physical properties.
The majority of VLM binaries known today are resolved systems identified and measured through high angular resolution imaging methods 
\citep{2018MNRAS.479.2702F},
and have angular separations $\rho \gtrsim 0\farcs1$, implying orbital semi-major axes $a \gtrsim 1$~au and periods $P \gtrsim 3$~yr at distances of 10~pc or larger \citep{2007prpl.conf..427B,2015AJ....150..163B}. While over two dozen 
of these systems
have partial or complete orbit determinations (e.g., \citealt{2010ApJ...711.1087K,2017ApJS..231...15D}), an unknown number of systems with shorter orbit periods remain undetected due to imaging resolution limits.
Other techniques to identify unresolved binaries, such as high-resolution spectroscopic monitoring {(e.g., \citealt{1999AJ....118.2460B,2007ApJ...666.1198B,2021ApJS..257...45H})}, 
astrometric monitoring ({e.g., \citealt{2020MNRAS.495.1136S})}, 
and the detection of overluminous sources {(e.g., \citealt{2013A&A...560A..52M,2019ApJS..240...19K})}
have uncovered several VLM binaries with smaller separations, including three eclipsing systems that provide radius measurements \citep{2006Natur.440..311S, 2015A&A...584A.128L, 2020NatAs...4..650T}. These methods are resource-intensive, however, {with low} yields
{($\lesssim$1--10\%; \citealt{Joergens2008,2014A&A...565A..20S})} and often incomplete determination of orbital and component parameters.

The distinct spectral characteristics of the late-type M, L, T, and Y dwarfs
that comprise the very low mass regime enable an alternative method for identifying unresolved binaries as spectral blend binaries. This method, also used in the detection of unresolved M dwarf plus white dwarf pairs 
\citep[e.g.,][]{2003AJ....125.2621R,2010MNRAS.402..620R,2012AJ....144...93M},
is applicable in the VLM regime due to the emergence of distinct molecular features in each spectral class. In particular, the {conversion of CO to CH$_4$} in the T dwarfs ({\teff} $\lesssim$ 1400~K) produces distinct {spectral} features 
in blended light spectra even {when components differ in brightness by} several magnitudes. The prototypical VLM spectral blend binary is 2MASS~J0518$-$2828AB, an L6 plus T4 system initially identified as a peculiar L dwarf, whose near-infrared spectrum could be explained as a blended light source \citep{Cruz2004}.
This system was subsequently confirmed {to be a binary by}  resolved Hubble Space Telescope imaging observations \citep{Burgasser2006HST}. {To date}, over 60 {VLM} spectral blend binary candidates and confirmed binary systems have been identified (e.g.,~\citealt{2010ApJ...710.1142B,2014ApJ...794..143B,2015ApJ...814..118B,2023AJ....166..226B}),
including sources with measured orbital periods $<$1~yr (e.g., \citealt{2008ApJ...678L.125B,2012ApJ...757..110B,2020MNRAS.495.1136S}).

\citet[hereafter B10]{2010ApJ...710.1142B} and \citet[hereafter B14]{2014ApJ...794..143B} developed spectral index-based methods for identifying {very low mass} spectral blend binaries. 
Both approaches rely on 
the contaminating signatures of CH$_4$ absorption in a combined-light spectrum {from a T dwarf companion}.
While these approaches have uncovered dozens of candidate and confirmed VLM binaries, 
they are based on an ad-hoc choice of indices and selection criteria that can result in
contamination from non-binary sources such as variable brown dwarfs \citep[e.g.][]{Radigan2014,Ashraf2022} and unusually blue L dwarfs (B14),
{and insensitivity to VLM pairs with equivalent spectral classification or extreme brightness ratios
These biases, and the degeneracies between age, mass, and spectral appearance for cooling brown dwarfs}. 
make it difficult to 
infer the {underlying VLM} binary fraction {from} samples of spectral blend systems \citep{2017PhDT.......291B}. 

{Recently}, supervised and unsupervised machine learning (ML) algorithms have been deployed to problems of astrophysical source identification and classification \citep{2019arXiv190407248B}. These algorithms can efficiently discern patterns in complex datasets, providing insight into underlying parameters or identifying clusters of sources with common traits.
One popular ML approach is the supervised learning method of random forests (RF),
which uses an ensemble of uncorrelated decision trees to classify or infer trends in complex data \citep{2001MachL..45....5B}. Each decision tree splits a dataset (features) through a series of threshold values (branches), which are iteratively updated and pruned by training on pre-classified (labeled) data. The results of the decision trees can be aggregated by averaging (regression) or majority voting (classification). RF models have been used extensively for the analysis of spectral data, to classify 
white dwarfs \citep{2023A&A...679A.127G} 
and VLM dwarfs \citep{2022MNRAS.513..516F,2024RNAAS...8..102Z}, and 
to infer the physical properties of galaxies \citep{2024A&A...690A.198A}, 
A-type stars \citep{2022RAA....22b5017C}, 
and VLM dwarfs \citep{2022ApJ...930..136L}, among other {applications}.
Relevant to this work, \citet{Aganze2022_1} demonstrated that ML algorithms are superior to index-based methods in identifying and classifying VLM dwarfs in large spectroscopic surveys,  
significantly reducing contamination, 
an important consideration when the sources of interest comprise a small fraction ($<$1\%) of the overall sample. 

Recognizing the benefits of machine learning methods to spectral classification problems, we examined the application of the random forest algorithm to the identification and characterization of very low mass spectral blend binaries.
This work follows a pilot study by \citet{Desai_2023} which demonstrated 95\% precision in identifying spectral blends of M7--L7 plus T1--T8 pairs. 
Here we expand this analysis to cover a broader set of binary combinations, and explore further details in the construction and training of the RF models.
This article is organized as follows:
Section~\ref{sec:design} describes the input spectral sample and training set for our models, and benchmarks our analysis by quantifying the precision and accuracy of the B10 and B14 index-based methods. 
Section~\ref{sec:RFmodel} describes the design and training of the RF models used for binary identification and component classification.
Section~\ref{sec:results} reviews performance of the models for 
various assumptions of input training data, data quality, and component composition.
Section~\ref{sec:application} further validates the performance of these models against a sample of confirmed binary spectra. 
{
Section~\ref{sec:discussion} discusses limitations to this study in terms of sample and model design, selection biases, and the typical compositions of VLM binaries in volume-limited samples.}
{Finally,} Section~\ref{sec:summary} summaries our findings {and prospects for VLM binary searches in future spectral surveys.}

\section{Model Training Data} \label{sec:design}

\subsection{Spectral Sample} \label{sec:sample}

Our training sample was constructed from low-resolution, near-infrared spectra of late-M, L and T dwarfs contained in the SpeX Prism Library Analysis Toolkit (SPLAT; \citealt{SPLAT}), obtained with the SpeX spectrograph on the NASA 3m Infrared Telescope Facility \citep{2003PASP..115..362R}. We started with approximately 1500 VLM dwarf spectra with published spectral types spanning M6 to T9. This initial sample was curated by removing previously reported binaries and spectral binary candidates, young brown dwarfs, subdwarfs, and misclassified giant stars. For sources with multiple spectral observations, we selected the spectrum with the highest signal-to-noise (S/N). We also conducted a visual inspection to remove other contaminants, and removed any spectra with undefined, infinite, or negative values 
in their flux or uncertainty arrays to ensure accurate S/N calculations. All spectra were interpolated onto a common wavelength scale spanning 0.9--2.4~$\mu$m, aligned with the pixel scale for a representative spectrum {with an average resolution of $\lambda/\Delta\lambda$ $\approx$ 100.}
Spectra were reclassified by direct comparison to near-infrared standards defined in \citet[][M and L dwarfs]{2010ApJS..190..100K} and \citet[][T dwarfs]{2006ApJ...637.1067B}. 
The final {template} sample of 1006 spectra is summarized in the Appendix.

\subsection{Synthetic Single Templates} \label{sec:singles}

\begin{figure*}[th]
\centering
\includegraphics[width=0.65\textwidth]{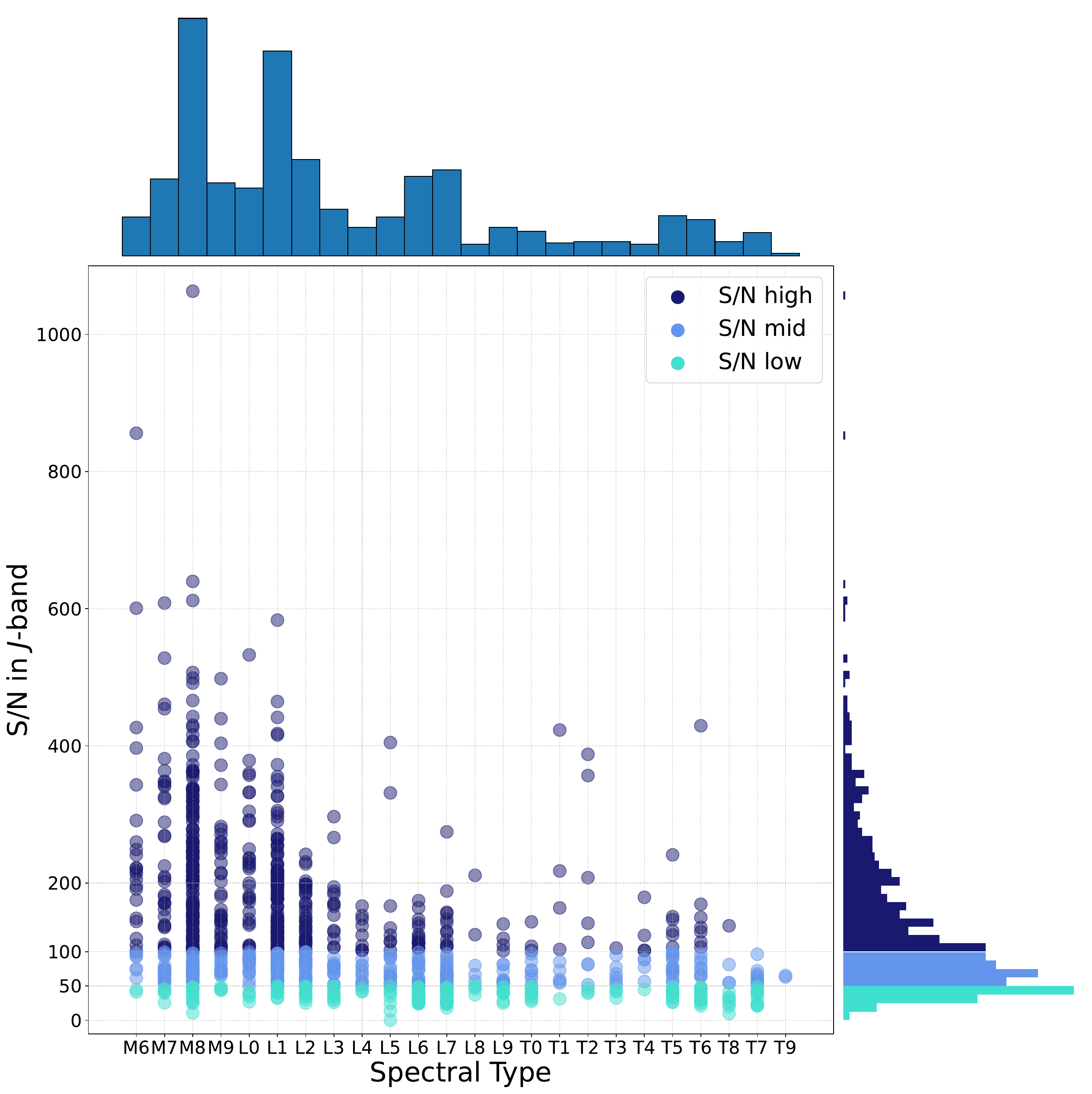}
\caption{$J$-band signal-to-noise (S/N) vs spectral type of our empirical spectral sample. Each individual spectrum is plotted and color-coded by their assignment to low (S/N $<$ 50{, light blue}), mid (50 $\leq$ S/N $<$ 100{, medium blue}), and high (S/N $\geq$ 100{, dark blue}) S/N groups. We show the marginalized distributions of the spectral type along the top axis, and for S/N along the right axis.
}  
\label{fig:sample}
\end{figure*}

Figure~\ref{fig:sample} displays the distributions of spectral types and signal-to-noise ratios for these template spectra, the latter calculated in the 1.20--1.35~$\micron$ $J$-band flux peak to accommodate the strong absorption features present in late-L and T dwarf spectra.
The distribution of spectral types is highly asymmetric due to the intrinsic brightnesses of the sources and the underlying luminosity function, both of which vary considerably with spectral type \citep{Luminosity_Bardalez,2024ApJS..271...55K}.
Our initial exploration of RF models indicated that such unbalanced samples can lead to significant biases in classification \citep{Desai_2023}. 
{In addition}, varying S/N is of particular concern as the subtle
absorption features from a companion spectrum can be obscured by noise in a low S/N spectrum.

To address these sample imbalances,
we constructed a synthetic dataset of single templates evenly-distributed across spectral types M6 to T9 and across signal-to-noise ratios through noise scaling and flux resampling.
We defined three $J$-band S/N groups: 
low (2 $\leq$ S/N $<$ 50), 
mid (50 $\leq$ S/N $<$ 100), 
and high (S/N $\geq$ 100), and developed separate RF models for each of these groups. 
For each spectral subtype, we 
first chose a subsample of SpeX spectra with $J$-band S/N equal to or greater than our target S/N range. 
We randomly selected spectra from this subsample with replacement, then varied the spectral fluxes using Monte Carlo methods, drawing a new flux value at each wavelength from a Gaussian distribution centered on the original flux value and with a width equal to the uncertainty. We generated lower S/N spectra in this process by scaling up the uncertainty before resampling. 
We re-computed the $J$-band S/N and the standard-match classification for each new synthetic spectrum. 
This process was repeated across all spectral subtypes until we had created a sample of 17,060 synthetic templates, distributed as
250 templates per subtype for 24 subtypes spanning M6 to T9 for the low and mid S/N groups (6,000 templates each); and 230 templates per subtype for 22 subtypes spanning M6 to T9 for the high S/N group, excluding T7 and T9 which lacked high S/N SpeX spectra (5,060 templates).


\subsection{Synthetic Binary Templates} \label{sec:binaries}

\begin{figure*}[th]
\plotone{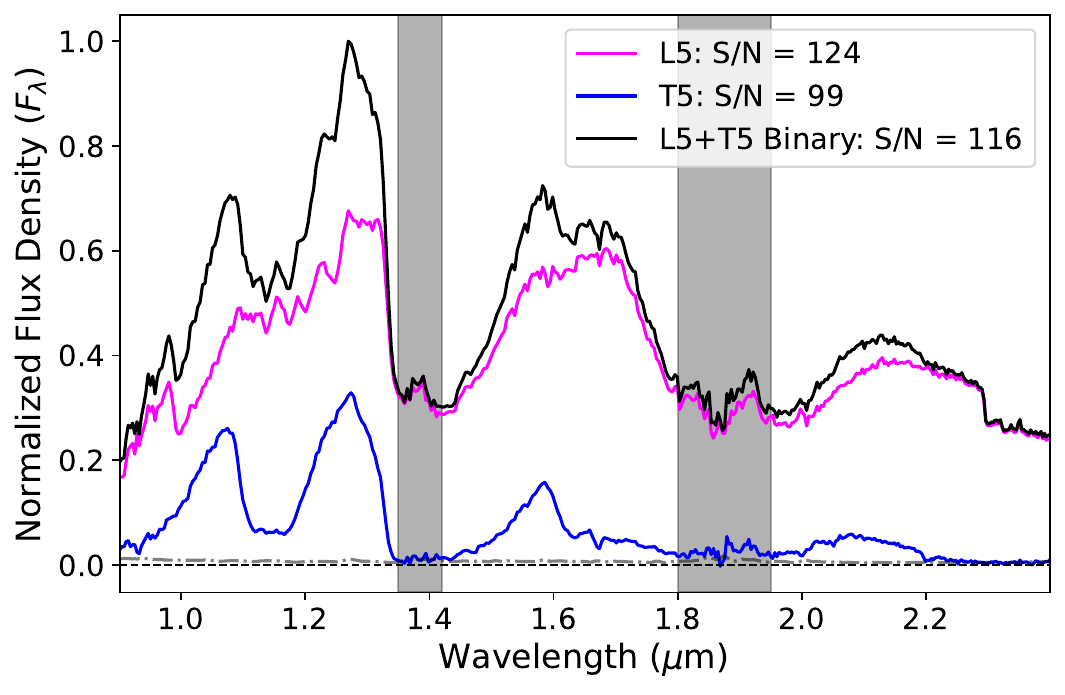}
\caption{
Example of a synthetic binary template with S/N = 116 (black line), constructed from a S/N = 124 L5 dwarf spectrum (magenta line) and a S/N = 99 T5 dwarf spectrum (blue line). 
The fluxes of the single components are scaled relative to each other {using} the \citet{2012ApJS..201...19D} {$M_J$}/spectral type relation, and to the combined spectrum which is normalized between 1.20--1.35~$\mu$m.
The uncertainty for the combined light spectrum is shown as a dash-dot line, and the vertical gray bands indicate {regions of strong} telluric absorption. }
\label{fig:binary}
\end{figure*}

We created an evenly-distributed sample of synthetic binary templates following a similar approach as our single templates, accounting for equal representation of primary and secondary spectral type combinations as well as signal-to-noise groupings. 
We first scaled all of the SpeX spectra to absolute fluxes using the \citet{2012ApJS..201...19D} absolute magnitude/spectral type relation for the 2MASS $J$-band filter.
{This relation has an inherent uncertainty of 0.4~mag. Our baseline analysis does not account for this uncertainty; however, we evaluated its influence in one set of binary identification models as discussed in Section~\ref{sec:BIdmodel}.}
We selected subsamples of spectra matching the intended primary and secondary subtypes (secondaries were required to have the same or later classifications), and with 
S/N values equal to or greater than our intended binary template S/N. 
We randomly paired primary and secondary templates from these subsamples, added the spectral fluxes (uncertainties were combined in quadrature), and re-normalized the combined template between
1.20--1.35~$\mu$m to mitigate selection biases \citep{Desai_2023}.
We expanded the number of spectra within each primary-secondary classification and S/N groupings using the same Monte Carlo approach as our single templates.
The resulting binary templates were then re-classified by comparison to spectral standards and the $J$-band S/N re-computed.
An example {of a synthetic} binary {template} for an L5 plus T5 combination in the mid S/N group is illustrated in Figure~\ref{fig:binary}. 
This process was repeated to generate 20 binaries for each of the 300 primary-secondary combinations, yielding 6,000 synthetic binary templates for each of the low and mid S/N groups.
For the high S/N group, the absence of high S/N T7 and T9 spectra reduced the number of possible pairings to 253, yielding 5,060 binary templates. The size of the synthetic binary template sample intentionally matches the size of the synthetic single template sample for each S/N group.

\subsection{Benchmarking Spectral Index Methods} \label{sec:index}

In order to assess the utility of our machine learning approach, we quantified the performance of the original index-based selection methods {of} B10 and B14 using our template sample.
For L dwarf plus T dwarf binaries, B10 used 
eight spectral indices and near-infrared spectral types to establish six selection criteria that segregate spectral binaries from single stars.
Objects satisfying at least three criteria were considered ``strong'' binary candidates, those satisfying only two criteria were considered ``weak'' candidates{, based on the sources known to be binary at the time of the study. B10} 
also compared the $\chi^2$ deviation between a candidate spectrum 
($S[\lambda]$) 
and both best-fit single and binary templates ($T[\lambda]$),
\begin{equation}
    \chi^2 \equiv \sum_{\lambda} \left[ \frac{S[\lambda] - \alpha T[\lambda]}{\sigma_S[\lambda]} \right]^2
\end{equation} 
where $\sigma_S[\lambda]$ is the uncertainty spectrum of the source and $\alpha$ is a scaling factor that minimizes $\chi^2$ \citep{2005ApJ...623.1115C}. 
A source was considered a spectral binary candidate if the binary template was a significantly better fit than the single template based on an F-test comparison of $\chi^2$ values.  
This comparison also allowed determination of component {spectral types}.
The B10 approach was defined for 
primaries classified L5--T2 and secondaries classified T2--T7, and applied to combined-light spectra with classifications of 
L8--T4.

B14 followed a similar procedure {for late-M/early-L dwarf plus T dwarf binaries}, evaluating 13 indices to define 17 selection criteria, with ``strong'' candidates satisfying at least eight criteria and ``weak'' candidates satisfying four to eight criteria. 
B14 also used a $\chi^2$ template comparison to eliminate spurious candidates, particularly blue L dwarfs, and to determine component classifications.
The B14 approach was defined for 
primaries classified M7--L7 and secondaries classified T1--T7, and applied to combined-light spectra with classifications of M7.5--L8 \citep{2015AJ....150..163B}.

We applied the 
B10 and B14 index criteria to the synthetic single and binary templates that matched the appropriate combined-light spectral types for the two methods.
To simplify our test,
we included both ``weak'' and ``strong''  binary candidates, and did not perform the $\chi^2$ test.  
To assess the performance of the indices, we measured 
the statistics of recall (R): 
\begin{equation}\label{eqn:recall}
    R \equiv \frac{\text{TP}}{\text{TP}+\text{FN}}
\end{equation}
precision (P): 
\begin{equation}\label{eqn:precision}
    P \equiv \frac{\text{TP}}{\text{TP}+\text{FP}}
\end{equation}
and F1-score:
\begin{equation}\label{eqn:f1}
    F1 \equiv \frac{2\times\text{TP}}{2\times\text{TP}+\text{FP}+\text{FN}}
\end{equation}
where 
TP is the number of true positives (binary templates selected as binaries),
TN is the number of true negatives (single templates rejected as binaries),
FP is the number of false positives (single templates selected as binaries), and
FN is the number of false negatives (binary templates rejected as binaries).
Recall measures the fraction of selected binaries among true binaries,
precision measures the fraction of true binaries among selected binaries,
and F1-score combines these statistics. 
Values range from 0 to 1, with the latter indicating perfect performance.

\begin{figure}[th]
\centering
\includegraphics[width=0.48\textwidth]{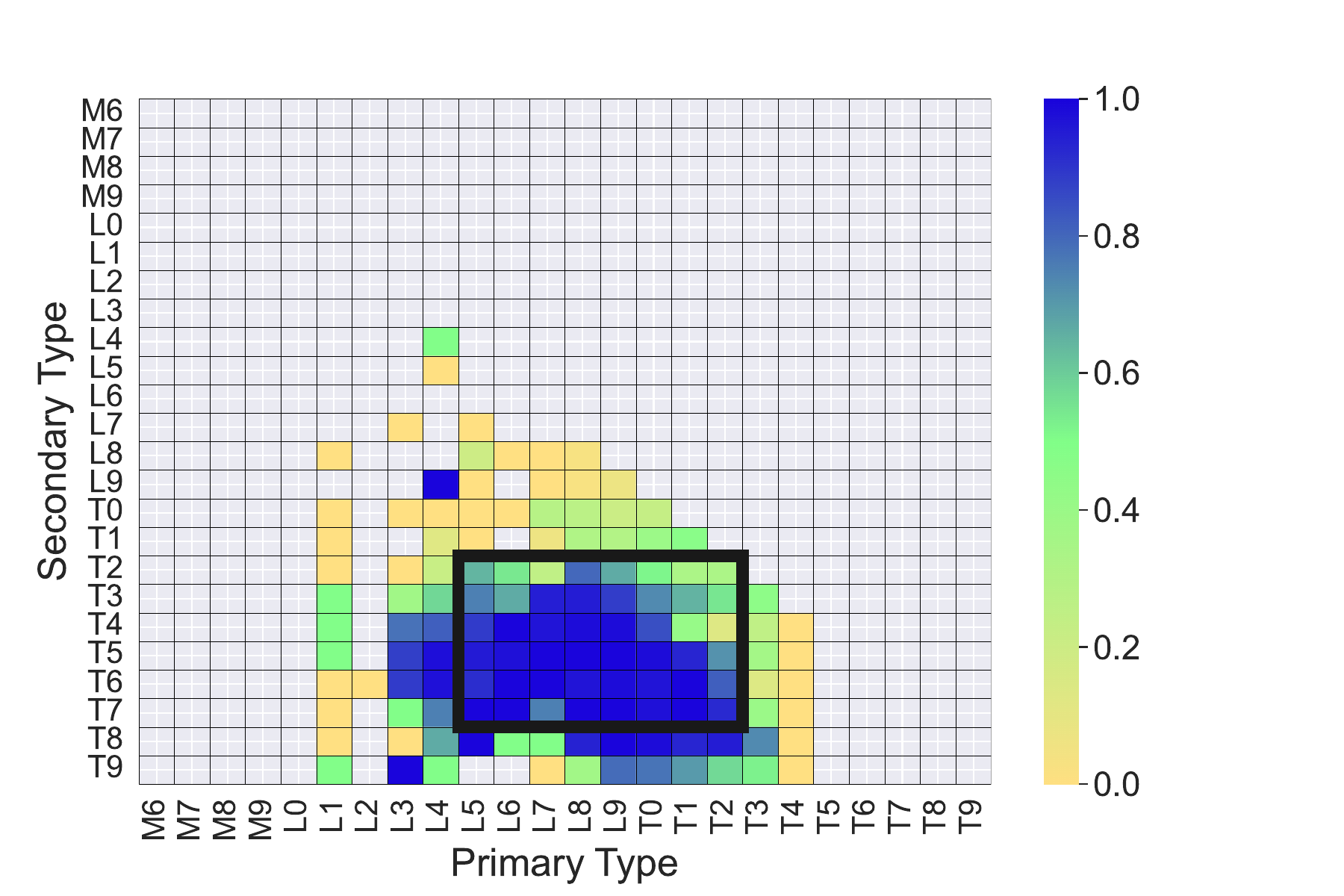}
\includegraphics[width=0.48\textwidth]{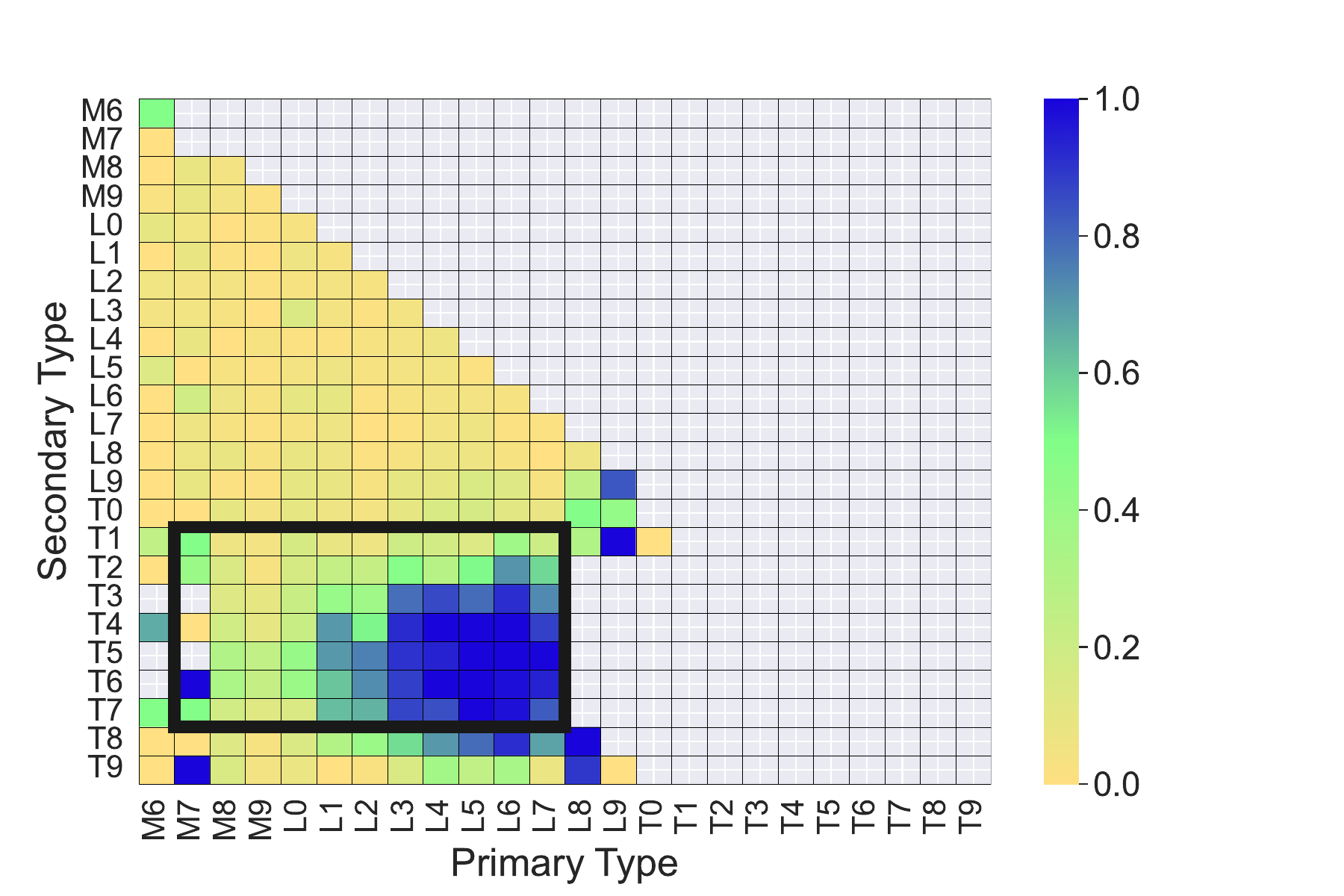}
\caption{Recall for the index-based methods of B10 (left) and B14 (right)
as a function of component classifications. Only those combinations whose combined-light spectra match the applicable spectral range of these methods (L8--T4 for B10, M8--L8 for B14; black boxes) are shown.
Each box represents an average of recall across all S/N groups.}
\label{fig:index_heatmap}
\end{figure}

Performance statistics as a function of combined-light spectral type are shown in Table~\ref{tab:index_stats}.
While both methods perform reasonably well in aggregate for identifying single sources 
(TN = 0.75 for B10, 0.95 for B14), 
they have relatively poor performance in identifying binary sources 
(TP = 0.64 for B10, 0.23 for B14).
Focusing on recall (Table~\ref{tab:index_lmh_binarycomb} and Figure~\ref{fig:index_heatmap}),
the B10 indices identify 62\%--63\% of true binaries, {and} 
the B14 indices identify 19\%--27\% of true binaries, depending on S/N.
These results suggest relatively poor performance; however, in select spectral compositions 
performance improves considerably.
For L5--T2 primaries and T2--T7 secondaries, the B10 recall increases to 0.79--0.85, while for M7--L7 primaries and T1--T7 secondaries, the B14 recall increases to 0.45--0.48, again depending on S/N.
Outside these optimal combinations, and particularly for systems with equivalent component types
(e.g., late-M + late-M, early-L + early-L, etc.),
recall declines to
R $<$ 0.1.\footnote{One exception is early-T + early-T combinations for the B10 index method, where the rapid evolution of spectral morphologies across the range results in a relatively high recall of R = 0.45.}
Hence, both approaches miss a significant fraction of binary systems 
when components are outside the spectral classification ranges that these methods are optimized for.
Note that S/N has a minor impact on performance statistics ($\Delta{R} \lesssim 0.1$).

\begin{deluxetable}{ccccccc}
\tablecaption{Performance Statistics of Index-based Methods for Spectral Binary Identification \label{tab:index_stats}}  
\tabletypesize{\scriptsize} 
\tablehead{
\colhead{Combined} & 
\multicolumn{3}{c}{B14} & 
\multicolumn{3}{c}{B10}
\\
\cline{2-4} \cline{5-7}
\colhead{Type} & 
\colhead{P} &  
\colhead{R} &
\colhead{F1} &
\colhead{P} &  
\colhead{R} &
\colhead{F1} 
}
\startdata
M8 & 0.69 & 0.11 & 0.18 & \nodata & \nodata & \nodata \\
M9 & 0.87 & 0.06 & 0.10 & \nodata & \nodata & \nodata \\
L0 & 0.68 & 0.12 & 0.20 & \nodata & \nodata & \nodata \\
L1 & 0.78 & 0.19 & 0.31 & \nodata & \nodata & \nodata \\
L2 & 0.83 & 0.14 & 0.24 & \nodata & \nodata & \nodata \\
L3 & 0.92 & 0.42 & 0.58 & \nodata & \nodata & \nodata \\
L4 & 0.76 & 0.45 & 0.56 & \nodata & \nodata & \nodata \\
L5 & 0.82 & 0.39 & 0.53 & \nodata & \nodata & \nodata \\
L6 & 0.89 & 0.18 & 0.29 & \nodata & \nodata & \nodata \\
L7 & 0.97 & 0.20 & 0.34 & \nodata & \nodata & \nodata \\
L8 & 0.86 & 0.41 & 0.56 & 0.97 & 0.23 & 0.38 \\
L9 & \nodata & \nodata & \nodata & 0.87 & 0.57 & 0.69 \\
T0 & \nodata & \nodata & \nodata & 0.75 & 0.77 & 0.76 \\
T1 & \nodata & \nodata & \nodata & 0.63 & 0.69 & 0.66 \\
T2 & \nodata & \nodata & \nodata & 0.61 & 0.76 & 0.68 \\
T3 & \nodata & \nodata & \nodata & 0.59 & 0.72 & 0.64 \\
T4 & \nodata & \nodata & \nodata & 0.97 & 0.16 & 0.27 \\
\enddata
\tablecomments{Performances statistics are precision (P), recall (R), and F1-score (F1). Templates satisfying both ``weak'' and ``strong'' criteria were assumed to be selected as binary candidates.}
\end{deluxetable}

\begin{deluxetable}{lcccccc}
\tablecaption{Recall Statistics for Index-based Methods \label{tab:index_lmh_binarycomb}}  
\tabletypesize{\scriptsize} 
\tablehead{
\colhead{Binary Combination} & 
\multicolumn{3}{c}{B14} & 
\multicolumn{3}{c}{B10} 
\\
\cline{2-4} \cline{5-7}
\colhead{(Primary + Secondary)} & 
\colhead{low} &  
\colhead{mid} &
\colhead{high} &
\colhead{low} &  
\colhead{mid} &
\colhead{high} 
}
\startdata
Full Sample & 0.27 & 0.22 & 0.19 & 0.63 & 0.66 & 0.62 \\
M7-L7 (P) + T1-T7 (S) & 0.48 & 0.47 & 0.45 & 0.78 & 0.77 & 0.79 \\
L5-T2 (P) + T2-T7 (S) & 0.79 & 0.87 & 0.90 & 0.79 & 0.85 & 0.79 \\
late M + late M & 0.14 & 0.01 & 0.01 & \nodata & \nodata & \nodata \\
late M + early L & 0.07 & 0.02 & 0.01 & \nodata & \nodata & \nodata \\
late M + late L & 0.08 & 0.05 & 0.01 & \nodata & \nodata & \nodata \\
late M + early T & 0.20 & 0.11 & 0.07 & \nodata & \nodata & \nodata \\
late M + late T & 0.21 & 0.12 & 0.16 & \nodata & \nodata & \nodata \\
early L + early L & 0.12 & 0.01 & 0.01 & \nodata & \nodata & \nodata \\
early L + late L & 0.12 & 0.03 & 0.03 & \nodata & \nodata & \nodata \\
early L + early T & 0.40 & 0.37 & 0.37 & 0.70 & 0.66 & 0.69 \\
early L + late T & 0.47 & 0.48 & 0.62 & 0.74 & 0.90 & 0.83 \\
late L + late L & 0.13 & 0.05 & 0.03 & 0.11 & 0.00 & 0.00 \\
late L + early T & 0.55 & 0.55 & 0.50 & 0.76 & 0.77 & 0.74 \\
late L + late T & 0.69 & 0.75 & 0.97 & 0.86 & 0.90 & 1.00 \\
early T + early T & \nodata & \nodata & \nodata & 0.46 & 0.52 & 0.45 \\
early T + late T & \nodata & \nodata & \nodata & 0.66 & 0.69 & 0.76 \\
\enddata
\tablecomments{Spectral type groupings are as follows:
late-M = M6--M9,
early-L = L0--L4,
late-L = L5--L9,
early-T = T0--T4, and
late-T = T5--T9.
Templates satisfying both ``weak'' and ``strong'' criteria were assumed to be selected as binary candidates. Only spectral type groupings with more than 10 templates are included.}
\end{deluxetable}

Finally, we note that it took 20.7 min to measure the indices and determine binary classifications for the 31,546 template spectra with the B10 and B14 methods on a laptop computer,\footnote{All computations in this investigation were run on an 11th Generation Intel Core i9-11900H processor at 2.50GHz, with 16.0 GB (15.7 GB usable) RAM and a 64-bit operating system with x64-based processor.} an average of 39 ms per spectrum.
The original classification of templates using a $\chi^2$ comparison was less computationally efficient, taking an average of 2.3 seconds per object.

\section{Random Forest Model Design} \label{sec:RFmodel}

Building on prior work \citep{Desai_2023}, {we evaluated a random forest model as} an alternative {approach} to {very low mass} binary identification and classification. 
We used the {\tt scikit-learn} package \citep{scikit-learn} to implement two algorithms:
a RF classifier ({\tt RandomForestClassifier}) which returns a binary class based on the
majority outcome of decision trees, and 
a RF regressor ({\tt RandomForestRegressor}) which returns a numerical value based on the 
average across decision trees.
Our models used the normalized flux points of the spectra and additional quantities computed from them (see below) as features, and return a decision on whether a source is a binary (classifier) and the component classifications (classifier and regressor).

Table~\ref{tab:RFparameters} {summarizes} the parameters {for} the architecture of our {random forest} models. For most parameters, we used default values defined in the {\tt scikit-learn} package, but 
optimized four parameters via iterative tuning: 
the number of estimators ({\tt n\_estimators}), equal to the total number of decision trees in the RF model; 
the maximum depth ({\tt max\_depth}) of each decision tree; 
the minimum number of samples used to split a leaf node ({\tt min\_samples\_split}); and 
the minimum samples per leaf ({\tt min\_samples\_leaf}), which determines when a decision tree branch ends. 
These parameters were optimized using the F1-score (Eqn.~\ref{eqn:f1}).
We found that {\tt n\_estimators} and {\tt max\_depth} had the largest impact on the F1-score, with larger values of each yielding larger F1-scores, albeit at a computational cost of training more decision trees. 
We also found that smaller {\tt min\_samples\_split} and {\tt min\_samples\_leaf} improved F1-scores slightly. 
We therefore chose parameter values of 
{\tt n\_estimators} = 50,
unlimited {\tt max\_depth}, 
{\tt min\_samples\_split} = 2, and 
{\tt min\_samples\_leaf} = 2.

\begin{deluxetable}{lp{5.7cm}cc}
\tablecaption{Random Forest Model Parameters \label{tab:RFparameters}}
\tabletypesize{\scriptsize} 
\tablehead{ 
& & \multicolumn{2}{c}{Adopted Values} \\
\cline{3-4}
\multicolumn{1}{c}{Parameter} & 
\multicolumn{1}{c}{Description} &
\multicolumn{1}{c}{Classifier} &
\multicolumn{1}{c}{Regressor}
}
\startdata
{\tt n\_estimators} & Number of trees in the forest & 50 & 50 \\
{\tt max\_depth} & Maximum number of nodes in a tree &  No limit & No limit \\ 
{\tt min\_samples\_split} & Minimum number of samples to divide a node & 2 & 2 \\ 
{\tt min\_samples\_leaf} & Minimum required samples at a leaf node & 2 & 2 \\
{\tt max\_leaf\_nodes} & Maximum number of leafs in a node & {\tt None} & {\tt None} \\
{\tt max\_features} & Maximum number of nodes considered when finding the best node division & {\tt sqrt} & all \\ 
{\tt criterion} & Evaluate the quality of the division of a node by minimizing a loss function & {\tt gini} & {\tt squared\_error} \\ 
{\tt min\_weight\_fraction\_leaf} & Minimum fraction of weights required at each node & {\tt None} & {\tt None} \\
{\tt min\_impurity\_decrease} & a Node division only occurs if the decrease in impurity is greater than this value & 0 & 0 \\ 
{\tt bootstrap} & Whether a random subsample from the larger dataset is chosen when building trees & {\tt True} & {\tt True} \\ 
{\tt oob\_score} & Metric for calculating the out-of-bag (oob) score that evaluates performance of a tree & {\tt accuracy\_score} & {\tt r2\_score} \\ 
{\tt n\_jobs} & Number of CPU cores & {\tt None} & {\tt None} \\ 
{\tt warm\_start} & Whether to use the previous solution to fit the forest & {\tt False} & {\tt False} \\ 
{\tt class\_weight} & Weights for each class & {\tt None} &  - \\ 
{\tt ccp\_alpha} & Parameter that controls pruning; 0 implies no pruning & 0 & 0 \\ 
{\tt max\_samples} & Number of samples to train each base estimator & {\tt n\_samples} & {\tt n\_samples} \\ 
{\tt monotonic\_cst} & Monotonicity constraint on features; {\tt None} implies no constraint & {\tt None} &  {\tt None} \\ 
\enddata
\end{deluxetable}

\subsection{Binary Identification Models} \label{sec:BIdmodel}

For our binary identification model (BId), we explored eight variations of RF models for each spectral S/N range (Table~\ref{tab:BIdmodels}).
{Our baseline model} BId1 used the 409 flux points in the wavelength range 0.9--2.4~$\mu$m as input features. 
Model BId2 included an additional 409 flux points of the difference spectrum between the input spectrum and its best-fit single template (Section~\ref{sec:index}), for a total of 818 flux features.  
Model BId3 used the input spectrum after masking out regions of strong telluric absorption over 1.35--1.42~${\mu}m$ and 1.80--1.95~${\mu}m$, resulting in 351 flux features. 
Model BId4 combined the flux samples from models BId2 and BId3, resulting in 702 flux features. 
Models BId5 and BId6 were trained on a subset of templates encompassing the spectral type ranges examined in B14, primary types M7--L7 and secondary types T1--T7, with data prepared as in models BId1 and BId2. 
Models BId7 and BId8 were trained on a subset of templates encompassing the spectral type ranges examined in B10,
primary types L5--T2 and secondary types T2--T7, again with data prepared as in models BId1 and BId2.
These last four models aimed to assess whether the RF models achieved better performance in the specific binary compositions explored in the index-based studies.
{Finally, model BId9 replicates BId1 but examines the impact of systematic uncertainty in the \citet{2012ApJS..201...19D} $M_J$/spectral type relation, by scaling the component spectra with magnitudes drawn from the relation value plus a 0.4~mag random gaussian uncertainty.}
Each model and S/N sample was trained using 70\% of  
our single and binary templates selected randomly,
and tested on the remaining 30\% of the templates.
Training times were consistently under 10~s for each model.

\begin{deluxetable}{lcccc}
\tablecaption{Binary Identification (BId) Models \label{tab:BIdmodels}}
\tabletypesize{\scriptsize} 
\tablehead{
\colhead{Model} & 
\colhead{Binary} &  
\colhead{Difference} &
\colhead{Telluric} &
\colhead{Training} \\
\colhead{Name} & 
\colhead{SpTs} &  
\colhead{Flux} &
\colhead{Masking} &
\colhead{Time (sec)} \\
}
\startdata
 BId1 & M6-T9 & No & No & 5.2\\
 BId2 & M6-T9 & Yes & No & 7.5\\
 BId3 & M6-T9 & No & Yes & 4.9\\
 BId4 & M6-T9 & Yes & Yes & 6.8\\
 BId5 & M7-L7 (P) T1-T7 (S) & No & No & 2.6\\
 BId6 & M7-L7 (P) T1-T7 (S) & Yes & No & 3.6\\
 BId7 & L5-T2 (P) T2-T7 (S) & No & No & 1.6\\
 BId8 & L5-T2 (P) T2-T7 (S) & Yes & No & 2.7\\
 {BId9} & M6-T9{\tablenotemark{a}} & No & No & 5.2\\
\enddata
\tablenotetext{a}{{This model accounts for uncertainties in the absolute magnitudes of individual components by accounting for the 0.4~mag {systematic} uncertainty in the \citet{2012ApJS..201...19D} $M_J$/spectral type relation.}}
\end{deluxetable}

\subsection{Component Classification Models} \label{sec:Bclassmodel}

Our second class of models determined the spectral classifications of the primary and secondary stars within a binary system, for which we used the {\tt scikit-learn} {\tt MultiOutputRegressor} model.  This implementations extends the functionality of single-output regressor models to allow for multiple outputs, with each output variable treated independently during training. 
As with our BId models, we explored eight different sets of models (BClass) for each S/N sample based on the input features (inclusion of difference spectrum, full or telluric masked input spectrum) and constraints on the binary compositions, and each model was trained and tested on 70\% and 30\% of the binary templates, respectively. Training times were longer for these models, but were still under 20~s for all cases explored.

\begin{deluxetable}{lcccc}
\tablecaption{Binary Subtype Classification (BClass) Models \label{tab:BClassmodels}}  
\tabletypesize{\scriptsize} 
\tablehead{
\colhead{Model} & 
\colhead{Binary} &  
\colhead{Difference} &
\colhead{Telluric} &
\colhead{Training} \\
\colhead{Name} & 
\colhead{SpTs} &  
\colhead{Flux} &
\colhead{Masking} &
\colhead{Time (sec)} \\
}
\startdata
 BClass1 & M6-T9 & No & No & 9.0\\
 BClass2 & M6-T9 & Yes & No & 18 \\
 BClass3 & M6-T9 & No & Yes & 8.0\\
 BClass4 & M6-T9 & Yes & Yes & 16 \\
 BClass5 & M7-L7 (P) T1-T7 (S) & No & No & 2.4\\
 BClass6 & M7-L7 (P) T1-T7 (S) & Yes & No & 4.5\\
 BClass7 & L5-T2 (P) T2-T7 (S) & No & No & 1.3\\
 BClass8 & L5-T2 (P) T2-T7 (S) & Yes & No & 2.4\\
\enddata
\end{deluxetable}

\subsection{Feature Importance} \label{sec:fi}

In addition to accurate identification and classification of binary spectra, we also sought
to identify the spectral {features} that indicate the presence of an unresolved companion.
The RF model enables this assessment through 
feature importance (FI), which quantifies the contribution of each feature to the performance of the model.
In scikit-learn, FI is calculated using the mean decrease in impurity (MDI)
when a feature is used to split a node. 
For our binary identification model, impurity is measured using the Gini index; for our binary classification model, impurity is measured using mean squared error. 

\section{Model Performance} \label{sec:results}

\subsection{Binary Identification} \label{sec:results-id}

\begin{figure}[th]
\centering
\includegraphics[width=0.65\textwidth]{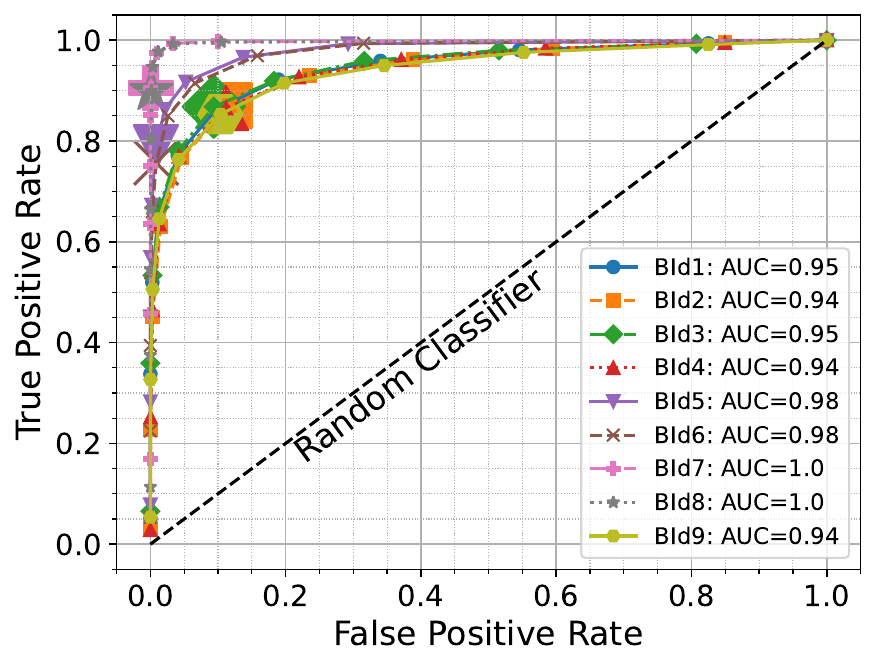}
\caption{{Receiver operating characteristic (ROC) curves for the nine binary identification (BId) models examined in this study, comparing the true positive rate (true binaries) to the false positive rate (singles identified as binaries) as a function of detection threshold. The threshold for binary identification, computed as the fraction of decision trees identifying a spectrum as binary, was allowed to vary from 0 to 1. Random selection would follow the dashed line, while the computed curves indicate high fidelity in identifying true binaries. The legend indicates the color, point style, and line style corresponding to the given BId model, and lists the area under the curve (AUC) metric for each model. Larger symbols correspond to a detection threshold of 0.5, corresponding to majority vote.}}
\label{fig:BId_roc}
\end{figure}

\begin{figure}[th]
\centering
\includegraphics[width=0.32\linewidth]{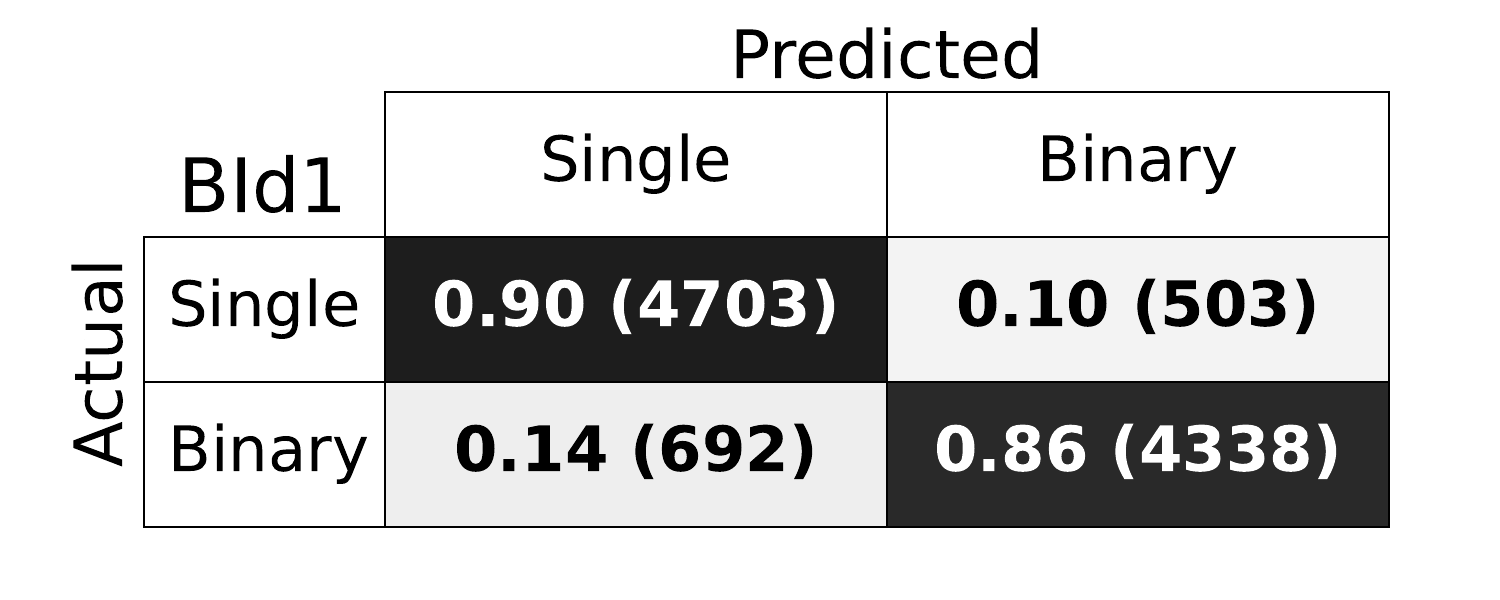}
\includegraphics[width=0.32\linewidth]{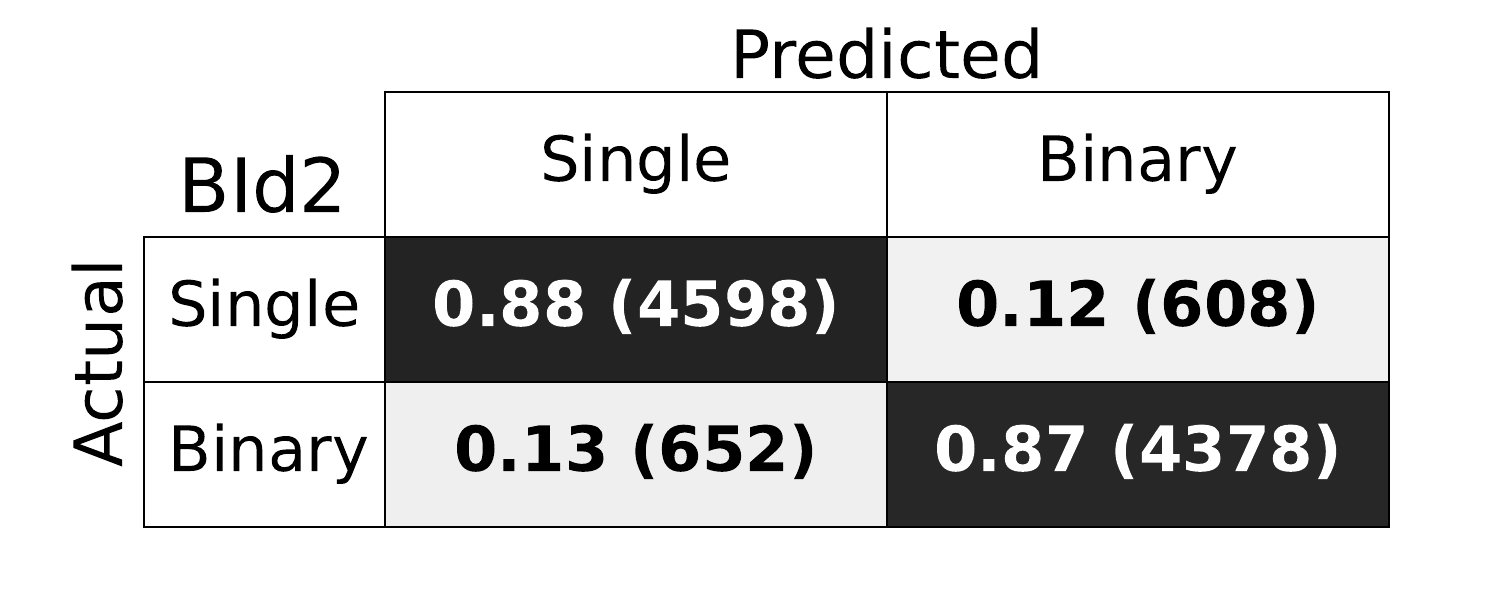}
\includegraphics[width=0.32\linewidth]{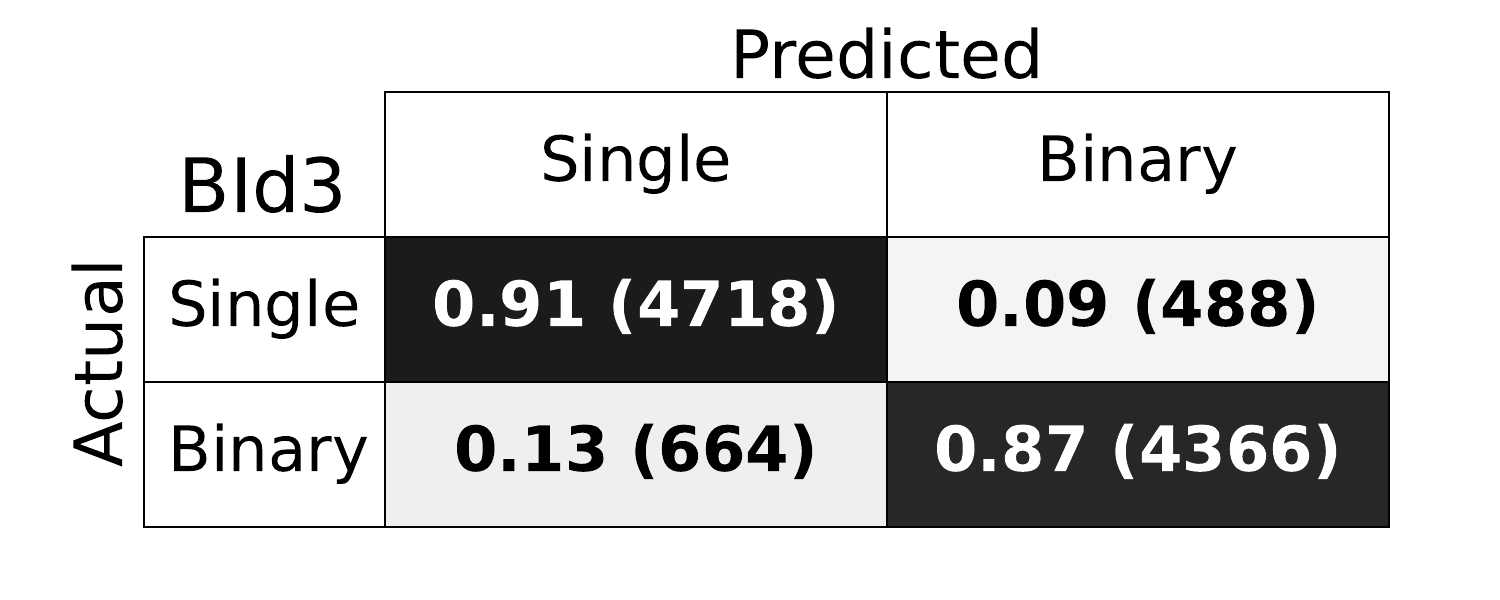} \\
\includegraphics[width=0.32\linewidth]{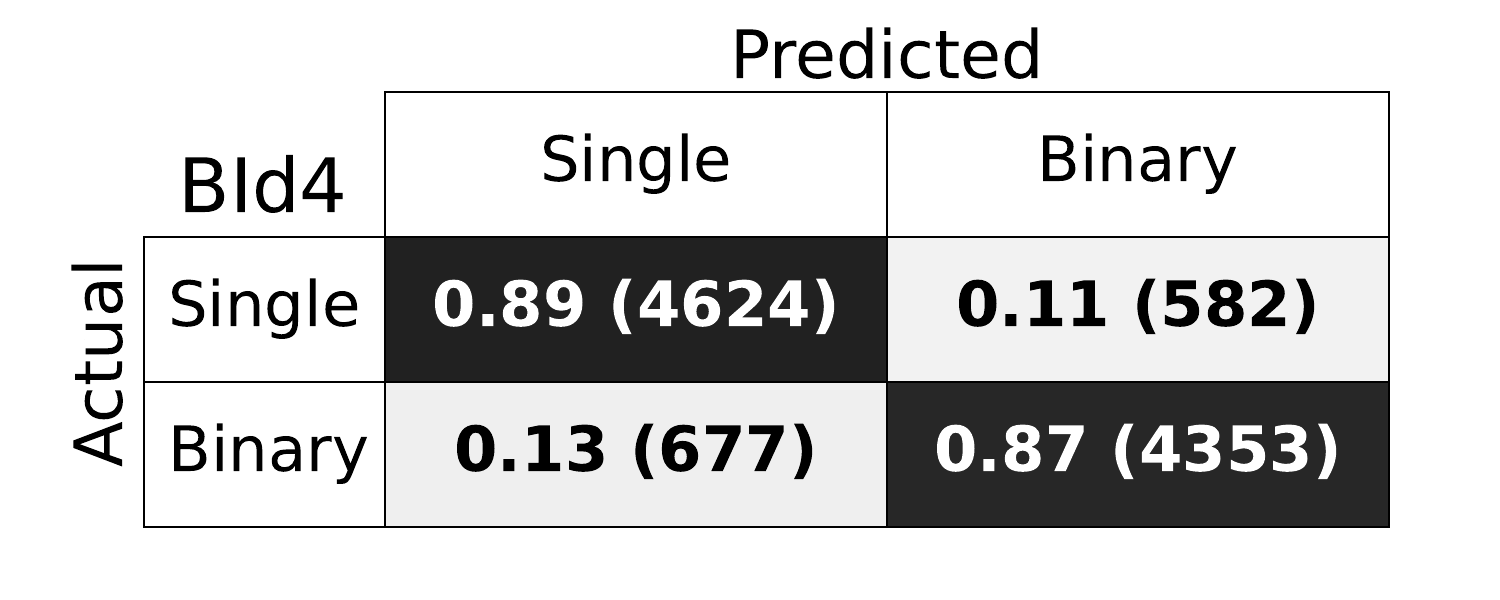}
\includegraphics[width=0.32\linewidth]{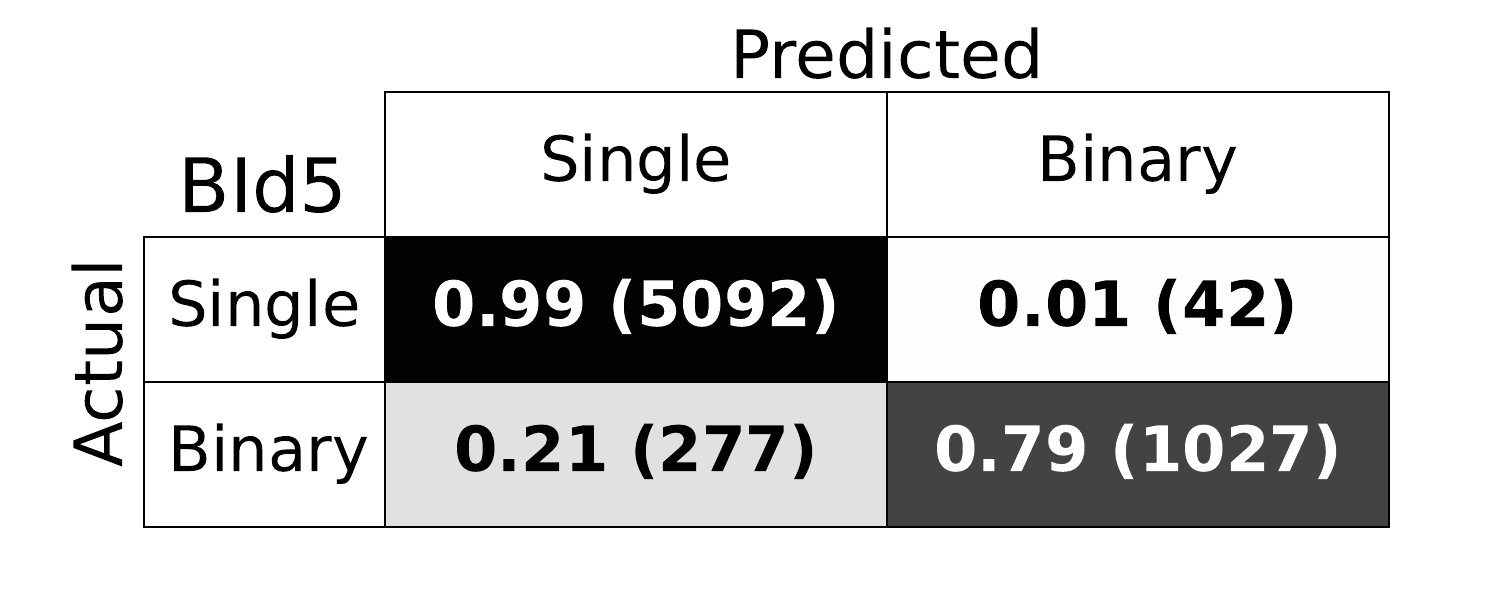}
\includegraphics[width=0.32\linewidth]{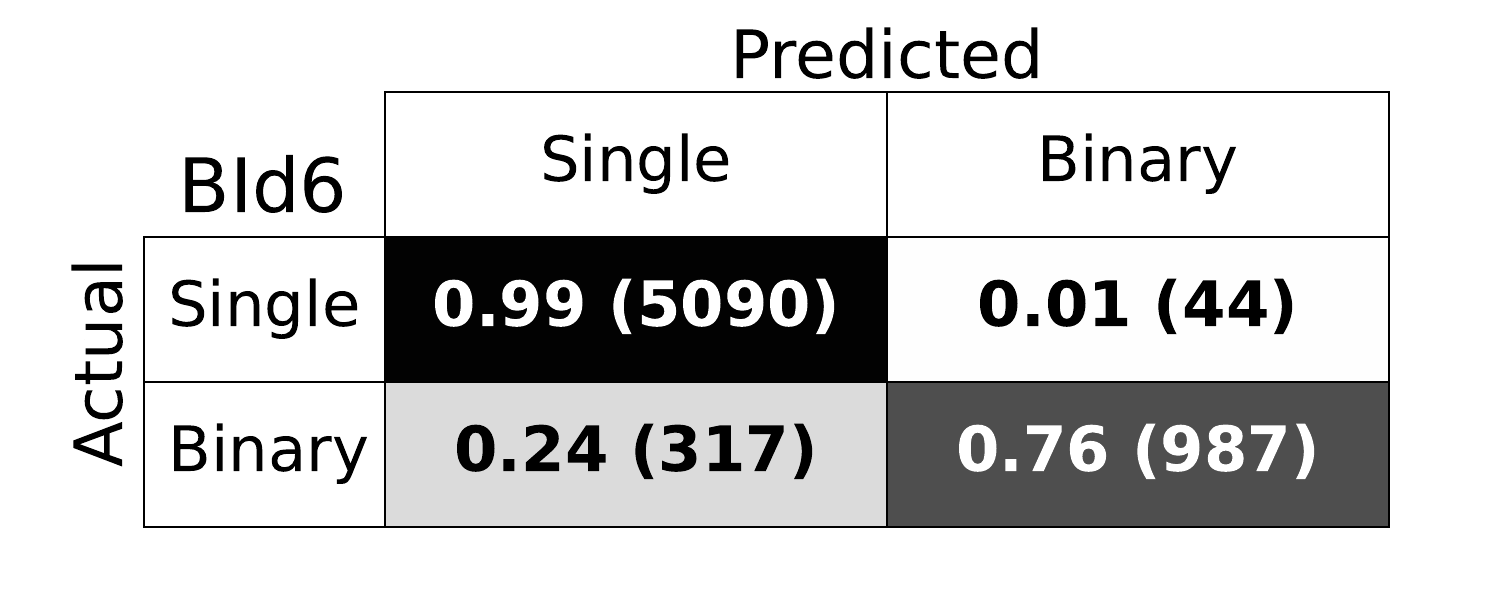} \\
\includegraphics[width=0.32\linewidth]{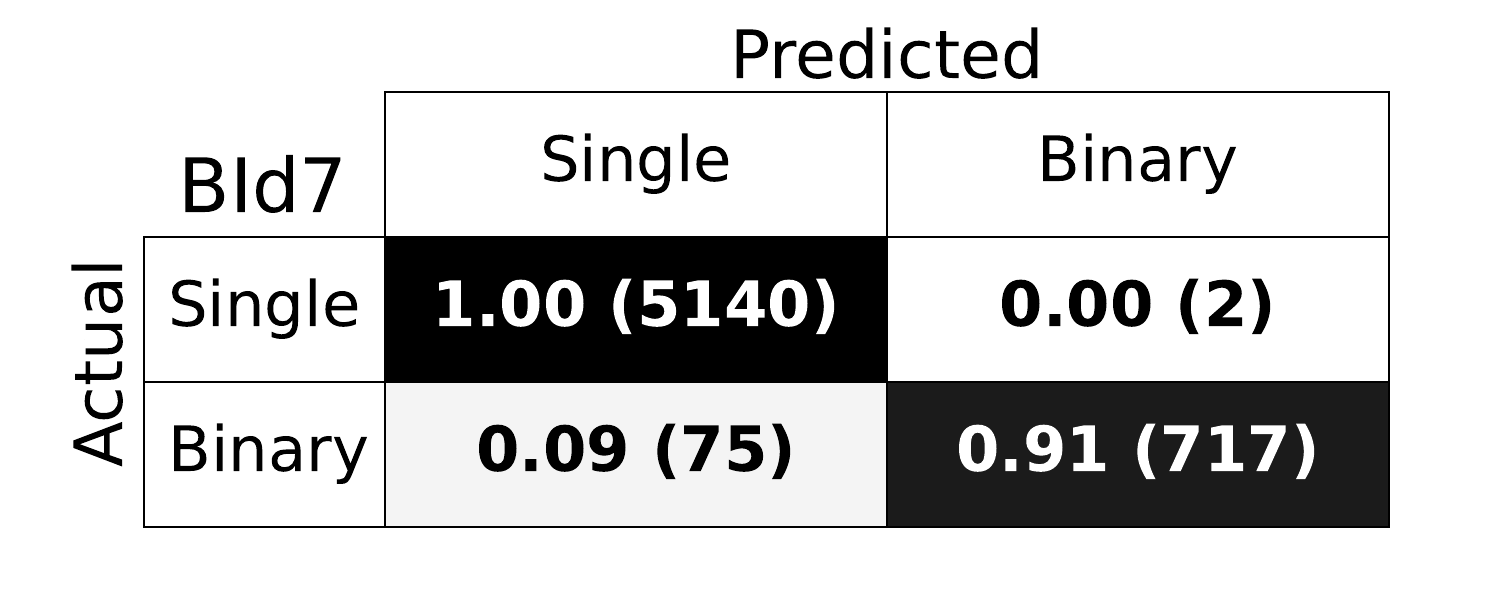}
\includegraphics[width=0.32\linewidth]{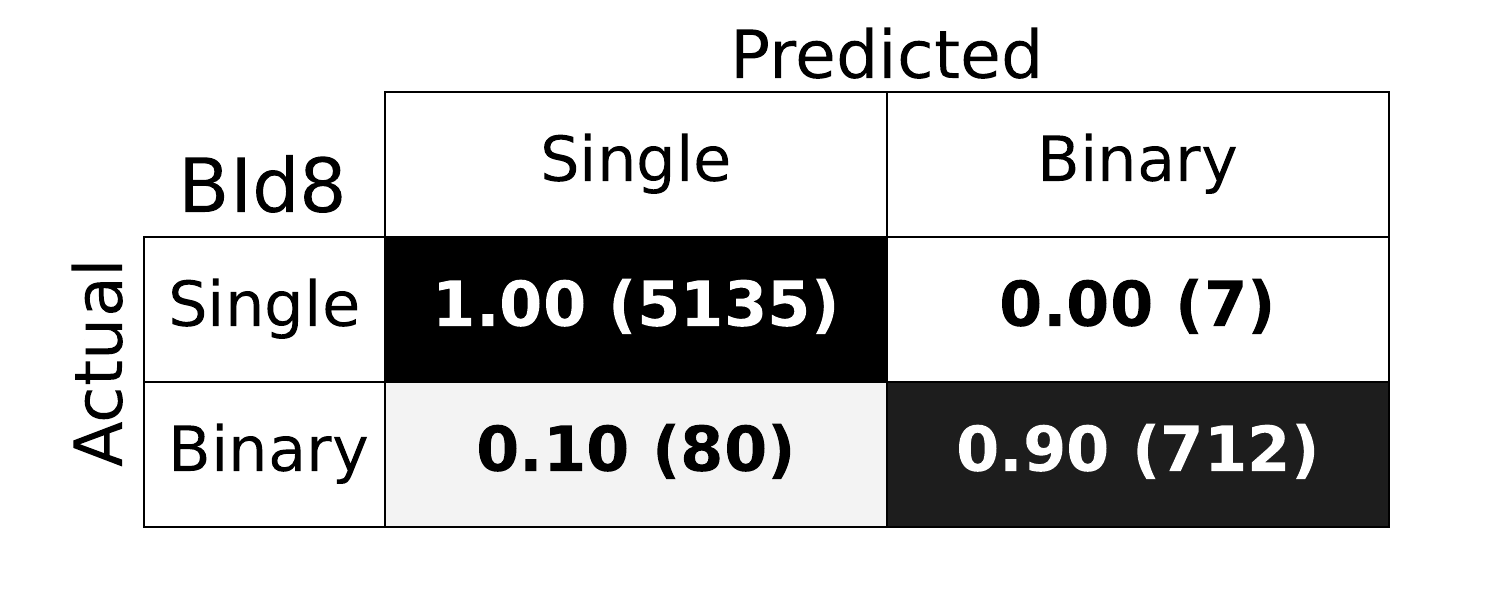}
\includegraphics[width=0.32\linewidth]{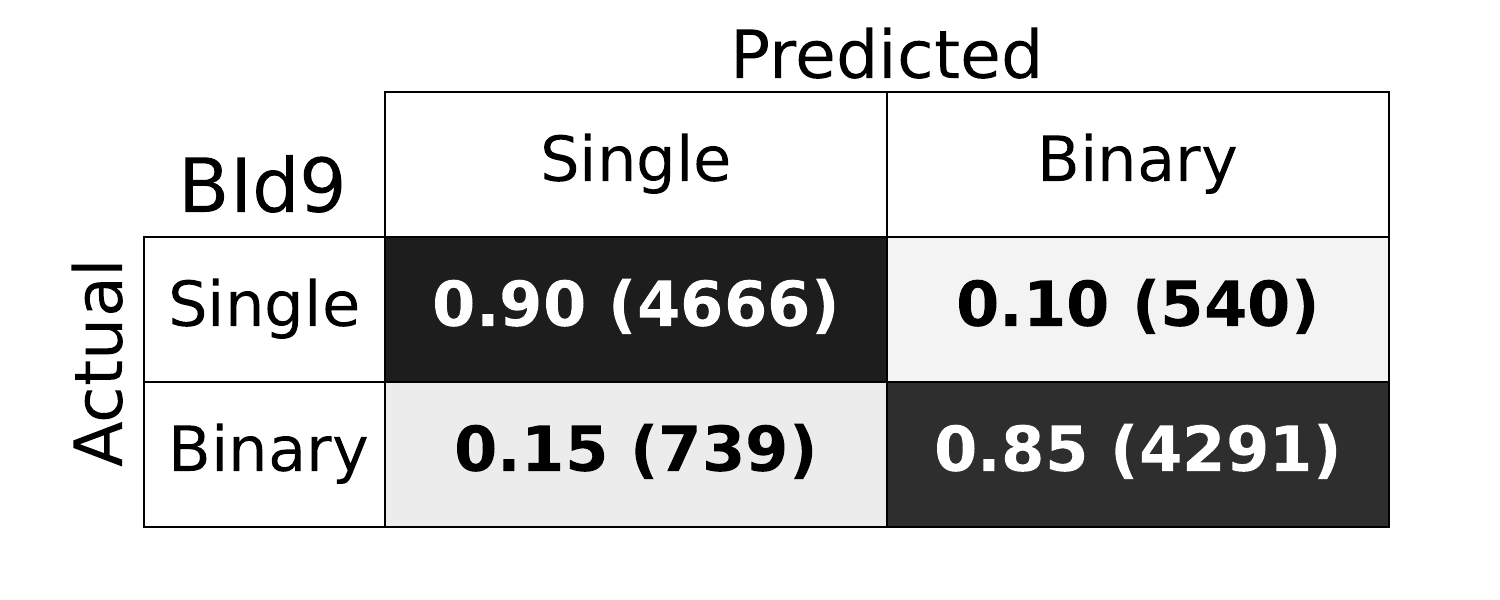} \\
\label{fig:BId_confusion}
\caption{{Confusion matrices for each of the nine binary identification (BId) models examined in this study, comparing actual binary status (rows) with predicted binary status (columns). Numbers in each square panel indicate the fraction and number of templates corresponding to each condition, while the shading indicates the fraction along a row from 0\% (white) to 100\% (black).}}
\end{figure}

Table~\ref{tab:BIdresults} summarizes the cumulative precision, recall, and F1-score performance for each of the {nine} binary identification models and three S/N groups ({27} models total),
{while Figures~\ref{fig:BId_roc} and~\ref{fig:BId_confusion}
display the receiver operating characteristic (ROC; \citealt{peterson1954theory}) curves and confusion matrices for these models.}
All models showed exceptional performance, with 
precisions $\geq$ 0.85,
recall $\geq$ 0.75,
and F1-scores $\geq$ 0.84; the majority of models had performance metrics $>$0.9. 
The {baseline} model, BId1, had a precision, recall, and F1-score of 0.91, 0.86, and 0.88 respectively, averaged across S/N groups.
Notably, performance metrics are inversely correlated with S/N, with the lowest S/N group showing the best performance.
The best overall performance is seen in the BId7 and BId8 models, based on the ``optimal'' spectral compositions for spectral binaries studied in B10, with all three metrics being $\gtrsim$0.9 for the mid- and high-S/N templates.
{The BId9 model, which accounts for variance in component absolute brightness, shows only slightly worsened performance (1--2\%) compared to our baseline model, indicating that absolute brightness variance is not a major factor in binary identification.
The excellent performance of all of the models can be see in the ROC curves, which show that beyond decision tree selection thresholds of 0.5 the models achieve $>$80\% true positive rates while maintaining $<$5\% false positive rates.  The areas under the curve (AUCs), a common performance metric for ROC curves,  are $\geq$0.95 for all but the BId2 model.}

\begin{deluxetable}{lccccccc}
\tablecaption{Precision (P), Recall (R), and F1-score for Binary Identification Models \label{tab:BIdresults}}  
\tabletypesize{\scriptsize} 
\tablehead{ \\
\colhead{Model} & 
\colhead{S/N} &  
\colhead{P} &
\colhead{R} &
\colhead{F1} &
\colhead{\# Singles} & 
\colhead{\# Binaries} & 
\colhead{Unc.}
}
\startdata
BId1 & low & 0.98 & 0.89 & 0.93 & 1827 & 1773 & 0.02 \\
BId1 & mid & 0.86 & 0.82 & 0.84 & 1827 & 1773 & 0.02 \\
BId1 & high & 0.85 & 0.88 & 0.87 & 1552 & 1484 & 0.03 \\
BId2 & low & 0.94 & 0.88 & 0.91 & 1827 & 1773 & 0.02 \\
BId2 & mid & 0.84 & 0.83 & 0.84 & 1827 & 1773 & 0.02 \\
BId2 & high & 0.86 & 0.89 & 0.87 & 1552 & 1484 & 0.03 \\
BId3 & low & 0.99 & 0.89 & 0.93 & 1827 & 1773 & 0.02 \\
BId3 & mid & 0.86 & 0.84 & 0.85 & 1827 & 1773 & 0.02 \\
BId3 & high & 0.85 & 0.88 & 0.87 & 1552 & 1484 & 0.03 \\
BId4 & low & 0.95 & 0.87 & 0.91 & 1827 & 1773 & 0.02 \\
BId4 & mid & 0.84 & 0.83 & 0.84 & 1827 & 1773 & 0.02 \\
BId4 & high & 0.86 & 0.89 & 0.87 & 1552 & 1484 & 0.03 \\
BId5 & low & 0.99 & 0.85 & 0.92 & 1798 & 506 & 0.04 \\
BId5 & mid & 0.94 & 0.71 & 0.81 & 1798 & 506 & 0.04 \\
BId5 & high & 0.91 & 0.75 & 0.82 & 1523 & 427 & 0.05 \\
BId6 & low & 1.00 & 0.84 & 0.91 & 1798 & 506 & 0.04 \\
BId6 & mid & 0.93 & 0.70 & 0.80 & 1798 & 506 & 0.04 \\
BId6 & high & 0.90 & 0.75 & 0.82 & 1523 & 427 & 0.05 \\
BId7 & low & 1.00 & 0.88 & 0.93 & 1804 & 284 & 0.06 \\
BId7 & mid & 1.00 & 0.89 & 0.94 & 1804 & 284 & 0.06 \\
BId7 & high & 1.00 & 0.96 & 0.98 & 1534 & 224 & 0.07 \\
BId8 & low & 0.98 & 0.83 & 0.90 & 1804 & 284 & 0.06 \\
BId8 & mid & 1.00 & 0.91 & 0.95 & 1804 & 284 & 0.06 \\
BId8 & high & 0.99 & 0.97 & 0.98 & 1534 & 224 & 0.07 \\
{BId9} & low & 0.98 & 0.87 & 0.93 & 1827 & 1773 & 0.02 \\
{BId9} & mid & 0.85 & 0.82 & 0.83 & 1827 & 1773 & 0.02 \\
{BId9} & high & 0.84 & 0.87 & 0.85 & 1827 & 1773 & 0.02 \\
\enddata
\tablecomments{Uncertainty (Unc.) on statistics is based on the Poisson {counting} uncertainty of the smaller of the {number of} synthetic single and binary templates.}
\end{deluxetable}

\begin{figure}[th]
\plotone{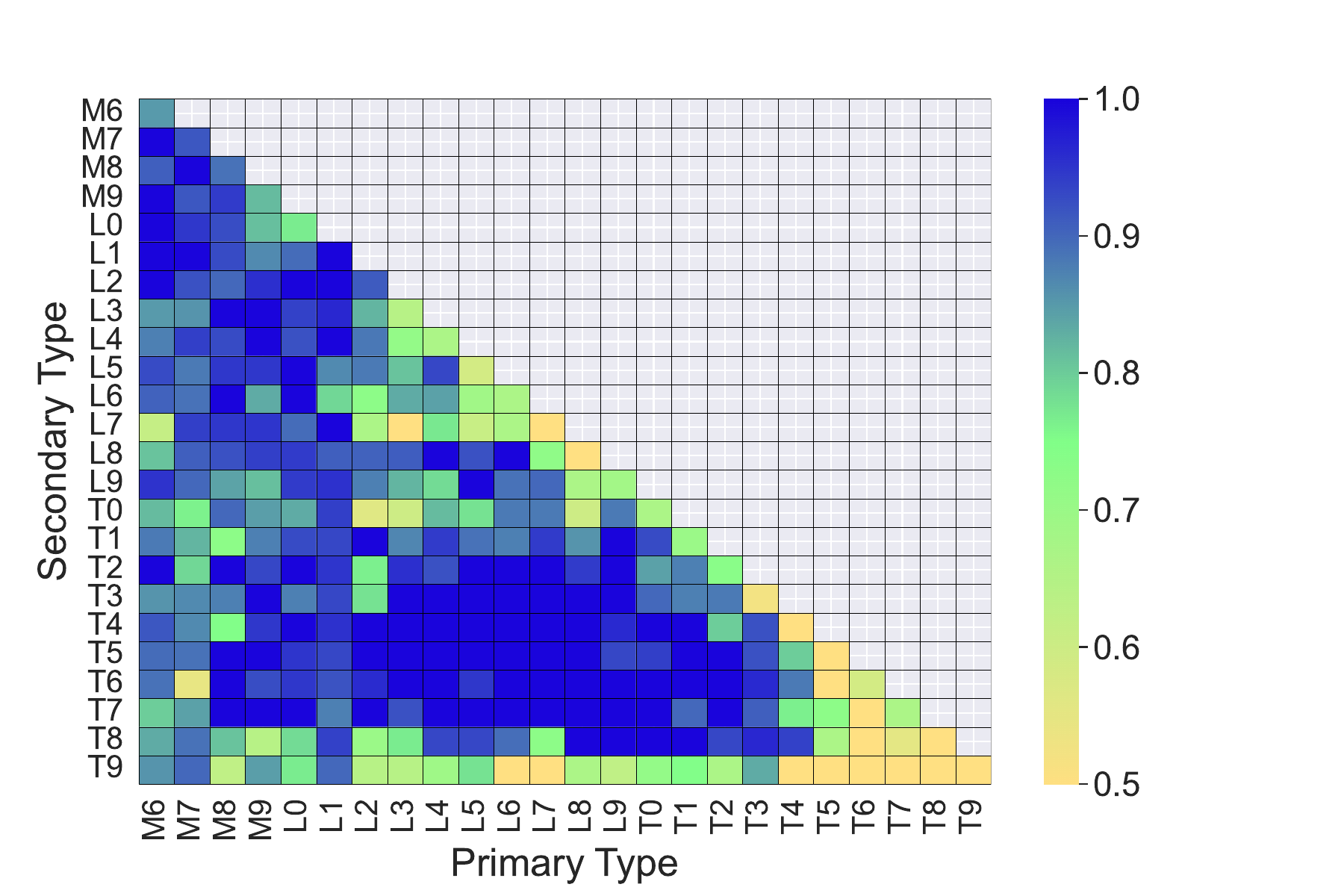}
\caption{Recall for the BId1 model for all spectral type combinations used for our synthetic binaries. Values represent the cumulative recall across all S/N ranges.
Note that the color scale is narrowed to span values of 0.5--1.0.}
\label{fig:BId1_heatmap}
\end{figure}

Figure~\ref{fig:BId1_heatmap} displays the recall performance of the BId1 binary identification model decomposed across all combinations of primary (P) and secondary (S) component types. 
The 
B10 range (P $\in$ \{L5,T2\}, S $\in$ \{T2,T7\})
have recall values close to unity, while binaries with equal component types (diagonal axis), those with a T8-T9 secondary, or those with T5 or later primaries have the smallest recall values of $\lesssim$0.5. These regions encompass component compositions where spectral peculiarities are minimal in the combined light spectrum. Notably, these metrics are both superior to and cover a broader range of spectral type combinations than the index-based methods of B10 and B14 (Figure~\ref{fig:index_heatmap}). More importantly,
the speed of classification is more than an order of magnitude faster than the index method. The test sample was classified at a rate of 441 spectra/second, or 2~ms per spectrum, compared to 39 ms per spectrum for the index methods.

\begin{figure*}[th]
\centering
\includegraphics[width=0.45\textwidth]{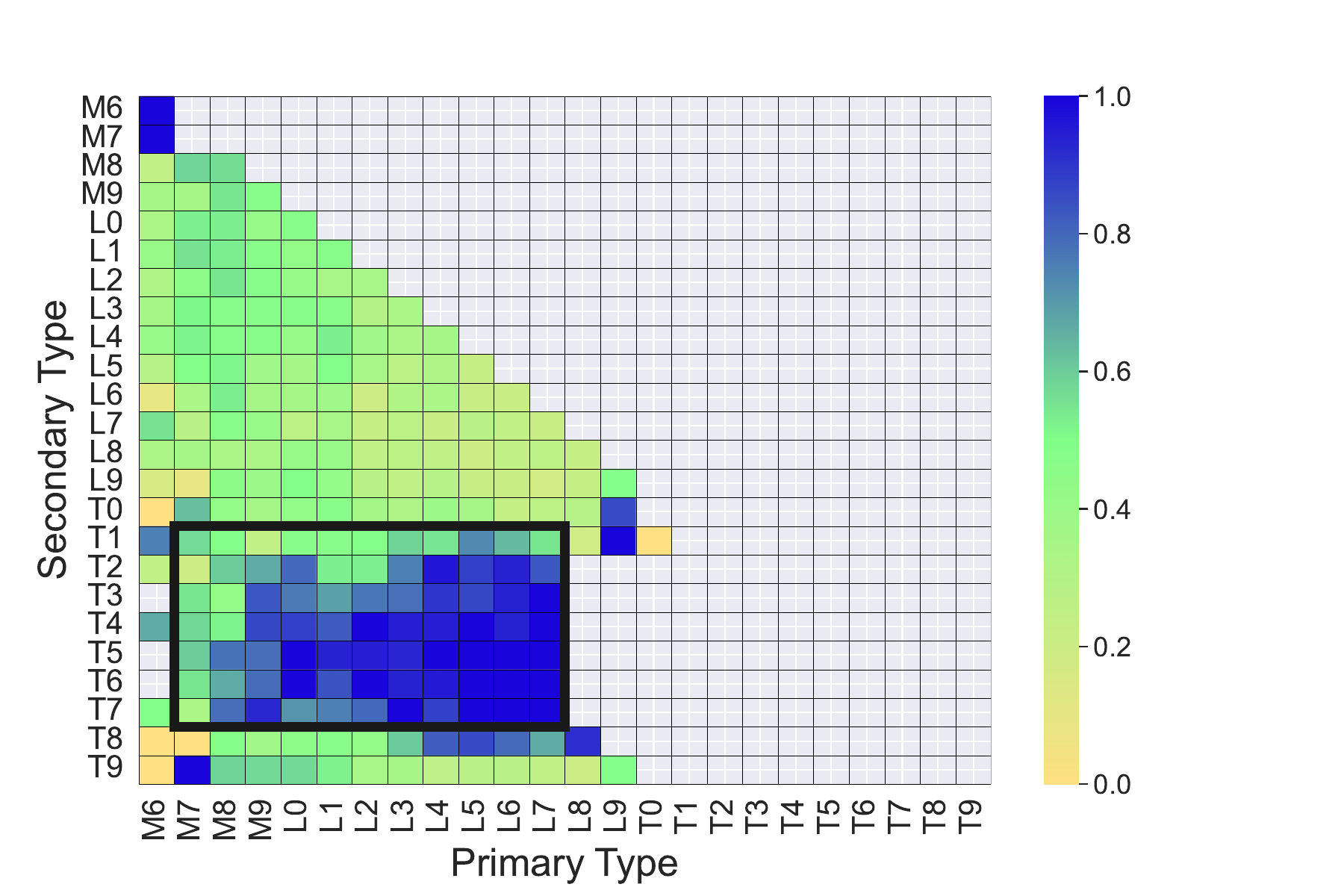}
\includegraphics[width=0.45\textwidth]{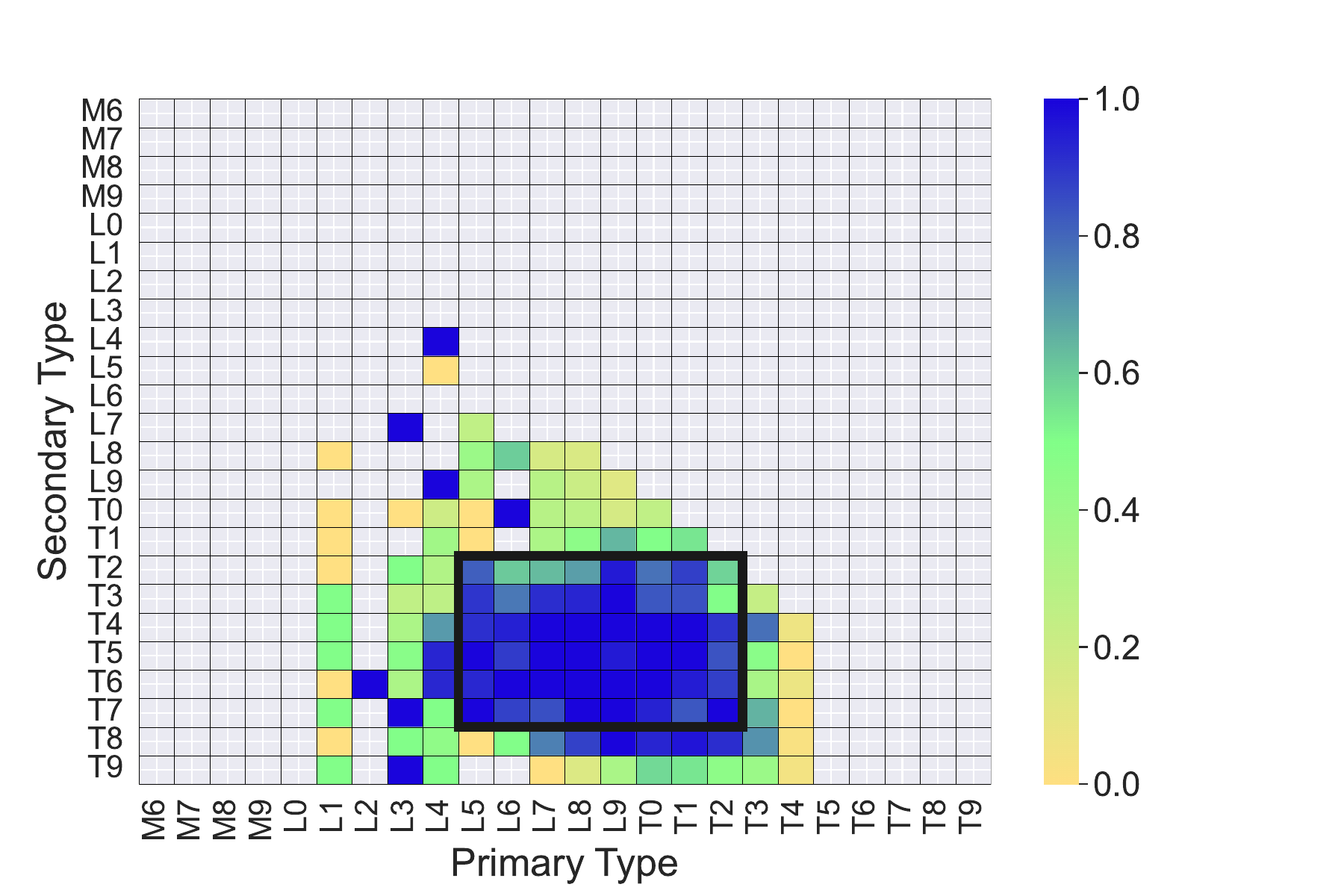}
\caption{Recall for the BId5 model (left) and the BId7 model (right) for all spectral type combinations used for our synthetic binaries. Values represent the cumulative recall across all S/N ranges. Note that the color scale spans values of 0.0--1.0, broader than the range shown in Figure~\ref{fig:BId1_heatmap}.}
\label{fig:BId5_BId7}
\end{figure*}

We conducted a similar analysis on our alternative models. 
Incorporating the difference flux (BId2 and BId4) or masking the telluric absorption regions (BId3 and BId4) did not significantly improve model recall over our baseline model BId1, 
even when broken down by component spectral types.
Models trained on templates encompassing the spectral type ranges used in the B10 and B14 studies (BId5--8) clearly outperform the index-based methods. The BId5 model shows a nearly factor of 4 improvement in recall as compared to the B14 index method, 0.60 versus 0.16 averaged over all relevant combinations and S/N values;
while the BId7 model had modestly higher recall compared to the B10 index method, 0.59 versus 0.56 (Figure~\ref{fig:BId5_BId7}).
However, these specialized models still underperformed relative to 
the BId1 model, even within their respective training ranges. Regions with high recall in the specialized models were already well-classified by BId1, resulting in no substantial gain in performance.

\begin{figure*}[th]
\centering
\includegraphics[width=0.65\textwidth]{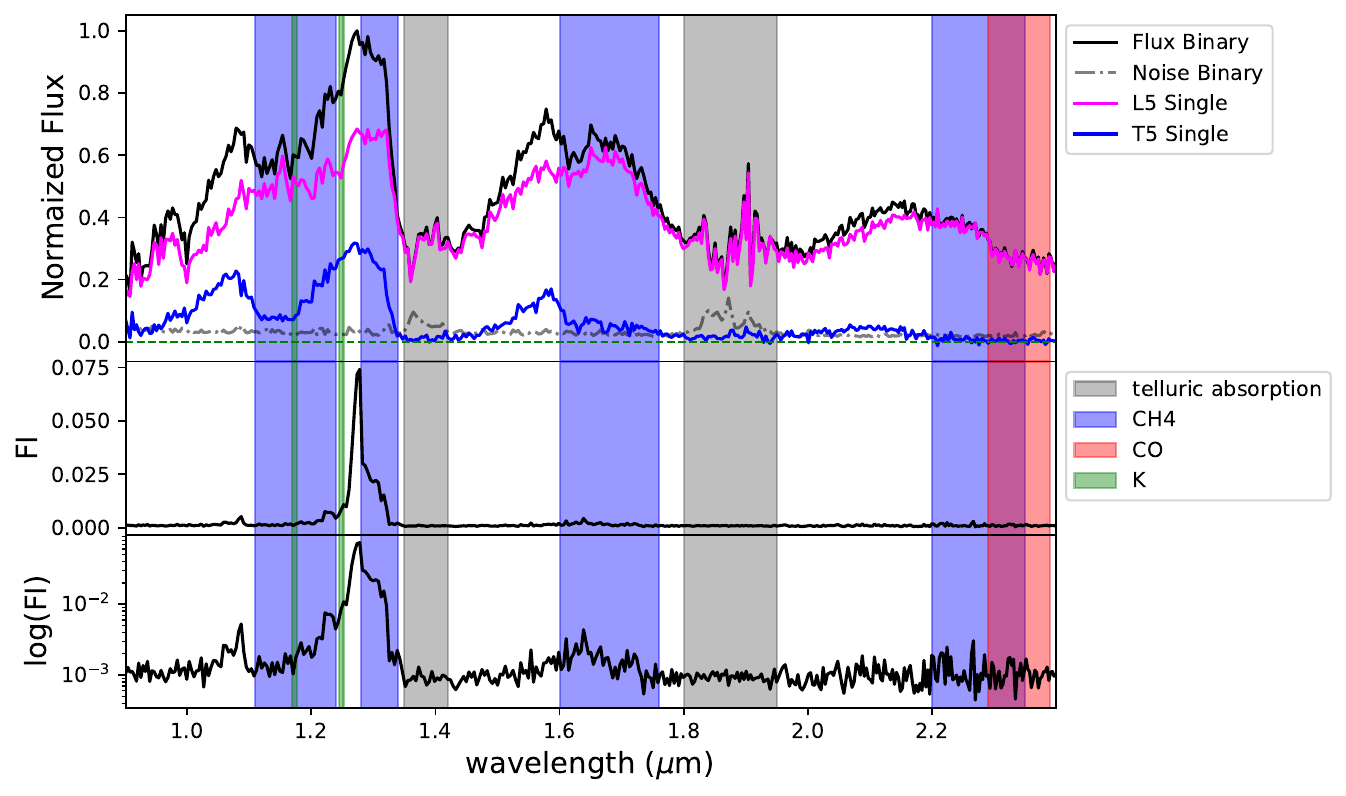}
\includegraphics[width=0.65\textwidth]{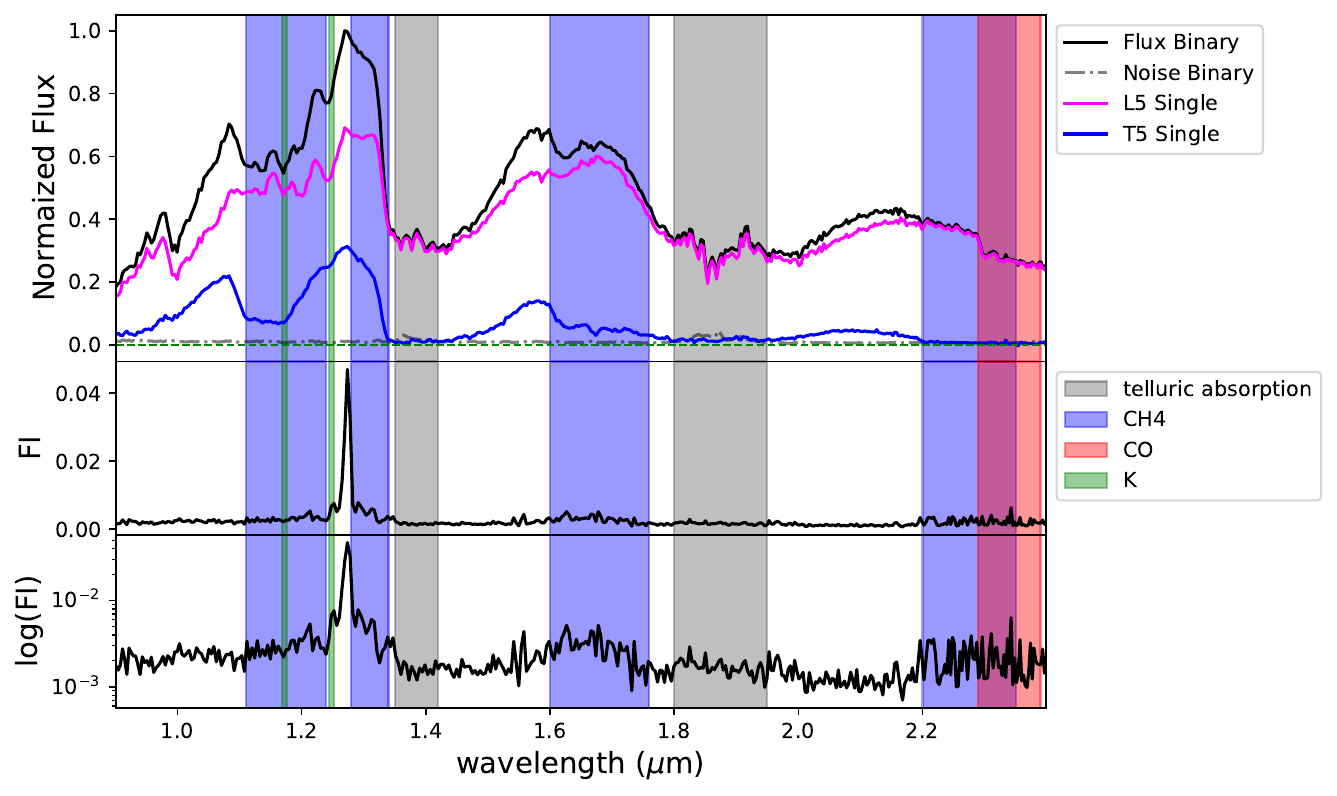}
\includegraphics[width=0.65\textwidth]{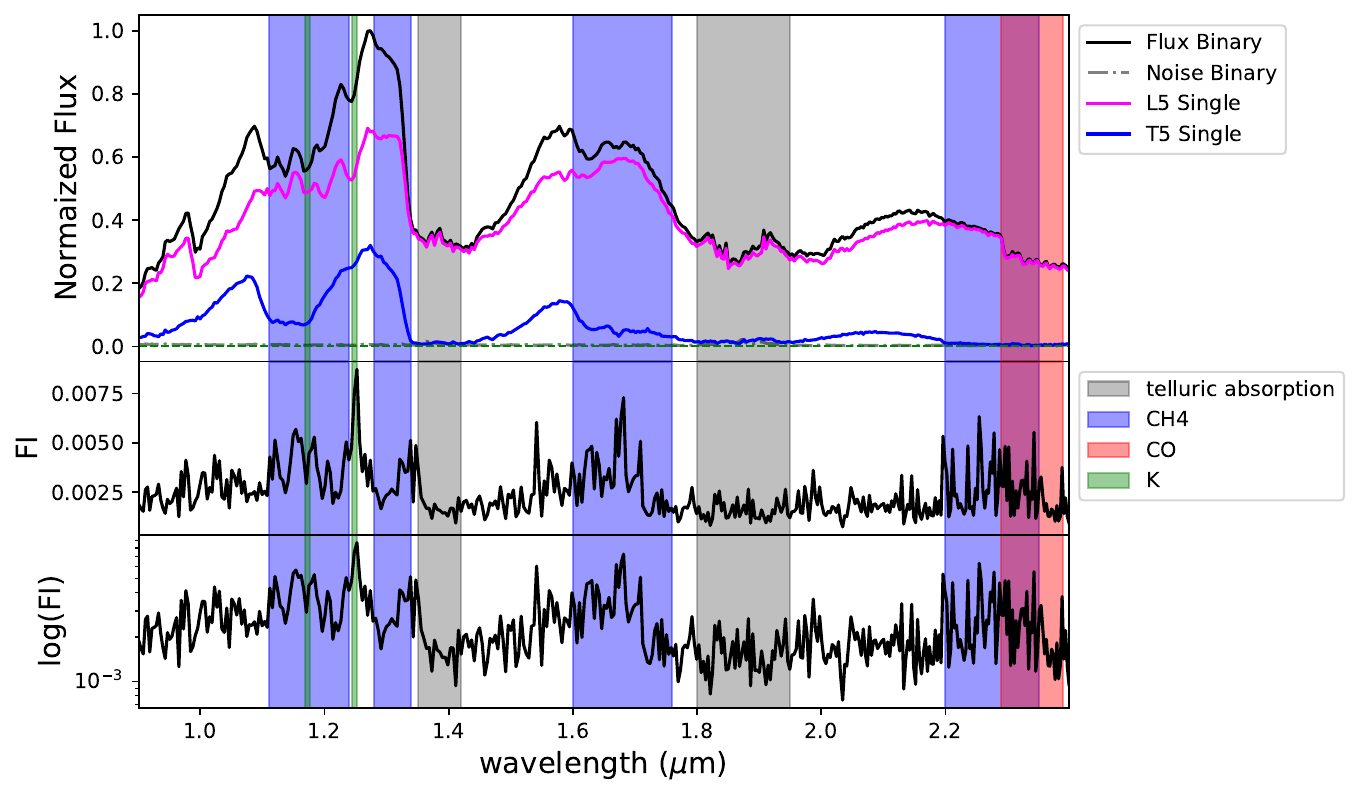}
\caption{Feature importance (FI) for the BId1 model for low (top), mid (middle) and high (bottom) S/N groups.
Each panel displays FI on linear (middle plot) and logarithmic scales (bottom plot), compared to an example of an L5 plus T5 binary spectrum 
with a S/N in the associated range (top plot).
Regions associated with telluric (grey), K~I (green), CH$_4$ (blue) and CO (red) absorption are indicated by vertical bands.}
\label{fig:FIRF}
\end{figure*}

Figure~\ref{fig:FIRF} displays feature importance as a function of wavelength on both linear and log scales for the BId1 model across the three S/N categories, compared to an L5 plus T5 binary combination to associate FI peaks to spectral {features}.
For the low S/N case, FI has a strong peak at 1.274~$\mu$m, extending around the $J$-band flux peak for T dwarfs.
Indeed, the logarithmic scaling of FI  
resembles the spectral structure of a T dwarf. 
We interpret this structure as the RF algorithm focusing on the emission features of T dwarf companions, 
as the S/N is too low to clearly distinguish absorption features from a faint companion.
In contrast, 
the high S/N FI shows more structure across the spectral band, and is more tuned to the 
absorption features present in L and T dwarf spectra, including the K~I lines 
at 1.25~$\mu$m, CH$_4$ molecular structure at 1.67~$\mu$m and 2.25~$\mu$m, and the CO band at 2.32~$\mu$m.
{The presence of significant structure in the FI distribution at the wavelengths of these features indicates that the binary model may be more selective of systems with late-L and T dwarf secondaries, likely due to their distinct spectral features as compared to late-M and early-L dwarfs.}

\subsection{Component Classification} \label{sec:results-class}

Table~\ref{tab:BClassresults} summarizes the performance of our binary component classification model, quantified by the difference ($\Delta$) and standard deviation ($\sigma_\Delta$) between the input and output primary and secondary classifications, averaged across the test data. We also provide estimates of the uncertainties on these values assuming Poisson {counting} statistics {based on} the number of {synthetic} binary spectra. 
The classification of all 25,684 binaries in the test sample took 162~s, or 6 ms per spectrum.

For all of the model and S/N groups, we find excellent agreement among the primary classifications, with absolute median classification errors of $\lesssim$0.1~subtypes and typical standard deviations of $\lesssim$0.5~subtypes. We also find small median classification errors for the secondary types, but higher standard deviations of $\lesssim$1~subtype.
In nearly all cases, performance improves for the higher S/N groups, particularly for primary classification.
We see minimal improvement in performance when difference flux or telluric masking is included (BClass2-4),
indicating that neither contributes significantly to determining component types.
However, there are notable improvements in performance for
models based on the B10 and B14 subtype pairings (BClass5 and BClass7), for which standard deviations of secondary classifications decrease by a factor of two to $\lesssim$0.5 subtypes. 

\begin{figure*}[th]
\centering
\includegraphics[width=0.46\textwidth]{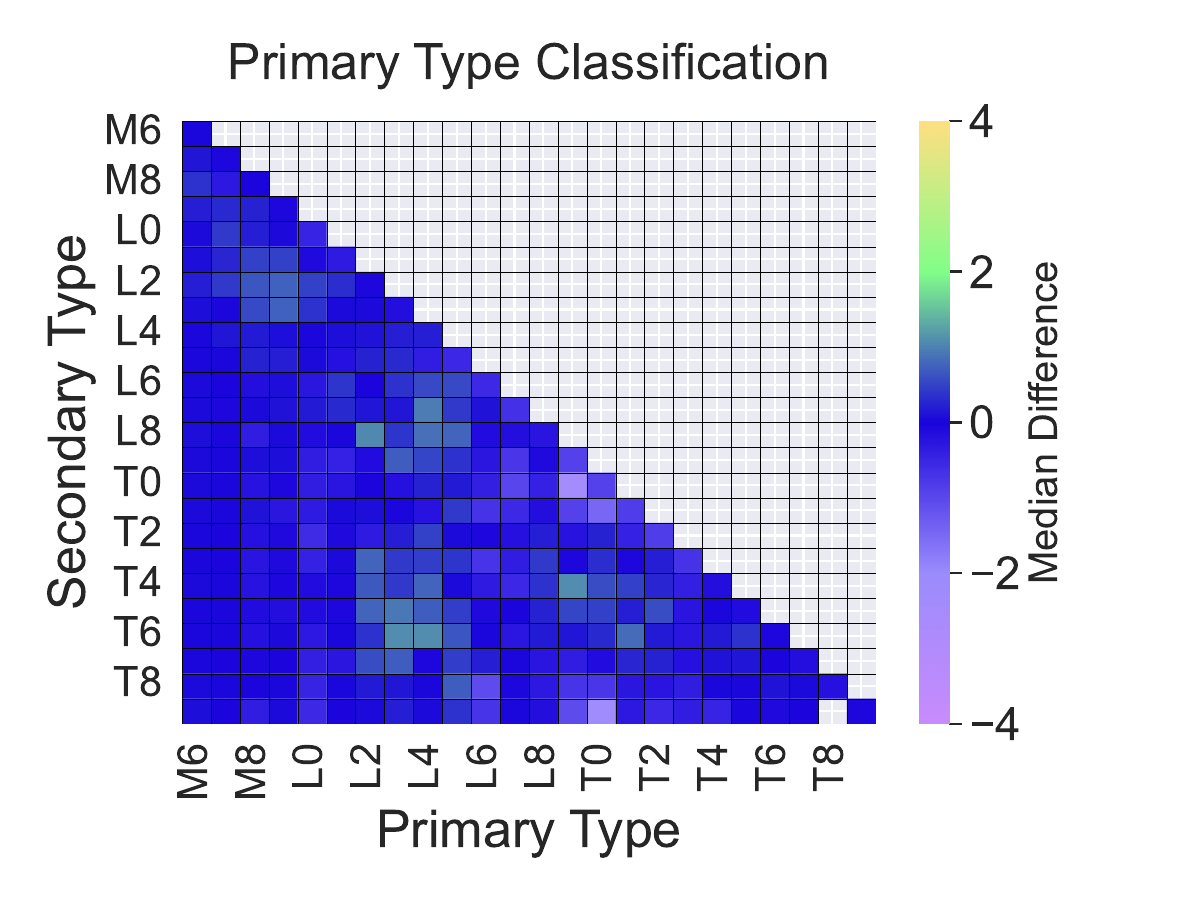}
\includegraphics[width=0.46\textwidth]{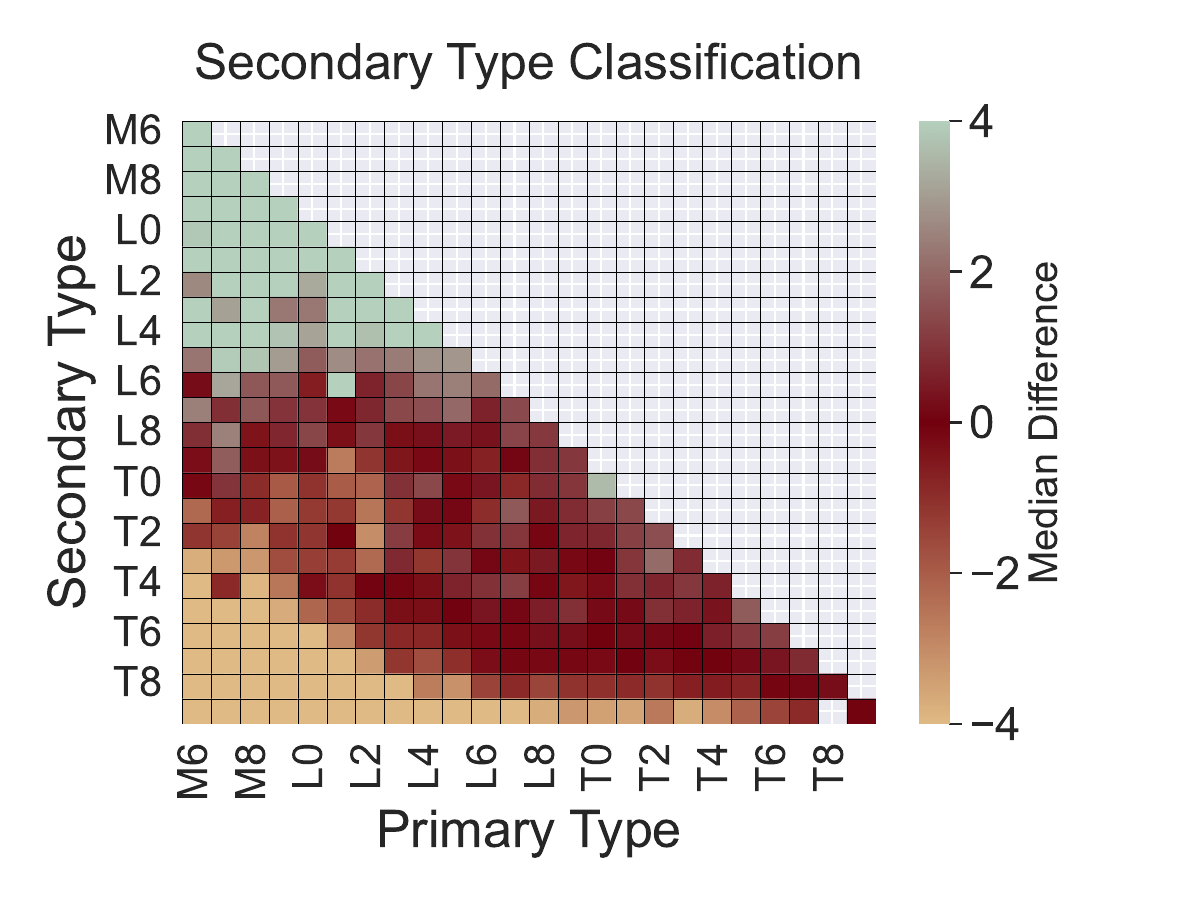}
\\
\includegraphics[width=0.46\textwidth]{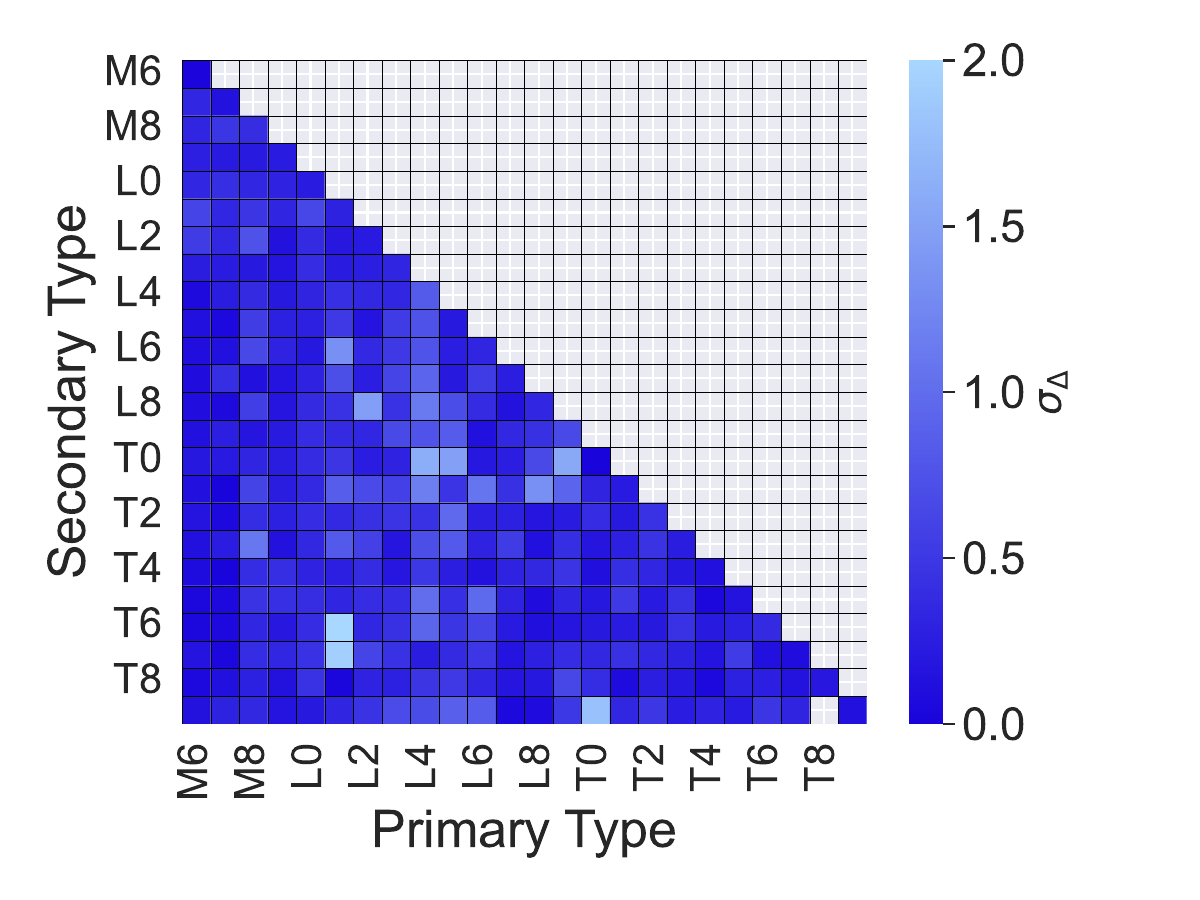}
\includegraphics[width=0.46\textwidth]{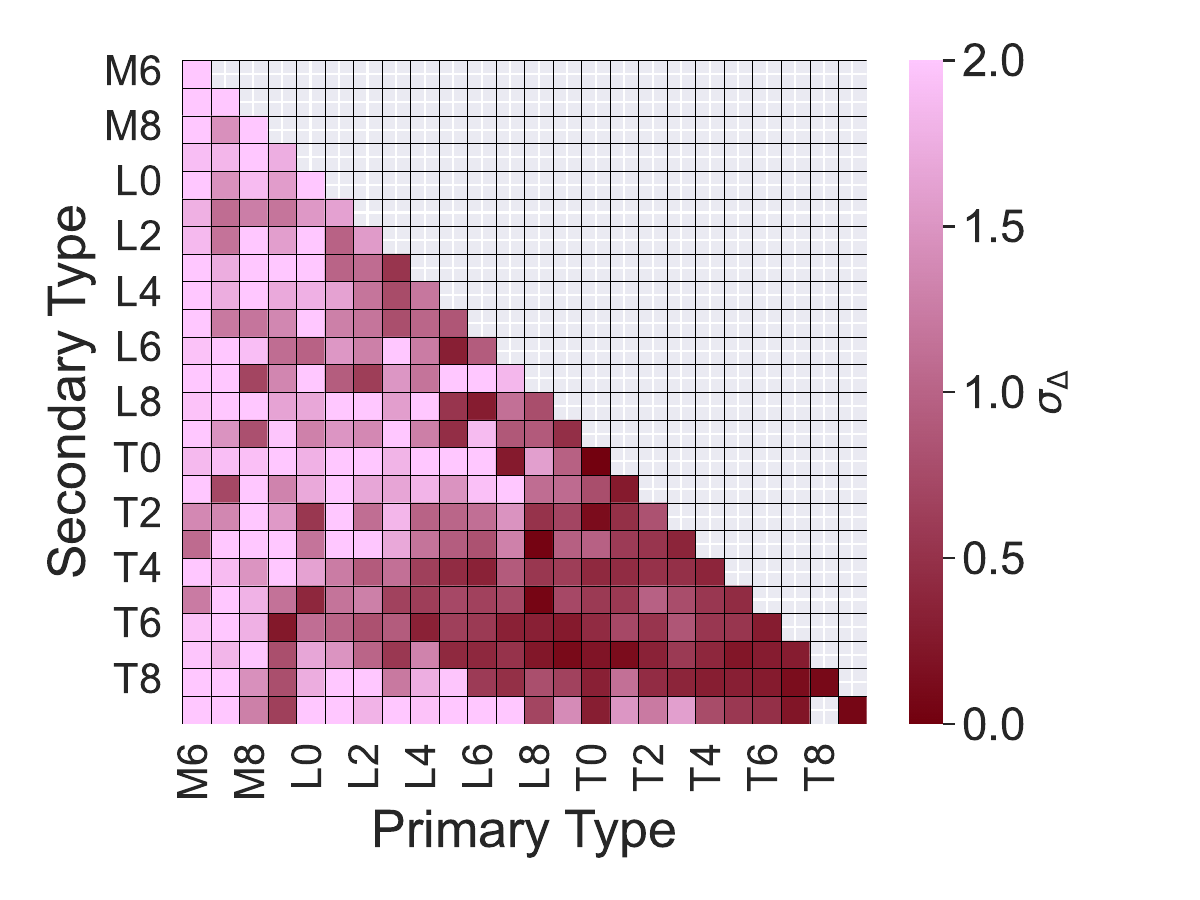}
\\
\includegraphics[width=0.46\textwidth]{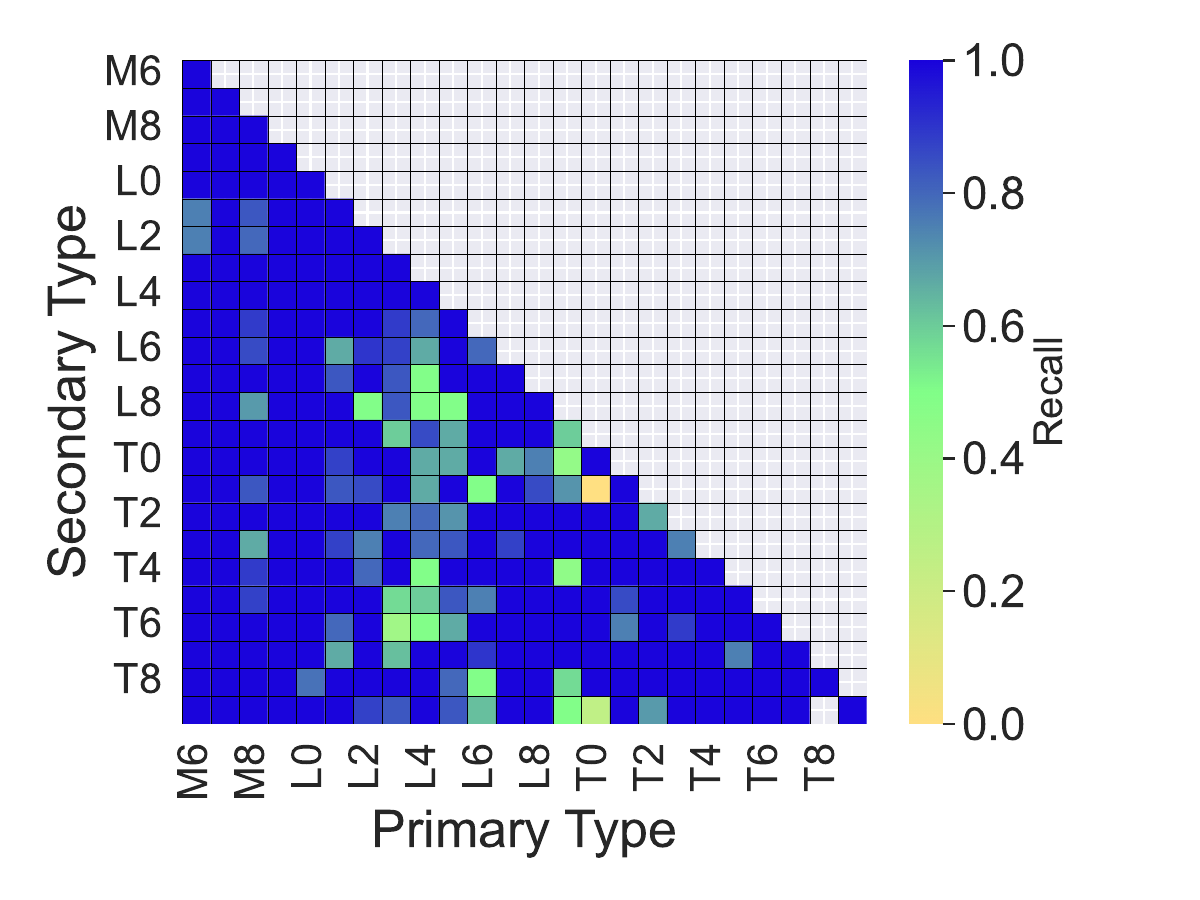}
\includegraphics[width=0.46\textwidth]{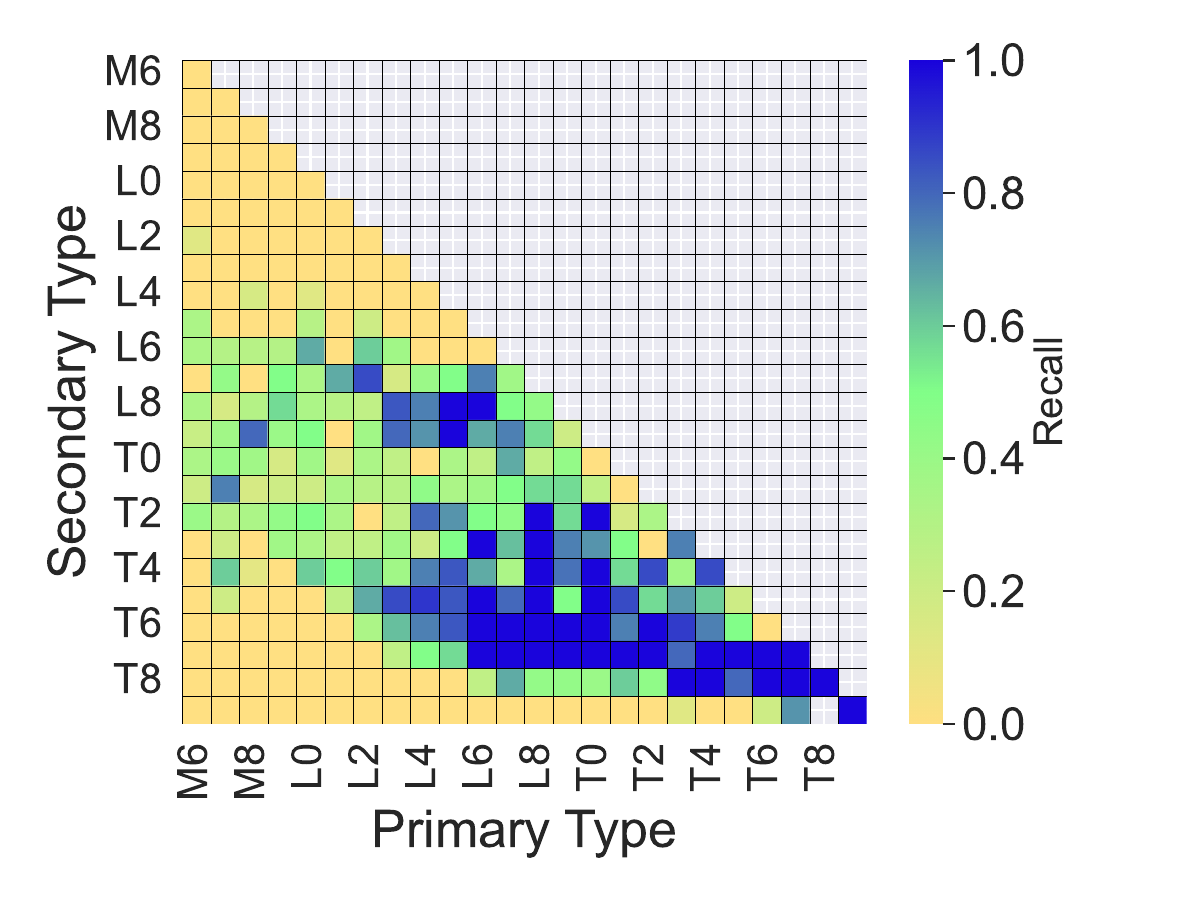}
\caption{Median classification error (top panels), standard deviations (middle panels), and recall (bottom panels) for primary (left) and secondary (right) classification
as a function of component type combination
for the BClass1 model and mid-S/N group.
Classification differences were calculated as predicted minus true. 
Recall is defined here as the percentage of sources classified within 1 subtype of their true values.
}
\label{fig:BClass1corner}
\end{figure*}

Figure~\ref{fig:BClass1corner} displays the distribution of median classification errors, standard deviations, and recall as a function of paired component types for the BClass1 model. 
Here, recall is defined as inferred spectral types {being} within one subtype of the input classification.
Primary classifications are nearly perfect for all spectral type combinations, with modest deviations ($\approx$1~subtype) for 
systems with mid-type L dwarf and L/T transition primaries.
The performance for secondary classification shows greater scatter, with significant skew (by up to $\sim$4~subtypes) toward later secondary types, 
likely due to the minimal contribution of the secondaries to the combined light spectrum.
A similar large skew is seen for late-M/L dwarf binaries with similar component classifications, likely due to similarities in spectral shape. 
We also see significant variance ($\sigma_\Delta\sim2$~subtypes) among secondary classifications for systems with late-M or early-L primaries.
On the other hand, a wide range of systems with T dwarf secondaries are accurately classified, with secondary classification recall {$\gtrsim$0.8} and scatter $<$1~subtype.

\begin{deluxetable}{lccccccc}
\tablecaption{Classification Errors and Standard Deviations for Binary Classification Models \label{tab:BClassresults}}  
\tabletypesize{\footnotesize} 
\tablehead{
& &
\multicolumn{2}{c}{Primary} &
\multicolumn{2}{c}{Secondary} &
\\
\cline{3-4} \cline{5-6}
\colhead{Model} & 
\colhead{S/N} &  
\colhead{$\Delta$} &
\colhead{$\sigma_\Delta$} &
\colhead{$\Delta$} &
\colhead{$\sigma_\Delta$} &
\colhead{\# Binaries} &
\colhead{Unc.} 
}
\startdata
BClass1 & low & $-$0.14 & 0.73 & $-$0.10 & 1.00 & 1800 & 0.02 \\
BClass1 & mid & $-$0.03 & 0.35 & +0.01 & 1.00 & 1800 & 0.02 \\
BClass1 & high & $-$0.02 & 0.27 & $-$0.07 & 0.93 & 1518 & 0.03 \\
BClass2 & low & $-$0.14 & 0.74 & $-$0.14 & 1.10 & 1800 & 0.02 \\
BClass2 & mid & $-$0.03 & 0.35 & $-$0.00 & 1.00 & 1800 & 0.02 \\
BClass2 & high & $-$0.03 & 0.28 & $-$0.05 & 0.90 & 1518 & 0.03 \\
BClass3 & low & $-$0.14 & 0.74 & $-$0.09 & 1.00 & 1800 & 0.02 \\
BClass3 & mid & $-$0.03 & 0.35 & $-$0.01 & 1.00 & 1800 & 0.02 \\
BClass3 & high & $-$0.03 & 0.28 & $-$0.07 & 0.92 & 1518 & 0.03 \\
BClass4 & low & $-$0.14 & 0.73 & $-$0.15 & 1.10 & 1800 & 0.02 \\
BClass4 & mid & $-$0.04 & 0.35 & $-$0.02 & 1.00 & 1800 & 0.02 \\
BClass4 & high & $-$0.04 & 0.29 & $-$0.05 & 0.91 & 1518 & 0.03 \\
BClass5 & low & $-$0.10 & 0.70 & +0.05 & 0.59 & 462 & 0.05 \\
BClass5 & mid & $-$0.02 & 0.35 & $-$0.03 & 0.52 & 462 & 0.05 \\
BClass5 & high & $-$0.02 & 0.34 & +0.07 & 0.45 & 396 & 0.05 \\
BClass6 & low & $-$0.08 & 0.70 & +0.02 & 0.59 & 462 & 0.05 \\
BClass6 & mid & $-$0.02 & 0.33 & $-$0.04 & 0.56 & 462 & 0.05 \\
BClass6 & high & $-$0.01 & 0.32 & +0.04 & 0.45 & 396 & 0.05 \\
BClass7 & low & $-$0.01 & 0.44 & $-$0.03 & 0.58 & 288 & 0.06 \\
BClass7 & mid & +0.01 & 0.34 & $-$0.03 & 0.51 & 288 & 0.06 \\
BClass7 & high & $-$0.00 & 0.21 & $-$0.00 & 0.23 & 240 & 0.06 \\
BClass8 & low & $-$0.03 & 0.43 & $-$0.01 & 0.59 & 288 & 0.06 \\
BClass8 & mid & +0.011 & 0.32 & $-$0.01 & 0.50 & 288 & 0.06 \\
BClass8 & high & $-$0.03 & 0.21 & +0.01 & 0.22 & 240 & 0.06 \\
\enddata
\tablecomments{Median error $\Delta$ is defined as true classification minus predicted classification. Both classification errors and standard deviations 
($\sigma_\Delta$)
are computed across all spectral type combinations within the template spectral type range for each model. 
Uncertainties (Unc.) on these statistics are based on Poisson {counting} uncertainty of the {number of binary spectra in the} sample.}
\end{deluxetable}

As with the binary identification models, incorporating difference spectra or masking telluric regions had no significant impact on the performance of primary or secondary classification compared to our baseline BClass1 model.
The B10- and B14-based models (BClass5-8) have similar measures of error and scatter as BClass1 for primary classifications, but yield a 2--4$\times$ improvement in the classification of the secondary component 
($\sigma_\Delta$ = 0.5--0.6 {subtypes} for BClass5 and BClass6, 
$\sigma_\Delta$ = 0.2--0.6 {subtypes} for BClass7 and BClass8), 
likely due to the prevalence of binaries with distinct component spectral types among the templates in the training and testing data. 

\begin{figure*}[th]
\centering
\includegraphics[width=0.65\textwidth]{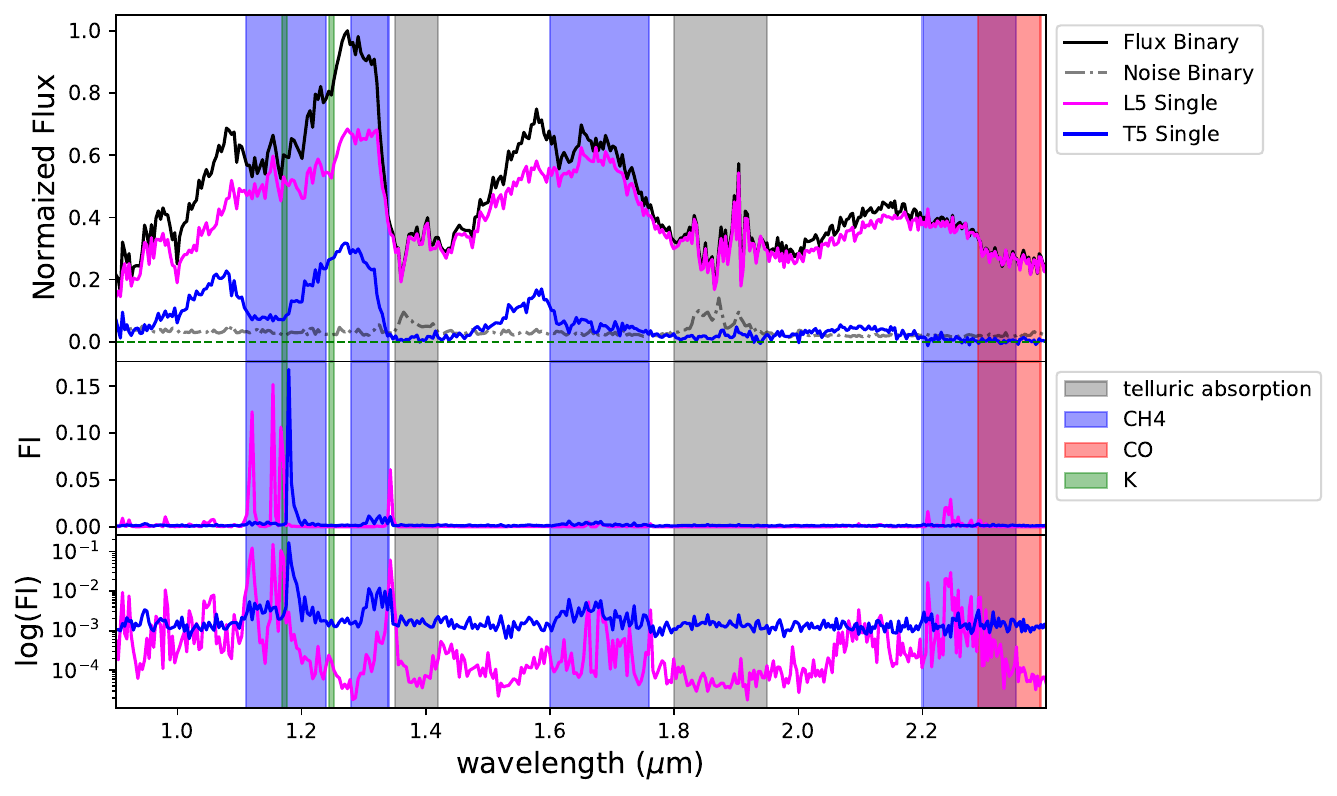}
\includegraphics[width=0.65\textwidth]{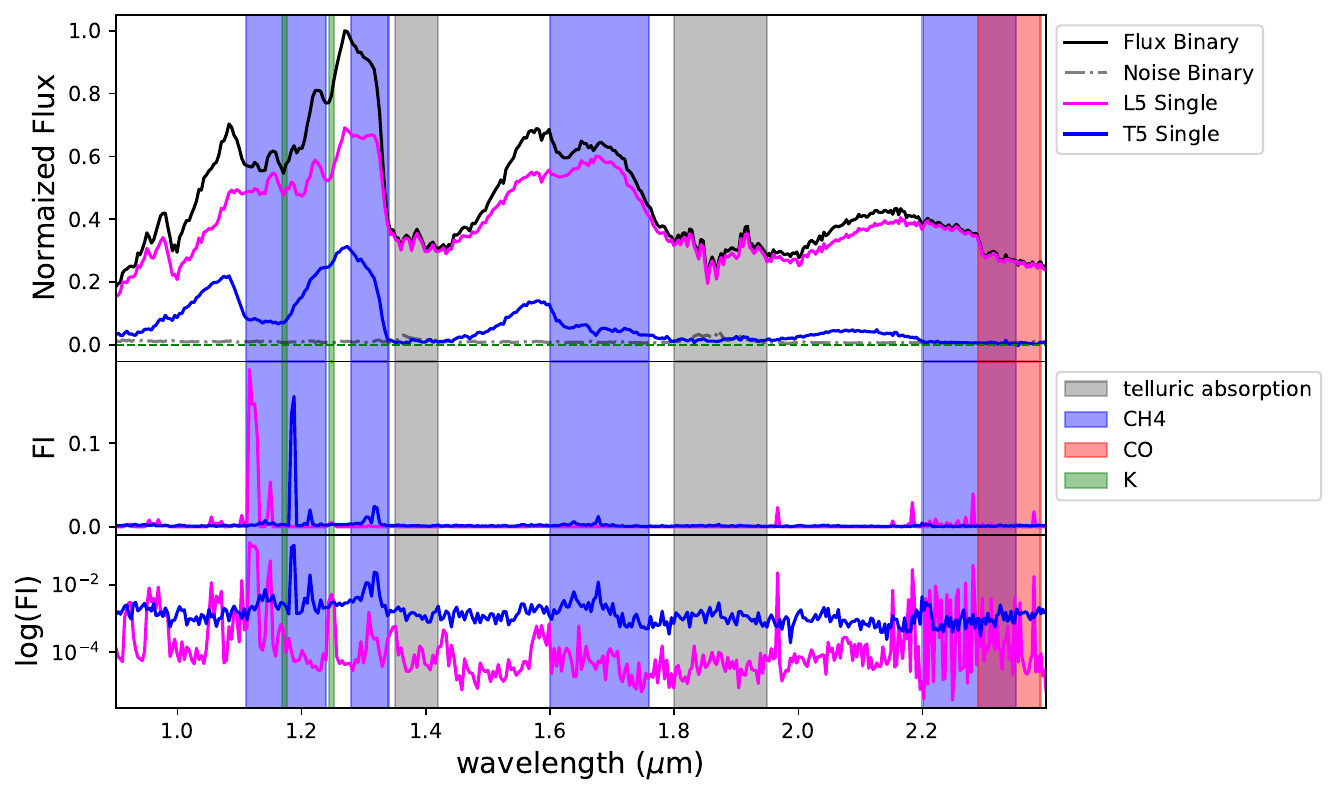}
\includegraphics[width=0.65\textwidth]{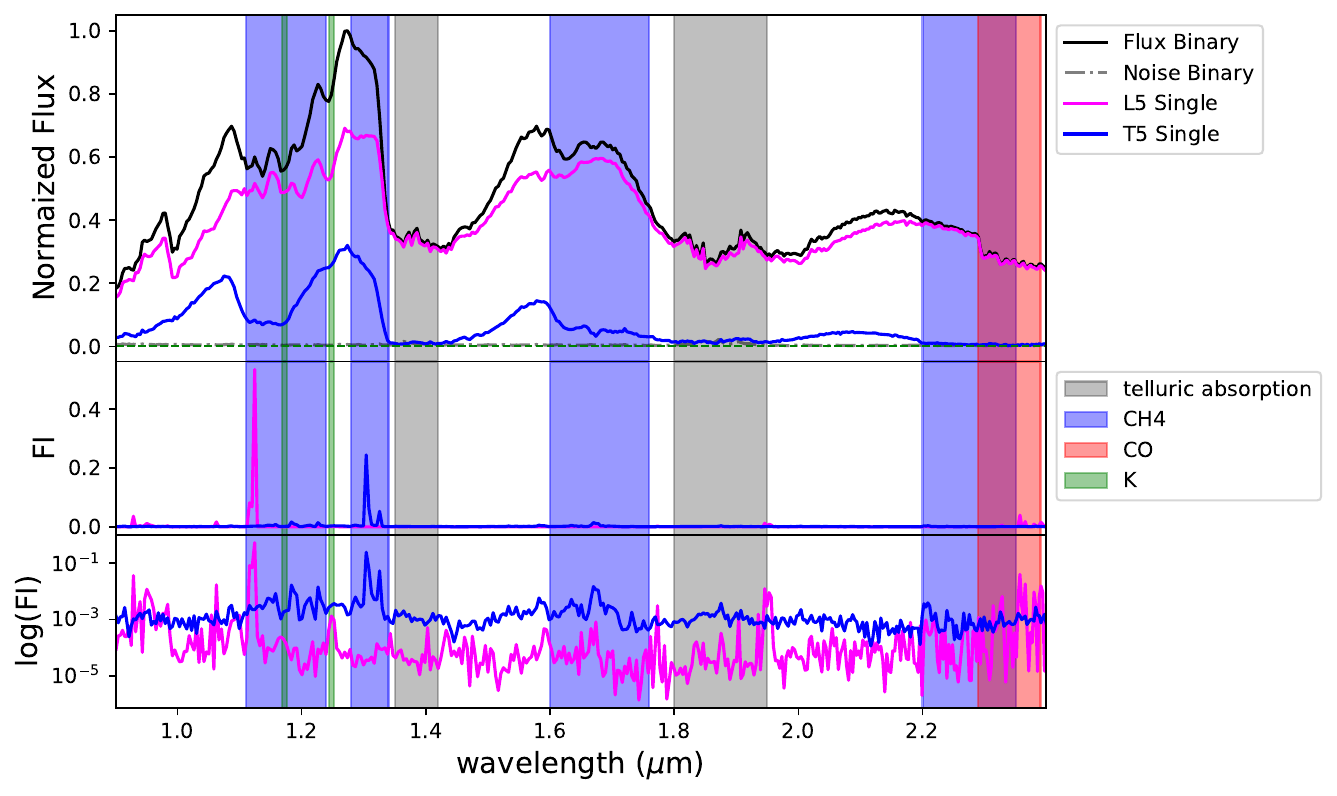}
\caption{Feature importance (FI) for the BClass1 model for low (top), mid (middle) and high (bottom) S/N groups (cf.\ Figure~\ref{fig:FIRF}). 
Each panel displays FI for primary (magenta) and secondary (blue) classification 
on linear (middle plot) and logarithmic scales (bottom plot), compared to an L5 plus T5 synthetic binary spectrum 
with a S/N in the associated range (top plot).
Regions associated with telluric (grey), K~I (green), CH$_4$ (blue) and CO (red) absorption are indicated by vertical bands.}
\label{fig:FIMO}
\end{figure*}

Figure~\ref{fig:FIMO} displays feature importance
for the BClass1 model for our three S/N groups, again compared to a spectrum of an L5 plus T5 synthetic binary to guide feature identification.
FI distributions for primary and secondary are {distinct} and evolve with S/N group. 
For the low S/N group, the primary classification FI shows spikes in the 1.10-1.24~$\mu$m region that encompasses both CH$_4$ and K~I absorption features, and at the interface of the CH$_4$ and CO bands at 2.2 and 2.3~$\mu$m.
The secondary classification FI also shows features in the 1.10-1.24~$\mu$m region, as well as at the 1.33~$\mu$m edge of the strong CH$_4$/H$_2$O band, and in the 1.6~$\mu$m CH$_4$ band.
These features are also present in the mid and high S/N groups, but become more concentrated at 1.12~$\mu$m for primary classification and at 1.25~$\mu$m for secondary classification. A new feature emerges in the primary FI for the mid- and high-S/N models at 1.95~$\mu$m, in the region of strong H$_2$O absorption and at the edge of the 1.80-1.95~$\mu$m telluric band. Beyond these spikes, the FI spectra are relatively flat across {other} wavelengths.

\section{Application to Known Binary Spectra} \label{sec:application}

Our analysis of synthesized binary spectra indicates robust determination of both unresolved binarity and component classifications spanning a wide range of very low mass dwarf component types. However, internal validation must be supported by {analysis} of actual binary spectra.
We curated a sample of known VLM binary spectra {for sources with primary or combined-light classifications of} M7 {from} the UltracoolSheet \citep{zenodo2024} {which have been} 
confirmed by resolved imaging, astrometric variability, or radial velocity variability.
{Of} 176 {known} binary systems, 43 have low-resolution IRTF/SpeX spectra available in the SPLAT database {with} resolved or combined-light spectroscopy
{(Table~\ref{tab:confirmed_binaries}). The} sources have primary spectral types spanning M8 to T8.5 and secondary spectral types spanning L1.5 to T9.5. 
{We evaluated the binary selection indices for those sources that meet the B10 and B14 index-based selection criteria, encompassing 14 and 25 of these binaries, respectively.}
We focused our evaluation on the correct identification and classification of these binaries (recall) for models 1, 5, and 7, using the models trained to the appropriate spectral data S/N.

\begin{longrotatetable}
    \begin{deluxetable}{lc|cc|cc|cc|ccc|cc|cc|cc|c}
\tablecaption{Random Forest Analysis of Confirmed Ultracool Dwarf Binaries \label{tab:confirmed_binaries}}   
\tabletypesize{\scriptsize} 
\tablehead{
& &
\multicolumn{2}{c|}{Prior} & 
\multicolumn{2}{c|}{Combined Light} &
\multicolumn{2}{c|}{Index Class} &
\multicolumn{3}{c|}{Binary Identification} &
\multicolumn{2}{c|}{BClass1} & 
\multicolumn{2}{c|}{BClass5} &
\multicolumn{2}{c|}{BClass7} &
\\
\cline{3-4} \cline{5-6} \cline{7-8} \cline{9-11} \cline{12-13} \cline{14-15} \cline{16-17}
\multicolumn{1}{c}{Name} &
\multicolumn{1}{c|}{Method\tablenotemark{a}} &
\multicolumn{1}{c}{P} &
\multicolumn{1}{c|}{S} &
\multicolumn{1}{c}{Type} &
\multicolumn{1}{c|}{S/N} &
\multicolumn{1}{c}{B14} &
\multicolumn{1}{c|}{B10} &
\multicolumn{1}{c}{BId1} &
\multicolumn{1}{c}{BId5} &
\multicolumn{1}{c|}{BId7} &
\multicolumn{1}{c}{P} &
\multicolumn{1}{c|}{S} &
\multicolumn{1}{c}{P} &
\multicolumn{1}{c|}{S} &
\multicolumn{1}{c}{P} &
\multicolumn{1}{c|}{S} &
\multicolumn{1}{c}{Ref} 
} 
\startdata
SDSS J000649.16-085246.3 & SP, SB1 & M8.5 & T5 & M8.0 & high & strong & \nodata & b & s & \nodata & M7.5 & T0.6 & M7.8 & T5.2 & L8.1 & T3.3 & 15 \\
2MASS J01303563-4445411 & V & M9 & L6 & M9.0 & high & single & \nodata & s & s & \nodata & M8.9 & T0.4 & M9.0 & T3.7 & L6.2 & T3.4 & 20 \\
2MASSW J0320284-044636 & SP, SB1 & M8.5 & T5 & M8.0 & high & strong & \nodata & b & s & \nodata & M8.1 & T3.7 & M8.3 & T5.4 & L8.1 & T3.4 & 9 \\
SDSS J034408.91+011124.9 & SP, A & L0.5 & T4 & L0.0 & high & weak & \nodata & b & s & \nodata & L0.6 & T3.1 & L0.9 & T4.6 & L8.1 & T3.3 & 3,4 \\
DENIS-P J035726.9-441730 & V & M9 & L1.5 & L2.0 & mid & single & \nodata & b & s & \nodata & L1.5 & L5.3 & L1.8 & T1.9 & L6.0 & T2.8 & 3,31 \\
SDSSp J042348.57-041403.5 & SP, V & L6.5 & T2 & T0.0 & mid & \nodata & weak & b & \nodata & b & L9.0 & T1.9 & L4.9 & T2.2 & L6.7 & T2.1 & 18,21 \\
WISEPA J045853.89+643452.9 & V & T8.5 & T9.5 & T8.0 & low & \nodata & \nodata & s & \nodata & \nodata & T8.0 & T8.2 & L5.9 & T5.3 & T1.7 & T5.8 & 30,14 \\
2MASS J05185995-2828372 & SP, V & L6 & T4 & L8.0 & mid & weak & strong & b & b & b & L7.1 & T5.6 & L6.8 & T5.5 & L7.0 & T5.5 & 18,21 \\
2MASS J05431887+6422528 & V & L0.5 & L1.5 & L2.0 & high & single & \nodata & b & s & \nodata & L1.5 & L7.9 & L1.7 & T1.5 & L5.8 & T2.7 & 3,19 \\
WISEPA J061135.13-041024.0 & V & L9 & T1.5 & T0.0 & low & \nodata & strong & b & \nodata & s & L9.4 & T2.8 & L5.0 & T3.9 & L8.6 & T2.8 & 26,24 \\
WISE J072003.20-084651.2 & SP, V, SB1 & M9 & T5 & M9.0 & high & single & \nodata & b & b & \nodata & M9.4 & T4.8 & M9.8 & T4.6 & L6.2 & T3.6 & 16 \\
SDSS J080531.84+481233.0 & SP, A, SB1 & L4 & T5 & T0.0 & low & \nodata & weak & s & \nodata & s & L3.6 & T4.7 & L3.5 & T5.3 & L7.3 & T3.8 & 17,21 \\
DENIS-P J0823031-491201 & A & L1 & L5 & L1.0 & high & single & \nodata & b & s & \nodata & L1.2 & L6.6 & L1.5 & T2.1 & L5.9 & T3.6 & 3,34 \\
2MASSW J0850359+105716 & V & L6.5 & L8.5 & L7.0 & mid & single & \nodata & b & s & \nodata & L6.4 & L9.8 & L6.7 & T1.6 & L6.4 & T2.4 & 13,21 \\
SDSSp J092615.38+584720.9 & SP, V & T3.5 & T5 & T4.0 & mid & \nodata & single & b & \nodata & s & T4.0 & T6.4 & L6.5 & T5.4 & T1.7 & T5.4 & 11,21 \\
2MASSI J1017075+130839 & V & L1.5 & L3 & L2.0 & high & single & \nodata & s & s & \nodata & L1.6 & L6.6 & L1.8 & T1.7 & L5.9 & T2.8 & 3,21 \\
SDSS J102109.69-030420.1 & SP, V & T0 & T5 & T3.0 & mid & \nodata & strong & b & \nodata & b & T1.7 & T5.1 & L6.2 & T5.5 & T1.4 & T5.1 & 7,21 \\
SDSS J105213.51+442255.7 & V, A & L6.5 & T1.5 & L9.0 & mid & \nodata & single & b & \nodata & b & L6.7 & T2.0 & L6.7 & T2.1 & L6.4 & T2.7 & 9,22 \\
2MASS J11061197+2754225 & SP, SB1 & T0 & T4.5 & T2.0 & high & \nodata & weak & b & \nodata & b & T0.7 & T3.9 & L5.9 & T4.5 & T0.5 & T3.6 & 29,11 \\
2MASS J11193254-1137466 & V & L7 & L7 & L7.0 & low & single & \nodata & s & s & \nodata & L7.0 & T4.4 & L6.8 & T3.5 & L7.0 & T4.3 & 25,5 \\
2MASSW J1146345+223053 & V & L3 & L3 & L3.0 & high & single & \nodata & b & s & \nodata & L2.1 & L5.7 & L2.4 & T1.5 & L6.2 & T2.5 & 11,21 \\
2MASSW J1207334-393254 & V & M8 & L3 & M8.0 & high & single & \nodata & s & s & \nodata & M7.5 & L5.2 & M8.0 & T3.7 & L6.3 & T3.4 & 28,2 \\
2MASS J12095613-1004008 & SP, V & T2.5 & T6.5 & T3.0 & low & \nodata & weak & s & \nodata & s & T2.2 & T6.6 & L5.8 & T5.4 & T1.8 & T5.7 & 6,21 \\
2MASS J12255432-2739466 & V & T5.5 & T8 & T6.0 & high & \nodata & \nodata & s & \nodata & \nodata & T5.4 & T7.0 & L6.5 & T4.7 & T1.7 & T5.8 & 6,21 \\
DENIS-P J122813.8-154711 & V & L5.5 & L5.5 & L4.0 & mid & single & \nodata & b & b & \nodata & L4.3 & L9.9 & L4.6 & T1.3 & L5.9 & T2.0 & 11,21 \\
2MASSI J1315309-264951 & SP, V & L3.5 & T7 & L5.0 & mid & strong & \nodata & b & s & \nodata & L5.6 & T5.1 & L5.8 & T1.4 & L6.3 & T2.4 & 12 \\
2MASS J13411160-3052505 & SP, V & L2.5 & T6.0 & L2.0 & high & weak & \nodata & b & s & \nodata & L1.9 & T2.2 & L1.8 & T3.6 & L6.4 & T2.8 & 3,4 \\
2MASS J14044941-3159329 & SP, V & L9 & T5 & T2.0 & mid & \nodata & strong & b & \nodata & b & T0.6 & T3.9 & L6.3 & T5.4 & T0.8 & T4.0 & 29,21 \\
SDSS J151114.66+060742.9 & SP, V & L5 & T5 & T0.0 & low & \nodata & strong & b & \nodata & b & L7.2 & T5.5 & L5.7 & T5.4 & L6.9 & T5.0 & 17,4 \\
2MASS J15200224-4422419 & V & L1.5 & L4.5 & L1.0 & high & single & \nodata & s & s & \nodata & L1.6 & T0.0 & L1.3 & T2.6 & L5.8 & T3.0 & 8 \\
SDSS J153417.05+161546.1 & SP, V & T0 & T5.5 & T3.0 & low & \nodata & strong & b & \nodata & b & T1.4 & T6.5 & L5.9 & T5.4 & T1.3 & T6.2 & 17,21 \\
2MASSI J1534498-295227 & V & T4.5 & T5 & T5.0 & high & \nodata & \nodata & s & \nodata & \nodata & T4.7 & T5.3 & L6.5 & T4.6 & T1.8 & T5.7 & 11,21 \\
2MASS J15500845+1455180 & V & L3.5 & L4 & L2.0 & high & single & \nodata & s & s & \nodata & L2.4 & L5.7 & L2.2 & T1.4 & L6.0 & T2.7 & 9,10 \\
2MASSI J1553022+153236 & V & T6.5 & T7.5 & T7.0 & high & \nodata & \nodata & s & \nodata & \nodata & T7.0 & T7.7 & L6.5 & T4.7 & T1.7 & T5.7 & 11,21 \\
2MASS J17072343-0558249 & V & M9 & L3 & L1.0 & low & single & \nodata & s & s & \nodata & L1.9 & L7.1 & L2.0 & T1.4 & L5.8 & T2.5 & 32 \\
WISEPA J171104.60+350036.8 & V & T8 & T9.5 & T7.0 & low & \nodata & \nodata & s & \nodata & \nodata & T7.6 & T8.5 & L5.7 & T5.2 & T1.7 & T6.0 & 26,27 \\
2MASSW J1728114+394859 & V & L5 & L7 & L7.0 & mid & single & \nodata & s & s & \nodata & L6.4 & T0.3 & L5.9 & T2.2 & L6.5 & T2.4 & {37},21 \\
2MASS J20261584-2943124 & SP, A & L0.5 & T6 & L1.0 & high & weak & \nodata & s & b & \nodata & L1.0 & T4.0 & L1.0 & T5.6 & L6.0 & T3.6 & 35,23 \\
SDSS J205235.31-160929.8 & SP, V & L8.5 & T1.5 & T1.0 & low & \nodata & weak & s & \nodata & s & L9.9 & T1.9 & L4.3 & T3.9 & L9.4 & T2.7 & 17,21 \\
2MASSW J2101154+175658 & V & L7 & L8 & L5.0 & low & single & \nodata & s & s & \nodata & L7.1 & L9.3 & L6.8 & T2.2 & L6.8 & T2.5 & 17,21 \\
2MASS J21321145+1341584 & V, A & L4.5 & L8.5 & L6.0 & mid & single & \nodata & b & s & \nodata & L6.0 & L7.9 & L4.9 & T2.4 & L6.5 & T2.4 & 36,21 \\
SDSSp J224953.45+004404.2 & V & L3 & L5 & L7.0 & mid & single & \nodata & b & s & \nodata & L6.7 & L8.5 & L5.8 & T2.2 & L6.6 & T2.7 & 9,1 \\
DENIS-P J225210.73-173013.4 & V & L4.5 & T3.5 & T0.0 & high & \nodata & weak & b & \nodata & s & L7.9 & T2.4 & L3.2 & T4.8 & L9.4 & T3.4 & 33,21 \\
\enddata
\tablecomments{Component primary (P) and secondary (S) classifications are from \citet{zenodo2024} and \citet{2015AJ....150..163B}. Combined-light {spectral} classifications were determined by comparison {to} spectral standards. S/N groups were assigned based on the 1.20--1.35~$\micron$ {(J-band)} S/N value.
For the BId columns, ``b'' indicates a binary classification and '`s'' indicates a single classification.} 
\tablenotetext{a}{Binary identification and confirmation methods: SP = spectral blend binary, V = resolved visual binary, A = astrometric variable, SB1 = radial velocity variable (primary motion).}
\tablerefs{
(1) \citet{2010ApJ...715..561A}; 
(2) \citet{2013ApJ...772...79A}; 
(3) \citet{2014ApJ...794..143B}; 
(4) \citet{2015AJ....150..163B}; 
(5) \citet{2017ApJ...843L...4B}; 
(6) \citet{2004AJ....127.2856B}; 
(7) \citet{2006ApJ...637.1067B}; 
(8) \citet{2007ApJ...658..557B}; 
(9) \citet{2008ApJ...681..579B}; 
(10) \citet{2009AJ....138.1563B}; 
(11) \citet{2010ApJ...710.1142B}; 
(12) \citet{2011ApJ...739...49B}; 
(13) \citet{2011AJ....141...70B};
(14) \citet{2012ApJ...745...26B}; 
(15) \citet{2012ApJ...757..110B}; 
(16) \citet{2015AJ....149..104B}; 
(17) \citet{2006AJ....131.2722C}; 
(18) \citet{2004ApJ...604L..61C}; 
(19) \citet{2017MNRAS.468.3499D}; 
(20) \citet{2011AJ....141....7D}; 
(21) \citet{2012ApJS..201...19D}; 
(22) \citet{2015ApJ...805...56D}; 
(23) \citet{2010AJ....140..110G}; 
(24) \citet{2014AJ....148....6G}; 
(25) \citet{2015AJ....150..182K}; 
(26) \citet{2011ApJS..197...19K}; 
(27) \citet{2012ApJ...758...57L}; 
(28) \citet{2007ApJ...669L..97L}; 
(29) \citet{2007AJ....134.1162L}; 
(30) \citet{2011ApJ...726...30M}; 
(31) \citet{2006AA...456..253M}; 
(32) \citet{2006AJ....132.2074M}; 
(33) \citet{2006ApJ...639.1114R}; 
(34) \citet{2015AA...579A..61S}; 
(35) \citet{2010AJ....139.1045S}; 
(36) \citet{2007AJ....133.2320S};
{(37) \citet{SPLAT}.} 
}
    \end{deluxetable}
\end{longrotatetable}

\begin{figure*}[th]
\plottwo{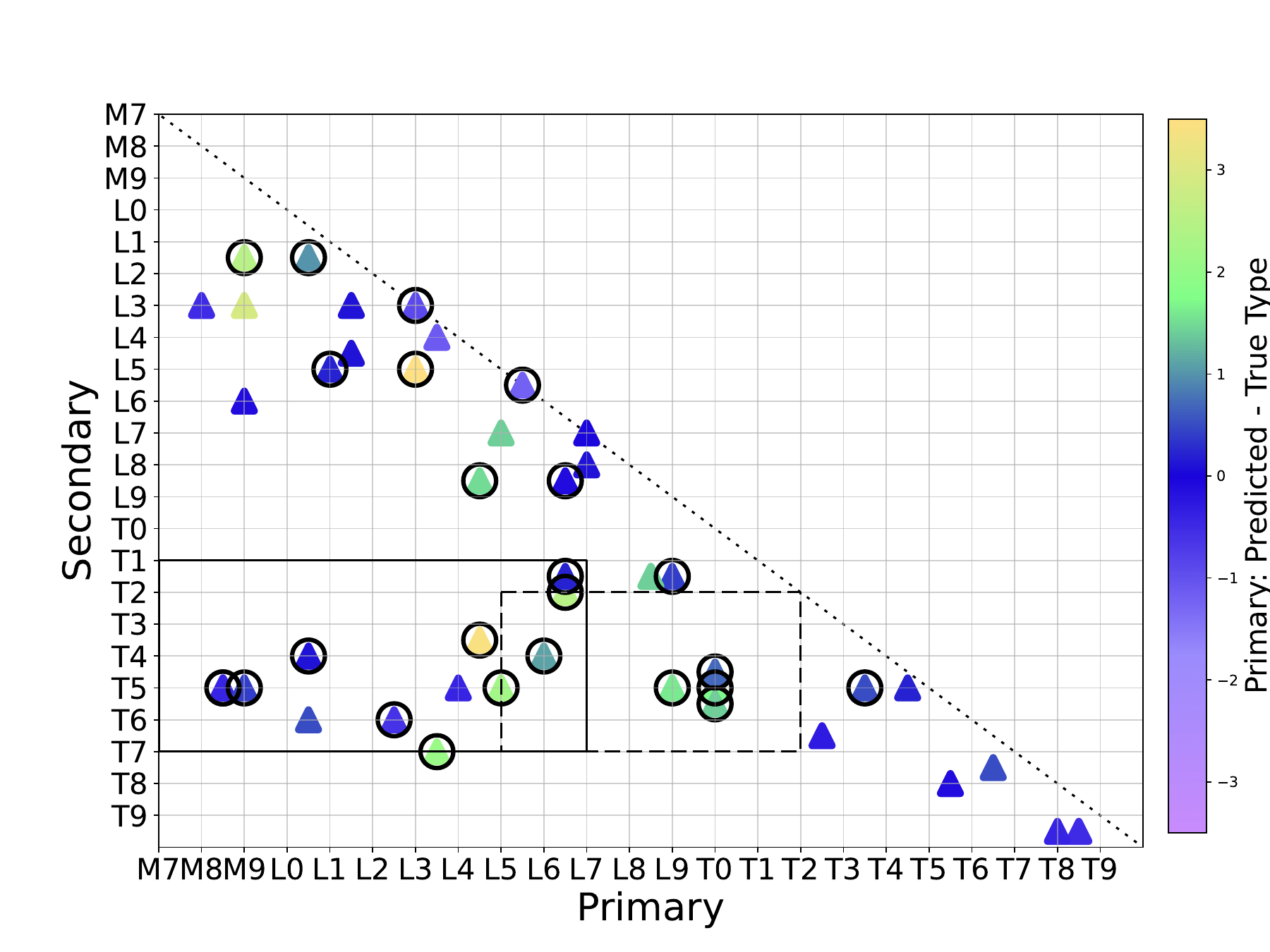}{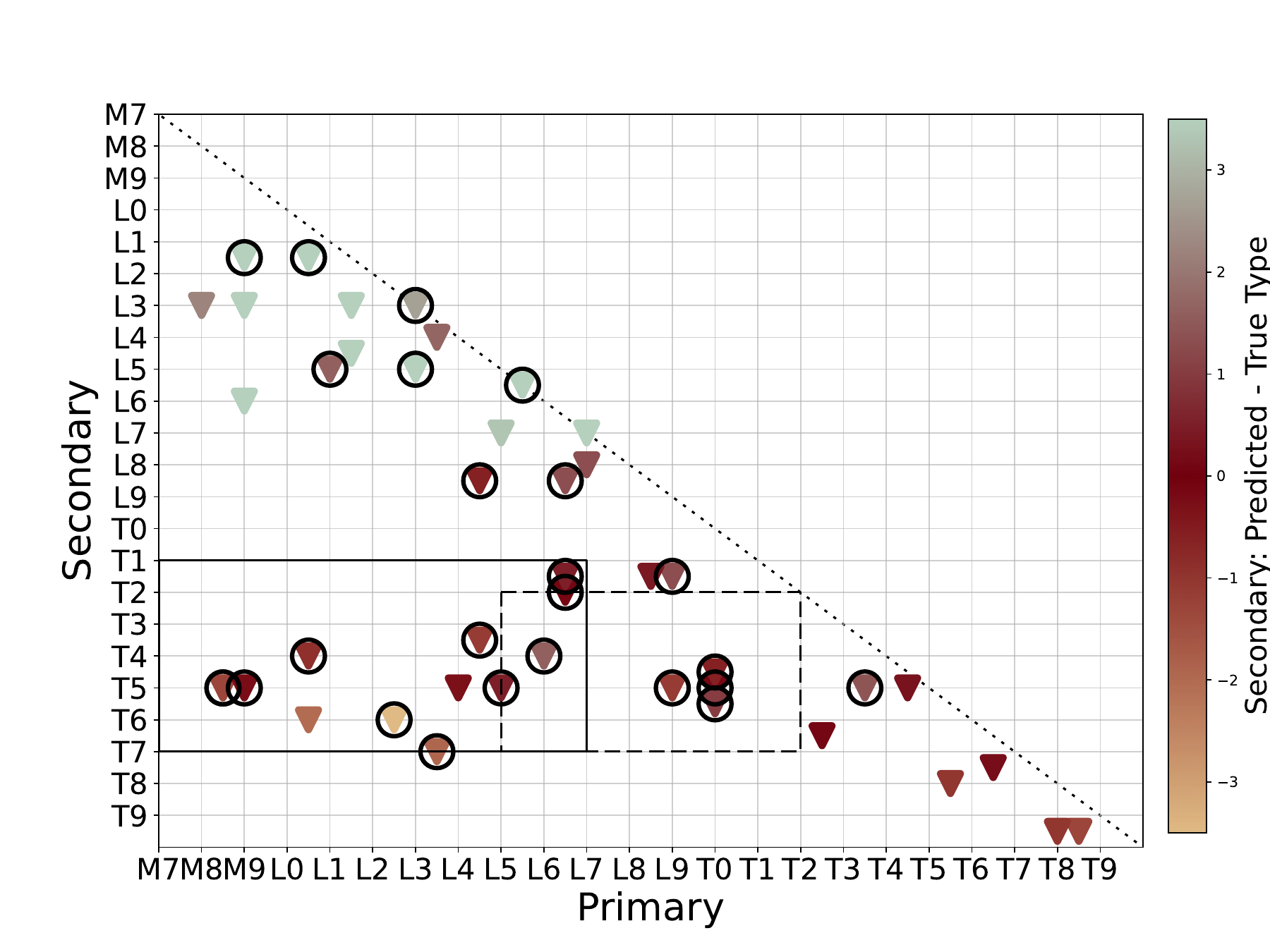}
\caption{Application of BId1 and BClass1 models for known binaries with SpeX spectral observations (Table~\ref{tab:confirmed_binaries}). 
The left panel compares primary classifications, the right panel compares secondary classifications.
Both panels map individual sources to their reported primary and secondary component types, with those correctly identified as a binary by the BId1 model indicated by large black circles.
The color shading indicates the difference between the predicted and reported component types.
We also highlight the component type spaces relevant to the B10 (solid line box) and B14 (dashed line box) index classification methods. 
}
\label{fig:ucs}
\end{figure*}

{Of the} 43 confirmed binaries with spectra, 25 are identified as binaries by the BId1 model, corresponding to a recall of R = 0.58. This performance is considerably poorer than the 0.83--0.89 recall scores from the testing data.
To gain insight on this difference, Figure~\ref{fig:ucs} maps the known component types of the binaries, with those identified as a binary by the BId1 model labeled. 
The binary sample is 
not uniformly distributed across component type space, with a higher faction of near equal-component types where all of the RF models show decreased performance (Figure~\ref{fig:index_heatmap}). 
Away from this edge, binary identification is more robust.
For the 23 systems with component classifications differing by more than three subtypes,
17 are identified as binary (R = 0.74), whereas only 8 of the 20 systems with more similar types are correctly identified (R = 0.40). 

The BId5 and BId7 models perform even more poorly in aggregate, 
with sample recall of 0.16 (4/25) and 0.57 (8/14), respectively.
These recall values are again 
far below the 0.76--0.85 and 0.89--0.94 values from our testing sample,
The models also have lower recall values compared to the BId1 model (0.60 and 0.79) and the {equivalent index-based approaches} (0.48 {for B14}, 0.86 {for B10}) for the same combined-light spectral ranges. 
Even if we downselect the binary sample to the {primary and secondary} classifications encompassed by the B14 (P $\in$ \{M7,L7\}, S $\in$ \{T1,T7\}) and B10 methods (P $\in$ \{L5,T2\}, S $\in$ \{T2,T7\}), we obtain only a modest improvement in recall to 0.38 and 0.64 {for the BId5 and BId7 models}, respectively, whereas the BId1 model (0.88 and 0.82) and indices (0.88 and 1.00) are highly robust in {these regimes}. 
Hence, the BId1 model appears to be as good or better than the B10 and B14 indices for binary identification, 
and far superior to the specialized BId5 and BId7 models.

For component classification, we quantified recall by requiring that primary and/or secondary types be within 1 subtype (2 subtypes) of the previously published value.
We can see visually in Figure~\ref{fig:ucs} that the majority of primary types are recovered by the BClass1 model with a recall of 0.65 (0.84), secondary types have a lower recall of 0.37 (0.67), and both components are recovered with a recall of 0.23 (0.58). Again, the substandard performance compared to test data can be attributed to the distribution of component classifications in the empirical sample, with misclassifications occurring for systems with similar late-M and L dwarf components (upper left quadrant of Figure~\ref{fig:ucs}) and those with large differences between component types (lower left quadrant of Figure~\ref{fig:ucs}).
For the 23 systems with at least three subtypes separating the primary and secondary component types, and with a T dwarf secondary, there is a modest gain in the accuracy of the secondary classification with a recall of 0.43 (0.70), but primary and combined component recall remains the same. 
In contrast to our binary identification models, we see clear improvement with the specialized BClass5 and BClass7 models.
The BClass5 model outperforms the BClass1 model for primary classification with a recall of 0.76 (0.84), but performs worse for secondary and combined classification recall. 
For the BClass7 model, if we relax the recall requirement to 2 subtypes, this model performs better in primary classification (R = 0.86), secondary classification (R = 0.93), and combined classification (R = 0.79) for the appropriate combined-light spectral types. Both of these models achieve near-perfect performance if the sample is further downselected to the B10 and B14 component classification ranges.

In summary, the BId1 model appears to perform as well as the B10 and B14 indices, and far better than the specialized RF models BId5 and BId7, in correctly identifying binary spectra, and is particularly robust for systems with differing component classifications (at least three subtypes in this experiment, with 74\% of systems recovered). The BClass1 model performs fairly well for primary classification and poorly for secondary classification, whereas the specialized BClass5 and BClass7 {models show} significantly {better performance} in their respective spectral type ranges, correctly classifying $\gtrsim$80\% of primaries and secondaries to within two subtypes. 
The lower recall statistics for this sample compared to the synthetic testing data can be explained by the 
non-uniform composition of real VLM binaries, which are skewed toward early-type and similar-type components.

\section{Limitations of the Random Forest Model} \label{sec:discussion}

{
Overall, we find our random forest models are superior to current index-based methods for identifying and characterizing very low mass spectral blend binaries in terms of performance, range of applicability, and speed. However, these models also suffer the same inherent limitations in identifying binaries with similar component spectral types or 
extreme flux ratio systems (e.g., late-M and late-T dwarf pairs). These limitations are relevant for existing binary samples
which are dominated by pairs with similar spectral types due to both selection biases (e.g., intrinsically brighter systems) 
and the intrinsic prevalence of near-equal mass systems among VLM systems \citep{2003ApJ...586..512B,2018MNRAS.479.2702F}. 
}

\begin{figure}[h]
\centering
\includegraphics[width=0.6\textwidth]{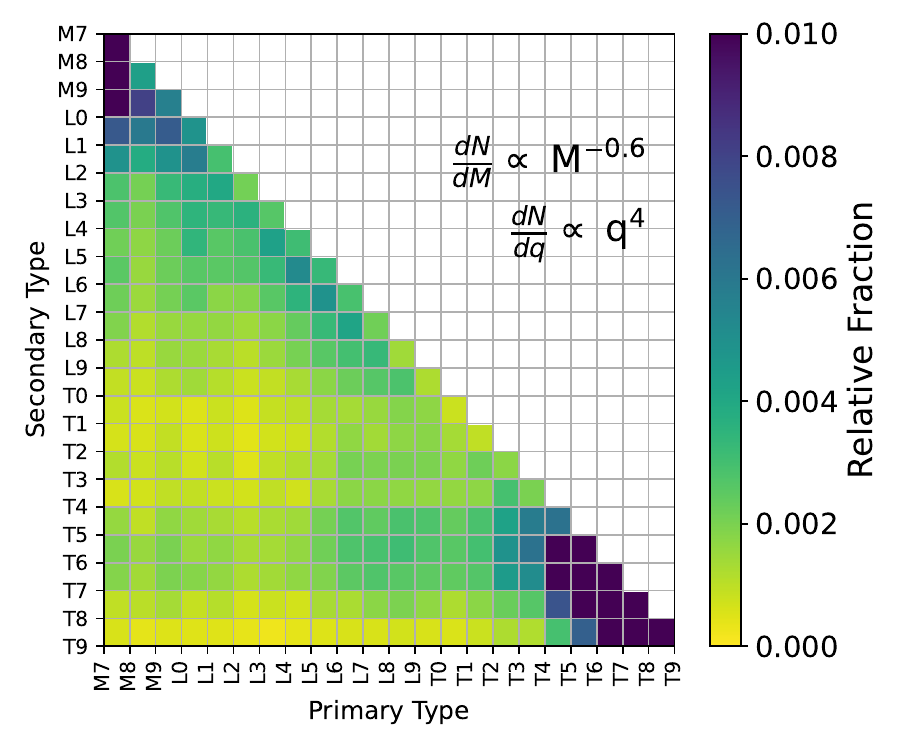}
\caption{{Component spectral type distributions for a simulated sample of VLM dwarf binaries assuming a uniform age distribution over 0.5--8~Gyr, a power-law mass distribution $\frac{dN}{dM} \propto M^{-0.6}$ for 0.01~{\msun} $\leq$ M $\leq$ 0.1~{\msun}, a power-law mass ratio distribution $\frac{dN}{dq} \propto q^4$ for 0.2 $\leq$ $q$ $\leq$ 1, \citet{2003A&A...402..701B} evolutionary models, and the \citet{2013ApJS..208....9P} spectral type/effective temperature relation. Color shading indicates the relative density distribution for each spectral type combination.}}
\label{fig:popsim}
\end{figure}

{
Even if we adhere to unbiased, volume-complete samples of very low mass binaries, a relatively small fraction of these systems are suited for detection as spectral blends \citep{2019ApJ...883..205B}. 
To illustrate this, we simulated 100,000 VLM binary systems in a volume-limited field sample assuming a uniform age distribution spanning 0.5--8~Gyr, a power-law mass function $\frac{dN}{dM} \propto M^{-0.6}$ for 0.01~{\msun} $\leq$ M $\leq$ 0.1~{\msun} \citep{2024ApJS..271...55K,2024ApJ...967..115B}, and a power-law mass ratio distribution $\frac{dN}{dq} \propto q^4$ for 0.2 $\leq$ $q \equiv M_2/M_1$ $\leq$ 1 \citep{2003ApJ...586..512B}. These simulated systems were evolved to present-day observables using the evolutionary models of \citet{2003A&A...402..701B}, and effective temperatures were converted to spectral types using the empirical relation of \citet{2013ApJS..208....9P}. 
Figure~\ref{fig:popsim} displays the resulting distribution of component types. A significant fraction of these systems (49\%) are mid- and late-T dwarf pairs, consistent with the long-term evolution of brown dwarf systems \citep{2004ApJS..155..191B,2024ApJS..271...55K}, and in a region where binary identification recall drops to $\lesssim$0.5 (Figure~\ref{fig:BId1_heatmap}). 
Another 16\% of the simulated sample is composed of late-M plus late-M dwarf pairs, primarily stellar-mass systems.
Spectral combinations in the B10 or B14 ranges comprise only 8\% of the simulated sample. Furthermore, these fractions are consistent to within a few percent over a broad range of assumptions for the mass function and mass ratio distribution. Hence, detectable spectral blend systems generally comprise a small fraction of the overall VLM binary population, as had been previously reported in volume-limited spectral surveys \citep{2019ApJ...883..205B}.
}

{
In addition to poor performance for both near-equal mass and low-temperature very low mass pairs, our random forest models have been trained on a relatively conservative dataset in which peculiar sources (e.g., metal-poor sources, young brown dwarfs, blue L dwarfs) have been excluded. These sources are nevertheless known to be contaminants in spectral blend binary searches (B14).
We have also ignored unresolved triple systems, which exist but at low frequencies \citep{2013ApJ...778...36R,2021MNRAS.500.5453S}.
}

\section{Summary and Next Steps} \label{sec:summary}

The key outcomes of this investigation are as follows:

\begin{itemize}
    \item We have developed a hierarchical set of random forest models that identify unresolved very low mass binary systems from blended light spectra and determine their component spectral types. Our training set was constructed from a sample of late-M, L and T dwarf near-infrared spectra that are used to synthesize both single and blended-light binary spectral templates.
    For both identification and classification, we tested a series of models to explore performance as a function of spectral signal-to-noise, binary composition, and the inclusion of difference spectra (after subtracting the best-fit standard spectrum) and masking telluric regions. 
    \item We benchmarked our RF models by assessing the performance of the index-based spectral binary identification methods of B10 and B14, finding these to be poorly suited for general binary combinations (recall = 0.62--0.66 and 0.19--0.27, respectively) but robust (recall $\gtrsim$ 0.9) for select component type combinations, specifically L dwarf primaries with T dwarf secondaries. 
    \item Our most reliable RF models for both identification and classification were the most general models, BId1 and BClass1, which are trained on the full spectral range of {M6} to T9 dwarfs and their binary combinations.
    The BId1 identification model achieves a median recall of 0.82--0.89 across the full training set, 
    while the BClass1 classification model yields median errors of $\lesssim$0.1~subtypes for all components,
    and systematic uncertainties of $\sim$0.5~subtype and $\sim$1 subtype for primary and secondary classifications, respectively. Binary identification model variants showed no significant improvement over the BId1 model,
    while classification models tuned to the B10 and B14 spectral type range improved secondary classification uncertainty to 0.5~subtypes.
    \item Detailed evaluation of model performance as a function of component spectral type shows that these models are more reliable for certain spectral type combinations, performing optimally for mid-L through mid-T primaries and T dwarf secondaries, overlapping with the optimal ranges for the B10 and B14 index methods. Classification models are generally robust for primaries across nearly all component types, while secondary classification is generally poor for late-M and L dwarf binaries (similar spectral morphologies) and late-M and mid-/late-T dwarf binaries (large flux differential).
    \item Feature importance plots indicate that these models are tuned to specific spectral features, notably emission peaks in T dwarf spectra and key VLM spectral absorption features (H$_2$O, CH$_4$, CO bands and K~I lines), largely in line with {the B10 and B14 spectral} index definitions.
    \item Validation of model performance on the combined-light spectra of known VLM dwarf binary systems indicates decreased performance compared to the training data, driven in large part by the higher frequency of systems with spectral type compositions the RF models are less sensitive too (e.g., M/L dwarf pairs, components with similar spectral types). For sources with larger differences in component types {($\gtrsim$3~subtypes)}, the BId1 binary identification model achieves robust performance (recall $>$0.8), comparable to or better than index-based methods. For classification, the BClass5 model (matched to the B14 index method) and the BClass7 model (matched to the B10 index method) have better performance, with robust recall of primary and secondary classifications to within 2~subtypes.
    \item The primary strength of the RF models is their speed, with training times measured in seconds and identification and classification of individual spectra measured in milliseconds on a typical laptop computer. 
    {On the other hand, these models have the same limitations inherent to the spectral blend method, namely low sensitivity to systems with similar component spectral types and those with extreme flux ratios. These limitations imply that only a small fraction ($\sim$8\%) of VLM binaries in a field population are likely to be identified by this approach.}
\end{itemize}

{
There are a number of ways that the machine learning approach described here can be improved for very low mass binary studies.
Expansion of the breadth of spectral data included in the modeling could improve performance for both identification and classification. Optical spectral data are more sensitive to the primary components of VLM binaries, while mid-infrared data 
improves the detection and characterization of low-temperature T dwarf companions (cf. \citealt{2008ApJ...689L..53B}). 
The flexibility of RF models also enables the use of non-spectral data as features.
For example, VLM dwarfs with known distances through parallax measurement, cluster membership, or association with co-moving companion could make use of absolute flux measurements to improve the detection of systems with similar masses and spectral types.
The training data used for these models can be also adjusted to more accurately reflect the underlying mass ratio, age distribution, and stellar/substellar evolution of the VLM population being studied, rather than equal distributions across spectral type pairs (Figure~\ref{fig:popsim}).
Moreover, while the specialized RF models tailored for the B10 and B14 index-based methods improved performance only for component classification,  
the speed of model training (seconds) suggests that individualized RF models could be constructed for specific combined-light or component spectral type groups; e.g., focusing on equal or extreme light-ratio systems.
Finally, our exploration of the binary identification and classification problem focused only on random forest models. Alternate ML architectures, such as support vector machines or neural networks, may be better suited for this task given the high dimensionality of data, particularly if non-spectral data are included \citep{2019arXiv190407248B}. 
}

{
A natural next step is to apply the machine learning framework explored in this study to additional spectral survey samples to search for new binaries. Beyond the SpeX prism sample investigated here, there are growing libraries of VLM infrared spectra obtained with space-based platforms such as
JWST \citep{2024ApJ...973..107B,2024ApJ...975..162L,2025arXiv251002026T,2025arXiv251101167M} and
Euclid \citep{2025arXiv250322497V,2025arXiv250322559M,2025ApJ...991...84D}, and future samples that are expected from SPHEREx \citep{spherex} and the Nancy Grace Roman Space Telescope \citep{2020JATIS...6d6001M}.
For these datasets, similar models could be deployed with the appropriate empirical or synthetic training sets. 
Our initial study demonstrates that these approaches can address both potential biases and bottlenecks in the identification of VLM binary systems for population studies.
}

\begin{acknowledgments}
JDDG and MD acknowledge funding from the UCSD Summer Undergraduate Research Program.
This work has benefited from The UltracoolSheet at \url{http://bit.ly/UltracoolSheet}, maintained by Will Best, Trent Dupuy, Michael Liu, Aniket Sanghi, Rob Siverd, and Zhoujian Zhang.
{We also thank our anonymous referee for their comments which improved the quality of this manscript.}
Data presented in this paper were obtained with the NASA Infrared Telescope Facility (IRTF) on the summit of Maunakea. 
{Spectral template data is available on Zenodo under an open-source 
Creative Commons Attribution license for SPLAT v1.11: 
\dataset[doi:10.5281/zenodo.17107529]{https://doi.org/10.5281/zenodo.17107529}.}
The authors recognize and acknowledge the significant cultural role
and reverence that the summit of Maunakea has with the
indigenous Hawaiian community, and that IRTF stands on Crown and Government Lands that the
State of Hawai’i is obligated to protect and preserve for future
generations of indigenous Hawaiians. 
Portions of this work were conducted at the University of California, San Diego,
which was built on the unceded territory of the Kumeyaay
Nation, whose people continue to maintain their political
sovereignty and cultural traditions as vital members of the San Diego community.
\end{acknowledgments}

\vspace{5mm}
\facilities{IRTF(SpeX)}

\software{
astropy \citep{2013A&A...558A..33A,2018AJ....156..123A,2022ApJ...935..167A},  
Matplotlib \citep{Hunter2007}, 
NumPy \citep{harris2020array}, 
pandas \citep{mckinney2010data}, 
scikitlearn \citep{scikit-learn},
SPLAT \citep{SPLAT},
}

\clearpage

\appendix

The training sample used in this work was constructed from low-resolution near-infrared spectra of late-M, L, and T dwarfs contained in the SpeX Prism Library Analysis Toolkit (SPLAT \citealt{SPLAT}), and obtained with the SpeX spectrograph on the NASA 3m Infrared Telescope Facility \citep{2003PASP..115..362R}
using its low-resolution prism-dispersed mode. 
Table~\ref{tab:templates} lists the 1006 spectra used in this study.
in order of increasing spectral type and S/N.
Classifications are based on the best $\chi^2$ match to near-infrared spectral standards defined in
\citet[][M and L dwarfs]{2010ApJS..190..100K} and \citet[][T dwarfs]{2006ApJ...637.1067B}.
S/N values were measured in the 1.2--1.35~$\mu$m $J$-band region and assigned to
low (2 $\leq$ S/N $<$ 50),
med (50 $\leq$ S/N $<$ 100),
and high (S/N $\geq$ 100) S/N groups.
References are given for the original data publication. 
All of the original spectra can be accessed from the 
SPLAT archive v1.11\footnote{\url{https://github.com/aburgasser/splat}} \citep{adam_burgasser_2025_17107529}, and we have made 
the specific training set for our models available 
{on Zenodo at \dataset[doi: 10.5281/zenodo.15611798]{https://doi.org/10.5281/zenodo.15611798}.}

\startlongtable
\begin{deluxetable}{lcccc}
    \tablecaption{SpeX Spectral Sample. \label{tab:templates}}  
    \tabletypesize{\scriptsize} 
    \tablehead{
    \colhead{SIMBAD Name} & 
    \colhead{SpT} &
    \colhead{S/N} &
    \colhead{S/N Class} &
    \colhead{Ref} } 
\startdata
2MASS J01141304+4354287 & M6 & 41 & low & 40 \\
2MASS J20033545+1158552 & M6 & 45 & low & 42 \\
LEHPM 2$-$1973 & M6 & 61 & mid & 3 \\
2MASS J00583814$-$1747311 & M6 & 73 & mid & 8 \\
2MASS J18244344+2937133 & M6 & 75 & mid & 8 \\
2MASS J01532750+3631482 & M6 & 93 & mid & 8 \\
2MASS J01405263+0453302 & M6 & 94 & mid & 42 \\
LHS 3566 & M6 & 98 & mid & 8 \\
\enddata
\end{deluxetable}

\bibliography{mlbinaries}{}

@article{scikit-learn,
 title={Scikit-learn: Machine Learning in {P}ython},
 author={Pedregosa, F. and Varoquaux, G. and Gramfort, A. and Michel, V.
         and Thirion, B. and Grisel, O. and Blondel, M. and Prettenhofer, P.
         and Weiss, R. and Dubourg, V. and Vanderplas, J. and Passos, A. and
         Cournapeau, D. and Brucher, M. and Perrot, M. and Duchesnay, E.},
 journal={Journal of Machine Learning Research},
 volume={12},
 pages={2825--2830},
 year={2011}
}

@ARTICLE{2010ApJ...711.1087K,
       author = {{Konopacky}, Q.~M. and {Ghez}, A.~M. and {Barman}, T.~S. and {Rice}, E.~L. and {Bailey}, J.~I., III and {White}, R.~J. and {McLean}, I.~S. and {Duch{\^e}ne}, G.},
        title = "{High-precision Dynamical Masses of Very Low Mass Binaries}",
      journal = {\apj},
         year = 2010,
        month = mar,
       volume = {711},
       number = {2},
        pages = {1087-1122},
          doi = {10.1088/0004-637X/711/2/1087},
archivePrefix = {arXiv},
       eprint = {1001.4800},
 primaryClass = {astro-ph.SR},
       adsurl = {https://ui.adsabs.harvard.edu/abs/2010ApJ...711.1087K}
}

@ARTICLE{2008ApJ...678L.125B,
       author = {{Blake}, Cullen H. and {Charbonneau}, David and {White}, Russel J. and {Torres}, Guillermo and {Marley}, Mark S. and {Saumon}, Didier},
        title = "{A Spectroscopic Binary at the M/L Transition}",
      journal = {\apjl},
         year = 2008,
        month = may,
       volume = {678},
       number = {2},
        pages = {L125},
          doi = {10.1086/588754},
archivePrefix = {arXiv},
       eprint = {0804.1364},
 primaryClass = {astro-ph},
       adsurl = {https://ui.adsabs.harvard.edu/abs/2008ApJ...678L.125B}
}

@ARTICLE{2010AJ....140..110G,
       author = {{Gelino}, Christopher R. and {Burgasser}, Adam J.},
        title = "{2MASS J20261584-2943124: an Unresolved L0.5 + T6 Spectral Binary}",
      journal = {\aj},
         year = 2010,
        month = jul,
       volume = {140},
       number = {1},
        pages = {110-118},
          doi = {10.1088/0004-6256/140/1/110},
archivePrefix = {arXiv},
       eprint = {1005.4080},
 primaryClass = {astro-ph.SR},
       adsurl = {https://ui.adsabs.harvard.edu/abs/2010AJ....140..110G}
}

@ARTICLE{2011ApJ...739...49B,
       author = {{Burgasser}, Adam J. and {Sitarski}, Breann N. and {Gelino}, Christopher R. and {Logsdon}, Sarah E. and {Perrin}, Marshall D.},
        title = "{The Hyperactive L Dwarf 2MASS J13153094-2649513: Continued Emission and a Brown Dwarf Companion}",
      journal = {\apj},
         year = 2011,
        month = sep,
       volume = {739},
       number = {1},
          eid = {49},
        pages = {49},
          doi = {10.1088/0004-637X/739/1/49},
archivePrefix = {arXiv},
       eprint = {1107.1484},
 primaryClass = {astro-ph.SR},
       adsurl = {https://ui.adsabs.harvard.edu/abs/2011ApJ...739...49B}
}

@ARTICLE{2012ApJ...757..110B,
       author = {{Burgasser}, Adam J. and {Luk}, Christopher and {Dhital}, Saurav and {Bardalez Gagliuffi}, Daniella and {Nicholls}, Christine P. and {Prato}, L. and {West}, Andrew A. and {L{\'e}pine}, S{\'e}bastien},
        title = "{Discovery of a Very Low Mass Triple with Late-M and T Dwarf Components: LP 704-48/SDSS J0006-0852AB}",
      journal = {\apj},
         year = 2012,
        month = oct,
       volume = {757},
       number = {2},
          eid = {110},
        pages = {110},
          doi = {10.1088/0004-637X/757/2/110},
archivePrefix = {arXiv},
       eprint = {1208.0352},
 primaryClass = {astro-ph.SR},
       adsurl = {https://ui.adsabs.harvard.edu/abs/2012ApJ...757..110B}
}

@ARTICLE{2015AJ....150..163B,
       author = {{Bardalez Gagliuffi}, Daniella C. and {Gelino}, Christopher R. and {Burgasser}, Adam J.},
        title = "{High Resolution Imaging of Very Low Mass Spectral Binaries: Three Resolved Systems and Detection of Orbital Motion in an L/T Transition Binary}",
      journal = {\aj},
         year = 2015,
        month = nov,
       volume = {150},
       number = {5},
          eid = {163},
        pages = {163},
          doi = {10.1088/0004-6256/150/5/163},
archivePrefix = {arXiv},
       eprint = {1510.00392},
 primaryClass = {astro-ph.SR},
       adsurl = {https://ui.adsabs.harvard.edu/abs/2015AJ....150..163B}
}

@dataset{zenodo2024,
  author       = {Best, William M. J. and
                  Dupuy, Trent J. and
                  Liu, Michael C. and
                  Sanghi, Aniket and
                  Siverd, Robert J. and
                  Zhang, Zhoujian},
  title        = {{The UltracoolSheet: Photometry, Astrometry, 
                   Spectroscopy, and Multiplicity for 4000+ Ultracool
                   Dwarfs and Imaged Exoplanets}},
  month        = feb,
  year         = 2024,
  publisher    = {Zenodo},
  version      = {2.0.0},
  doi          = {10.5281/zenodo.10573247},
  url          = {https://doi.org/10.5281/zenodo.10573247}
}

@article{2013A&A...558A..33A,
	adsurl = {http://adsabs.harvard.edu/abs/2013A%26A...558A..33A},
	archiveprefix = {arXiv},
	author = {{Astropy Collaboration} and {Robitaille}, T.~P. and {Tollerud}, E.~J. and {Greenfield}, P. and {Droettboom}, M. and {Bray}, E. and {Aldcroft}, T. and {Davis}, M. and {Ginsburg}, A. and {Price-Whelan}, A.~M. and {Kerzendorf}, W.~E. and {Conley}, A. and {Crighton}, N. and {Barbary}, K. and {Muna}, D. and {Ferguson}, H. and {Grollier}, F. and {Parikh}, M.~M. and {Nair}, P.~H. and {Unther}, H.~M. and {Deil}, C. and {Woillez}, J. and {Conseil}, S. and {Kramer}, R. and {Turner}, J.~E.~H. and {Singer}, L. and {Fox}, R. and {Weaver}, B.~A. and {Zabalza}, V. and {Edwards}, Z.~I. and {Azalee Bostroem}, K. and {Burke}, D.~J. and {Casey}, A.~R. and {Crawford}, S.~M. and {Dencheva}, N. and {Ely}, J. and {Jenness}, T. and {Labrie}, K. and {Lim}, P.~L. and {Pierfederici}, F. and {Pontzen}, A. and {Ptak}, A. and {Refsdal}, B. and {Servillat}, M. and {Streicher}, O.},
	doi = {10.1051/0004-6361/201322068},
	eid = {A33},
	eprint = {1307.6212},
	journal = {\aap},
	month = oct,
	pages = {A33},
	primaryclass = {astro-ph.IM},
	title = {{Astropy: A community Python package for astronomy}},
	volume = 558,
	year = 2013}

@inproceedings{SPLAT,
	adsurl = {http://adsabs.harvard.edu/abs/2017ASInC..14....7B},
	archiveprefix = {arXiv},
	author = {{Burgasser}, A.~J. and {Splat Development Team}},
	booktitle = {Astronomical Society of India Conference Series},
	eprint = {1707.00062},
	pages = {7-12},
	primaryclass = {astro-ph.SR},
	series = {Astronomical Society of India Conference Series},
	title = {{The SpeX Prism Library Analysis Toolkit (SPLAT): A Data Curation Model}},
	volume = 14,
	year = 2017}

@article{Desai_2023,
doi = {10.3847/2515-5172/acb54a},
url = {https://dx.doi.org/10.3847/2515-5172/acb54a},
year = {2023},
month = {jan},
publisher = {The American Astronomical Society},
volume = {7},
number = {1},
pages = {13},
author = {Malina Desai and Juan Diego Draxl Giannoni and Camille Dunning and Luke McDermott and Christian Aganze and Christopher A. Theissen and Adam J. Burgasser},
title = {Identifying Ultracool Binary Systems using Machine Learning Methods},
journal = {Research Notes of the AAS}
}

@ARTICLE{2018AJ....156..123A,
       author = {{Astropy Collaboration} and {Price-Whelan}, A.~M. and {Sip{\H{o}}cz}, B.~M. and {G{\"u}nther}, H.~M. and {Lim}, P.~L. and {Crawford}, S.~M. and {Conseil}, S. and {Shupe}, D.~L. and {Craig}, M.~W. and {Dencheva}, N. and {Ginsburg}, A. and {VanderPlas}, J.~T. and {Bradley}, L.~D. and {P{\'e}rez-Su{\'a}rez}, D. and {de Val-Borro}, M. and {Aldcroft}, T.~L. and {Cruz}, K.~L. and {Robitaille}, T.~P. and {Tollerud}, E.~J. and {Ardelean}, C. and {Babej}, T. and {Bach}, Y.~P. and {Bachetti}, M. and {Bakanov}, A.~V. and {Bamford}, S.~P. and {Barentsen}, G. and {Barmby}, P. and {Baumbach}, A. and {Berry}, K.~L. and {Biscani}, F. and {Boquien}, M. and {Bostroem}, K.~A. and {Bouma}, L.~G. and {Brammer}, G.~B. and {Bray}, E.~M. and {Breytenbach}, H. and {Buddelmeijer}, H. and {Burke}, D.~J. and {Calderone}, G. and {Cano Rodr{\'\i}guez}, J.~L. and {Cara}, M. and {Cardoso}, J.~V.~M. and {Cheedella}, S. and {Copin}, Y. and {Corrales}, L. and {Crichton}, D. and {D'Avella}, D. and {Deil}, C. and {Depagne}, {\'E}. and {Dietrich}, J.~P. and {Donath}, A. and {Droettboom}, M. and {Earl}, N. and {Erben}, T. and {Fabbro}, S. and {Ferreira}, L.~A. and {Finethy}, T. and {Fox}, R.~T. and {Garrison}, L.~H. and {Gibbons}, S.~L.~J. and {Goldstein}, D.~A. and {Gommers}, R. and {Greco}, J.~P. and {Greenfield}, P. and {Groener}, A.~M. and {Grollier}, F. and {Hagen}, A. and {Hirst}, P. and {Homeier}, D. and {Horton}, A.~J. and {Hosseinzadeh}, G. and {Hu}, L. and {Hunkeler}, J.~S. and {Ivezi{\'c}}, {\v{Z}}. and {Jain}, A. and {Jenness}, T. and {Kanarek}, G. and {Kendrew}, S. and {Kern}, N.~S. and {Kerzendorf}, W.~E. and {Khvalko}, A. and {King}, J. and {Kirkby}, D. and {Kulkarni}, A.~M. and {Kumar}, A. and {Lee}, A. and {Lenz}, D. and {Littlefair}, S.~P. and {Ma}, Z. and {Macleod}, D.~M. and {Mastropietro}, M. and {McCully}, C. and {Montagnac}, S. and {Morris}, B.~M. and {Mueller}, M. and {Mumford}, S.~J. and {Muna}, D. and {Murphy}, N.~A. and {Nelson}, S. and {Nguyen}, G.~H. and {Ninan}, J.~P. and {N{\"o}the}, M. and {Ogaz}, S. and {Oh}, S. and {Parejko}, J.~K. and {Parley}, N. and {Pascual}, S. and {Patil}, R. and {Patil}, A.~A. and {Plunkett}, A.~L. and {Prochaska}, J.~X. and {Rastogi}, T. and {Reddy Janga}, V. and {Sabater}, J. and {Sakurikar}, P. and {Seifert}, M. and {Sherbert}, L.~E. and {Sherwood-Taylor}, H. and {Shih}, A.~Y. and {Sick}, J. and {Silbiger}, M.~T. and {Singanamalla}, S. and {Singer}, L.~P. and {Sladen}, P.~H. and {Sooley}, K.~A. and {Sornarajah}, S. and {Streicher}, O. and {Teuben}, P. and {Thomas}, S.~W. and {Tremblay}, G.~R. and {Turner}, J.~E.~H. and {Terr{\'o}n}, V. and {van Kerkwijk}, M.~H. and {de la Vega}, A. and {Watkins}, L.~L. and {Weaver}, B.~A. and {Whitmore}, J.~B. and {Woillez}, J. and {Zabalza}, V. and {Astropy Contributors}},
        title = "{The Astropy Project: Building an Open-science Project and Status of the v2.0 Core Package}",
      journal = {\aj},
         year = 2018,
        month = sep,
       volume = {156},
       number = {3},
          eid = {123},
        pages = {123},
          doi = {10.3847/1538-3881/aabc4f},
archivePrefix = {arXiv},
       eprint = {1801.02634},
 primaryClass = {astro-ph.IM},
       adsurl = {https://ui.adsabs.harvard.edu/abs/2018AJ....156..123A}
}

@Article{Hunter2007,
  Author    = {Hunter, J. D.},
  Title     = {Matplotlib: A 2D graphics environment},
  Journal   = {Comp.\ Sci.\ \& Eng.},
  Volume    = {9},
  Number    = {3},
  Pages     = {90--95},
  publisher = {IEEE COMPUTER SOC},
  doi       = {10.1109/MCSE.2007.55},
  year      = 2007
}

@ARTICLE{Joergens2008,
       author = {{Joergens}, V.},
        title = "{Binary frequency of very young brown dwarfs at separations smaller than 3 AU}",
      journal = {\aap},
         year = 2008,
        month = dec,
       volume = {492},
       number = {2},
        pages = {545-555},
          doi = {10.1051/0004-6361:200810413},
archivePrefix = {arXiv},
       eprint = {0809.3001},
 primaryClass = {astro-ph},
       adsurl = {https://ui.adsabs.harvard.edu/abs/2008A&A...492..545J}
}

@ARTICLE{Cruz2004,
       author = {{Cruz}, Kelle L. and {Burgasser}, Adam J. and {Reid}, I. Neill and {Liebert}, James},
        title = "{2MASS J05185995-2828372: Discovery of an Unresolved L/T Binary}",
      journal = {\apjl},
         year = 2004,
        month = mar,
       volume = {604},
       number = {1},
        pages = {L61-L64},
          doi = {10.1086/383415},
archivePrefix = {arXiv},
       eprint = {astro-ph/0402172},
 primaryClass = {astro-ph},
       adsurl = {https://ui.adsabs.harvard.edu/abs/2004ApJ...604L..61C}
}

@ARTICLE{Burgasser2006HST,
       author = {{Burgasser}, Adam J. and {Kirkpatrick}, J. Davy and {Cruz}, Kelle L. and {Reid}, I. Neill and {Leggett}, Sandy K. and {Liebert}, James and {Burrows}, Adam and {Brown}, Michael E.},
        title = "{Hubble Space Telescope NICMOS Observations of T Dwarfs: Brown Dwarf Multiplicity and New Probes of the L/T Transition}",
      journal = {\apjs},
         year = 2006,
        month = oct,
       volume = {166},
       number = {2},
        pages = {585-612},
          doi = {10.1086/506327},
archivePrefix = {arXiv},
       eprint = {astro-ph/0605577},
 primaryClass = {astro-ph},
       adsurl = {https://ui.adsabs.harvard.edu/abs/2006ApJS..166..585B}
}

@Article{         harris2020array,
 title         = {Array programming with {NumPy}},
 author        = {Charles R. Harris and K. Jarrod Millman and St{\'{e}}fan J.
                 van der Walt and Ralf Gommers and Pauli Virtanen and David
                 Cournapeau and Eric Wieser and Julian Taylor and Sebastian
                 Berg and Nathaniel J. Smith and Robert Kern and Matti Picus
                 and Stephan Hoyer and Marten H. van Kerkwijk and Matthew
                 Brett and Allan Haldane and Jaime Fern{\'{a}}ndez del
                 R{\'{i}}o and Mark Wiebe and Pearu Peterson and Pierre
                 G{\'{e}}rard-Marchant and Kevin Sheppard and Tyler Reddy and
                 Warren Weckesser and Hameer Abbasi and Christoph Gohlke and
                 Travis E. Oliphant},
 year          = {2020},
 month         = sep,
 journal       = {Nature},
 volume        = {585},
 number        = {7825},
 pages         = {357--362},
 doi           = {10.1038/s41586-020-2649-2},
 publisher     = {Springer Science and Business Media {LLC}},
 url           = {https://doi.org/10.1038/s41586-020-2649-2}
}

@inproceedings{mckinney2010data,
  title={Data structures for statistical computing in python},
  author={McKinney, Wes and others},
  booktitle={Proceedings of the 9th Python in Science Conference},
  volume={445},
  pages={51--56},
  year={2010},
  organization={Austin, TX}
}

@ARTICLE{Aganze2022_1,
       author = {{Aganze}, Christian and {Burgasser}, Adam J. and {Malkan}, Mathew and {Theissen}, Christopher A. and {Tejada Arevalo}, Roberto A. and {Hsu}, Chih-Chun and {Bardalez Gagliuffi}, Daniella C. and {Ryan}, Russell E. and {Holwerda}, Benne},
        title = "{Beyond the Local Volume. I. Surface Densities of Ultracool Dwarfs in Deep HST/WFC3 Parallel Fields}",
      journal = {\apj},
         year = 2022,
        month = jan,
       volume = {924},
       number = {2},
          eid = {114},
        pages = {114},
          doi = {10.3847/1538-4357/ac35ea},
archivePrefix = {arXiv},
       eprint = {2110.07672},
 primaryClass = {astro-ph.SR},
       adsurl = {https://ui.adsabs.harvard.edu/abs/2022ApJ...924..114A}
}

@ARTICLE{Radigan2014,
       author = {{Radigan}, Jacqueline and {Lafreni{\`e}re}, David and {Jayawardhana}, Ray and {Artigau}, Etienne},
        title = "{Strong Brightness Variations Signal Cloudy-to-clear Transition of Brown Dwarfs}",
      journal = {\apj},
         year = 2014,
        month = oct,
       volume = {793},
       number = {2},
          eid = {75},
        pages = {75},
          doi = {10.1088/0004-637X/793/2/75},
archivePrefix = {arXiv},
       eprint = {1404.3247},
 primaryClass = {astro-ph.SR},
       adsurl = {https://ui.adsabs.harvard.edu/abs/2014ApJ...793...75R}
}

@ARTICLE{2014ApJ...794..143B,
       author = {{Bardalez Gagliuffi}, Daniella C. and {Burgasser}, Adam J. and {Gelino}, Christopher R. and {Looper}, Dagny L. and {Nicholls}, Christine P. and {Schmidt}, Sarah J. and {Cruz}, Kelle and {West}, Andrew A. and {Gizis}, John E. and {Metchev}, Stanimir},
        title = "{SpeX Spectroscopy of Unresolved Very Low Mass Binaries. II. Identification of 14 Candidate Binaries with Late-M/Early-L and T Dwarf Components}",
      journal = {\apj},
         year = 2014,
        month = oct,
       volume = {794},
       number = {2},
          eid = {143},
        pages = {143},
          doi = {10.1088/0004-637X/794/2/143},
archivePrefix = {arXiv},
       eprint = {1408.3089},
 primaryClass = {astro-ph.SR},
       adsurl = {https://ui.adsabs.harvard.edu/abs/2014ApJ...794..143B}
}

@ARTICLE{2006Natur.440..311S,
       author = {{Stassun}, Keivan G. and {Mathieu}, Robert D. and {Valenti}, Jeff A.},
        title = "{Discovery of two young brown dwarfs in an eclipsing binary system}",
      journal = {\nat},
         year = 2006,
        month = mar,
       volume = {440},
       number = {7082},
        pages = {311-314},
          doi = {10.1038/nature04570},
       adsurl = {https://ui.adsabs.harvard.edu/abs/2006Natur.440..311S}
}

@ARTICLE{2015A&A...584A.128L,
       author = {{Lodieu}, N. and {Alonso}, R. and {Gonz{\'a}lez Hern{\'a}ndez}, J.~I. and {Sanchis-Ojeda}, R. and {Narita}, N. and {Kawashima}, Y. and {Kawauchi}, K. and {Su{\'a}rez Mascare{\~n}o}, A. and {Deeg}, H. and {Prieto Arranz}, J. and {Rebolo}, R. and {Pall{\'e}}, E. and {B{\'e}jar}, V.~J.~S. and {Ferragamo}, A. and {Rubi{\~n}o-Mart{\'\i}n}, J.~A.},
        title = "{An eclipsing double-line spectroscopic binary at the stellar/substellar boundary in the Upper Scorpius OB association}",
      journal = {\aap},
         year = 2015,
        month = dec,
       volume = {584},
          eid = {A128},
        pages = {A128},
          doi = {10.1051/0004-6361/201527464},
archivePrefix = {arXiv},
       eprint = {1511.03083},
 primaryClass = {astro-ph.SR},
       adsurl = {https://ui.adsabs.harvard.edu/abs/2015A&A...584A.128L}
}

@ARTICLE{2020NatAs...4..650T,
       author = {{Triaud}, Amaury H.~M.~J. and {Burgasser}, Adam J. and {Burdanov}, Artem and {Kunovac Hod{\v{z}}i{\'c}}, Vedad and {Alonso}, Roi and {Bardalez Gagliuffi}, Daniella and {Delrez}, Laetitia and {Demory}, Brice-Olivier and {de Wit}, Julien and {Ducrot}, Elsa and {Hessman}, Frederic V. and {Husser}, Tim-Oliver and {Jehin}, Emmanu{\"e}l and {Pedersen}, Peter P. and {Queloz}, Didier and {McCormac}, James and {Murray}, Catriona and {Sebastian}, Daniel and {Thompson}, Samantha and {Van Grootel}, Val{\'e}rie and {Gillon}, Micha{\"e}l},
        title = "{An eclipsing substellar binary in a young triple system discovered by SPECULOOS}",
      journal = {Nature Astronomy},
         year = 2020,
        month = mar,
       volume = {4},
        pages = {650-657},
          doi = {10.1038/s41550-020-1018-2},
archivePrefix = {arXiv},
       eprint = {2001.07175},
 primaryClass = {astro-ph.SR},
       adsurl = {https://ui.adsabs.harvard.edu/abs/2020NatAs...4..650T}
}

@ARTICLE{2019arXiv190407248B,
       author = {{Baron}, Dalya},
        title = "{Machine Learning in Astronomy: a practical overview}",
      journal = {arXiv e-prints},
         year = 2019,
        month = apr,
          eid = {arXiv:1904.07248},
        pages = {arXiv:1904.07248},
          doi = {10.48550/arXiv.1904.07248},
archivePrefix = {arXiv},
       eprint = {1904.07248},
 primaryClass = {astro-ph.IM},
       adsurl = {https://ui.adsabs.harvard.edu/abs/2019arXiv190407248B}
}

@ARTICLE{2022RAA....22b5017C,
       author = {{Chen}, Shu-Xin and {Sun}, Wei-Min and {He}, Ying},
        title = "{Application of Random Forest Regressions on Stellar Parameters of A-type Stars and Feature Extraction}",
      journal = {Research in Astronomy and Astrophysics},
         year = 2022,
        month = feb,
       volume = {22},
       number = {2},
          eid = {025017},
        pages = {025017},
          doi = {10.1088/1674-4527/ac41c5},
       adsurl = {https://ui.adsabs.harvard.edu/abs/2022RAA....22b5017C}
}

@ARTICLE{2023A&A...679A.127G,
       author = {{Garc{\'\i}a-Zamora}, Enrique Miguel and {Torres}, Santiago and {Rebassa-Mansergas}, Alberto},
        title = "{White dwarf Random Forest classification through Gaia spectral coefficients}",
      journal = {\aap},
         year = 2023,
        month = nov,
       volume = {679},
          eid = {A127},
        pages = {A127},
          doi = {10.1051/0004-6361/202347601},
archivePrefix = {arXiv},
       eprint = {2308.07090},
 primaryClass = {astro-ph.SR},
       adsurl = {https://ui.adsabs.harvard.edu/abs/2023A&A...679A.127G}
}

@ARTICLE{2001MachL..45....5B,
       author = {{Breiman}, Leo},
        title = "{Random Forests.}",
      journal = {Machine Learning},
         year = 2001,
        month = jan,
       volume = {45},
        pages = {5-32},
          doi = {10.1023/A:1010933404324},
       adsurl = {https://ui.adsabs.harvard.edu/abs/2001MachL..45....5B}
}

@ARTICLE{2012ApJ...745...26B,
       author = {{Burgasser}, Adam J. and {Gelino}, Christopher R. and {Cushing}, Michael C. and {Kirkpatrick}, J. Davy},
        title = "{Resolved Spectroscopy of a Brown Dwarf Binary at the T Dwarf/Y Dwarf Transition}",
      journal = {\apj},
         year = 2012,
        month = jan,
       volume = {745},
       number = {1},
          eid = {26},
        pages = {26},
          doi = {10.1088/0004-637X/745/1/26},
archivePrefix = {arXiv},
       eprint = {1110.4664},
 primaryClass = {astro-ph.SR},
       adsurl = {https://ui.adsabs.harvard.edu/abs/2012ApJ...745...26B}
}

@ARTICLE{2008ApJ...681..579B,
       author = {{Burgasser}, Adam J. and {Liu}, Michael C. and {Ireland}, Michael J. and {Cruz}, Kelle L. and {Dupuy}, Trent J.},
        title = "{Subtle Signatures of Multiplicity in Late-type Dwarf Spectra: The Unresolved M8.5 + T5 Binary 2MASS J03202839-0446358}",
      journal = {\apj},
         year = 2008,
        month = jul,
       volume = {681},
       number = {1},
        pages = {579-593},
          doi = {10.1086/588379},
archivePrefix = {arXiv},
       eprint = {0803.0295},
 primaryClass = {astro-ph},
       adsurl = {https://ui.adsabs.harvard.edu/abs/2008ApJ...681..579B}
}

@ARTICLE{2004AJ....127.2856B,        author = {{Burgasser}, Adam J. and {McElwain}, Michael W. and {Kirkpatrick}, J. Davy and {Cruz}, Kelle L. and {Tinney}, Chris G. and {Reid}, I. Neill},         title = "{The 2MASS Wide-Field T Dwarf Search. III. Seven New T Dwarfs and Other Cool Dwarf Discoveries}",       journal = {\aj},          year = 2004,         month = may,        volume = {127},        number = {5},         pages = {2856-2870},           doi = {10.1086/383549}, archivePrefix = {arXiv},        eprint = {astro-ph/0402325},  primaryClass = {astro-ph},        adsurl = {https://ui.adsabs.harvard.edu/abs/2004AJ....127.2856B} }

@ARTICLE{2006AJ....131.2722C,        author = {{Chiu}, K. and {Fan}, X. and {Leggett}, S.~K. and {Golimowski}, D.~A. and {Zheng}, W. and {Geballe}, T.~R. and {Schneider}, D.~P. and {Brinkmann}, J.},         title = "{Seventy-One New L and T Dwarfs from the Sloan Digital Sky Survey}",       journal = {\aj},          year = 2006,         month = jun,        volume = {131},        number = {5},         pages = {2722-2736},           doi = {10.1086/501431}, archivePrefix = {arXiv},        eprint = {astro-ph/0601089},  primaryClass = {astro-ph},        adsurl = {https://ui.adsabs.harvard.edu/abs/2006AJ....131.2722C} }

@ARTICLE{2004ApJ...604L..61C,        author = {{Cruz}, Kelle L. and {Burgasser}, Adam J. and {Reid}, I. Neill and {Liebert}, James},         title = "{2MASS J05185995-2828372: Discovery of an Unresolved L/T Binary}",       journal = {\apjl},          year = 2004,         month = mar,        volume = {604},        number = {1},         pages = {L61-L64},           doi = {10.1086/383415}, archivePrefix = {arXiv},        eprint = {astro-ph/0402172},  primaryClass = {astro-ph},        adsurl = {https://ui.adsabs.harvard.edu/abs/2004ApJ...604L..61C} }

@ARTICLE{2015AJ....150..182K,        author = {{Kellogg}, Kendra and {Metchev}, Stanimir and {Gei{\ss}ler}, Kerstin and {Hicks}, Shannon and {Kirkpatrick}, J. Davy and {Kurtev}, Radostin},         title = "{A Targeted Search for Peculiarly Red L and T Dwarfs in SDSS, 2MASS, and WISE: Discovery of a Possible L7 Member of the TW Hydrae Association}",       journal = {\aj},          year = 2015,         month = dec,        volume = {150},        number = {6},           eid = {182},         pages = {182},           doi = {10.1088/0004-6256/150/6/182}, archivePrefix = {arXiv},        eprint = {1510.08464},  primaryClass = {astro-ph.SR},        adsurl = {https://ui.adsabs.harvard.edu/abs/2015AJ....150..182K} }

@ARTICLE{2011ApJS..197...19K,        author = {{Kirkpatrick}, J. Davy and {Cushing}, Michael C. and {Gelino}, Christopher R. and {Griffith}, Roger L. and {Skrutskie}, Michael F. and {Marsh}, Kenneth A. and {Wright}, Edward L. and {Mainzer}, A. and {Eisenhardt}, Peter R. and {McLean}, Ian S. and {Thompson}, Maggie A. and {Bauer}, James M. and {Benford}, Dominic J. and {Bridge}, Carrie R. and {Lake}, Sean E. and {Petty}, Sara M. and {Stanford}, S.~A. and {Tsai}, Chao-Wei and {Bailey}, Vanessa and {Beichman}, Charles A. and {Bloom}, Joshua S. and {Bochanski}, John J. and {Burgasser}, Adam J. and {Capak}, Peter L. and {Cruz}, Kelle L. and {Hinz}, Philip M. and {Kartaltepe}, Jeyhan S. and {Knox}, Russell P. and {Manohar}, Swarnima and {Masters}, Daniel and {Morales-Calder{\';o}n}, Maria and {Prato}, Lisa A. and {Rodigas}, Timothy J. and {Salvato}, Mara and {Schurr}, Steven D. and {Scoville}, Nicholas Z. and {Simcoe}, Robert A. and {Stapelfeldt}, Karl R. and {Stern}, Daniel and {Stock}, Nathan D. and {Vacca}, William D.},         title = "{The First Hundred Brown Dwarfs Discovered by the Wide-field Infrared Survey Explorer (WISE)}",       journal = {\apjs},          year = 2011,         month = dec,        volume = {197},        number = {2},           eid = {19},         pages = {19},           doi = {10.1088/0067-0049/197/2/19}, archivePrefix = {arXiv},        eprint = {1108.4677},  primaryClass = {astro-ph.SR},        adsurl = {https://ui.adsabs.harvard.edu/abs/2011ApJS..197...19K} }

@ARTICLE{2007AJ....134.1162L,        author = {{Looper}, Dagny L. and {Kirkpatrick}, J. Davy and {Burgasser}, Adam J.},         title = "{Discovery of 11 New T Dwarfs in the Two Micron All Sky Survey, Including a Possible L/T Transition Binary}",       journal = {\aj},          year = 2007,         month = sep,        volume = {134},        number = {3},         pages = {1162-1182},           doi = {10.1086/520645}, archivePrefix = {arXiv},        eprint = {0706.1601},  primaryClass = {astro-ph},        adsurl = {https://ui.adsabs.harvard.edu/abs/2007AJ....134.1162L} }

@ARTICLE{2006AJ....132.2074M,        author = {{McElwain}, Michael W. and {Burgasser}, Adam J.},         title = "{Resolved Spectroscopy of M Dwarf/L Dwarf Binaries. II. 2MASS J17072343-0558249AB}",       journal = {\aj},          year = 2006,         month = nov,        volume = {132},        number = {5},         pages = {2074-2081},           doi = {10.1086/508199}, archivePrefix = {arXiv},        eprint = {astro-ph/0608143},  primaryClass = {astro-ph},        adsurl = {https://ui.adsabs.harvard.edu/abs/2006AJ....132.2074M} }

@INPROCEEDINGS{2007prpl.conf..427B,
       author = {{Burgasser}, A.~J. and {Reid}, I.~N. and {Siegler}, N. and {Close}, L. and {Allen}, P. and {Lowrance}, P. and {Gizis}, J.},
        title = "{Not Alone: Tracing the Origins of Very-Low-Mass Stars and Brown Dwarfs Through Multiplicity Studies}",
    booktitle = {Protostars and Planets V},
         year = 2007,
       editor = {{Reipurth}, Bo and {Jewitt}, David and {Keil}, Klaus},
        month = jan,
        pages = {427},
          doi = {10.48550/arXiv.astro-ph/0602122},
archivePrefix = {arXiv},
       eprint = {astro-ph/0602122},
 primaryClass = {astro-ph},
       adsurl = {https://ui.adsabs.harvard.edu/abs/2007prpl.conf..427B}
}

@ARTICLE{2018MNRAS.479.2702F,
       author = {{Fontanive}, Cl{\'e}mence and {Biller}, Beth and {Bonavita}, Mariangela and {Allers}, Katelyn},
        title = "{Constraining the multiplicity statistics of the coolest brown dwarfs: binary fraction continues to decrease with spectral type}",
      journal = {\mnras},
         year = 2018,
        month = sep,
       volume = {479},
       number = {2},
        pages = {2702-2727},
          doi = {10.1093/mnras/sty1682},
archivePrefix = {arXiv},
       eprint = {1806.08737},
 primaryClass = {astro-ph.SR},
       adsurl = {https://ui.adsabs.harvard.edu/abs/2018MNRAS.479.2702F}
}

@ARTICLE{2017ApJS..231...15D,
       author = {{Dupuy}, Trent J. and {Liu}, Michael C.},
        title = "{Individual Dynamical Masses of Ultracool Dwarfs}",
      journal = {\apjs},
         year = 2017,
        month = aug,
       volume = {231},
       number = {2},
          eid = {15},
        pages = {15},
          doi = {10.3847/1538-4365/aa5e4c},
archivePrefix = {arXiv},
       eprint = {1703.05775},
 primaryClass = {astro-ph.SR},
       adsurl = {https://ui.adsabs.harvard.edu/abs/2017ApJS..231...15D}
}

@ARTICLE{2021ApJS..257...45H,
       author = {{Hsu}, Chih-Chun and {Burgasser}, Adam J. and {Theissen}, Christopher A. and {Gelino}, Christopher R. and {Birky}, Jessica L. and {Diamant}, Sharon J.~M. and {Bardalez Gagliuffi}, Daniella C. and {Aganze}, Christian and {Blake}, Cullen H. and {Faherty}, Jacqueline K.},
        title = "{The Brown Dwarf Kinematics Project (BDKP). V. Radial and Rotational Velocities of T Dwarfs from Keck/NIRSPEC High-resolution Spectroscopy}",
      journal = {\apjs},
         year = 2021,
        month = dec,
       volume = {257},
       number = {2},
          eid = {45},
        pages = {45},
          doi = {10.3847/1538-4365/ac1c7d},
archivePrefix = {arXiv},
       eprint = {2107.01222},
 primaryClass = {astro-ph.SR},
       adsurl = {https://ui.adsabs.harvard.edu/abs/2021ApJS..257...45H}
}

@article{2007ApJ...666.1198B,
	adsurl = {http://adsabs.harvard.edu/abs/2007ApJ...666.1198B},
	archiveprefix = {arXiv},
	author = {{Blake}, C.~H. and {Charbonneau}, D. and {White}, R.~J. and {Marley}, M.~S. and {Saumon}, D.},
	doi = {10.1086/520124},
	eprint = {0705.3901},
	journal = {\apj},
	month = sep,
	pages = {1198-1204},
	title = {{Multiepoch Radial Velocity Observations of L Dwarfs}},
	volume = 666,
	year = 2007}

@article{1999AJ....118.2460B,
	adsurl = {http://adsabs.harvard.edu/abs/1999AJ....118.2460B},
	author = {{Basri}, G. and {Mart{\'{\i}}n}, E.~L.},
	doi = {10.1086/301079},
	eprint = {arXiv:astro-ph/9908015},
	journal = {\aj},
	month = nov,
	pages = {2460-2465},
	title = {{PPL 15: The First Brown Dwarf Spectroscopic Binary}},
	volume = 118,
	year = 1999}

@ARTICLE{2015ApJ...814..118B,
       author = {{Best}, William M.~J. and {Liu}, Michael C. and {Magnier}, Eugene A. and {Deacon}, Niall R. and {Aller}, Kimberly M. and {Redstone}, Joshua and {Burgett}, W.~S. and {Chambers}, K.~C. and {Draper}, P. and {Flewelling}, H. and {Hodapp}, K.~W. and {Kaiser}, N. and {Metcalfe}, N. and {Tonry}, J.~L. and {Wainscoat}, R.~J. and {Waters}, C.},
        title = "{A Search for L/T Transition Dwarfs with Pan-STARRS1 and WISE. II. L/T Transition Atmospheres and Young Discoveries}",
      journal = {\apj},
         year = 2015,
        month = dec,
       volume = {814},
       number = {2},
          eid = {118},
        pages = {118},
          doi = {10.1088/0004-637X/814/2/118},
archivePrefix = {arXiv},
       eprint = {1612.02824},
 primaryClass = {astro-ph.SR},
       adsurl = {https://ui.adsabs.harvard.edu/abs/2015ApJ...814..118B}
}

@PHDTHESIS{2017PhDT.......291B,
       author = {{Bardalez Gagliuffi}, Daniella Carolina},
        title = "{Spectral binaries hold the key to the true ultracool binary fraction}",
       school = {University of California, San Diego},
         year = 2017,
        month = jan,
       adsurl = {https://ui.adsabs.harvard.edu/abs/2017PhDT.......291B}
}

@ARTICLE{2006ApJ...637.1067B,        author = {{Burgasser}, Adam J. and {Geballe}, T.~R. and {Leggett}, S.~K. and {Kirkpatrick}, J. Davy and {Golimowski}, David A.},         title = "{A Unified Near-Infrared Spectral Classification Scheme for T Dwarfs}",       journal = {\apj},          year = 2006,         month = feb,        volume = {637},        number = {2},         pages = {1067-1093},           doi = {10.1086/498563}, archivePrefix = {arXiv},        eprint = {astro-ph/0510090},  primaryClass = {astro-ph},        adsurl = {https://ui.adsabs.harvard.edu/abs/2006ApJ...637.1067B} }

@ARTICLE{2007ApJ...658..557B,        author = {{Burgasser}, Adam J. and {Looper}, Dagny L. and {Kirkpatrick}, J. Davy and {Liu}, Michael C.},         title = "{Discovery of a High Proper Motion L Dwarf Binary: 2MASS J15200224-4422419AB}",       journal = {\apj},          year = 2007,         month = mar,        volume = {658},        number = {1},         pages = {557-568},           doi = {10.1086/511518}, archivePrefix = {arXiv},        eprint = {astro-ph/0611697},  primaryClass = {astro-ph},        adsurl = {https://ui.adsabs.harvard.edu/abs/2007ApJ...658..557B} }

@ARTICLE{2010ApJ...710.1142B,        author = {{Burgasser}, Adam J. and {Cruz}, Kelle L. and {Cushing}, Michael and {Gelino}, Christopher R. and {Looper}, Dagny L. and {Faherty}, Jacqueline K. and {Kirkpatrick}, J. Davy and {Reid}, I. Neill},         title = "{SpeX Spectroscopy of Unresolved Very Low Mass Binaries. I. Identification of 17 Candidate Binaries Straddling the L Dwarf/T Dwarf Transition}",       journal = {\apj},          year = 2010,         month = feb,        volume = {710},        number = {2},         pages = {1142-1169},           doi = {10.1088/0004-637X/710/2/1142}, archivePrefix = {arXiv},        eprint = {0912.3808},  primaryClass = {astro-ph.SR},        adsurl = {https://ui.adsabs.harvard.edu/abs/2010ApJ...710.1142B} }

@ARTICLE{2011AJ....141...70B,        author = {{Burgasser}, Adam J. and {Bardalez-Gagliuffi}, Daniella C. and {Gizis}, John E.},         title = "{Hubble Space Telescope Imaging and Spectral Analysis of Two Brown Dwarf Binaries at the L Dwarf/T Dwarf Transition}",       journal = {\aj},          year = 2011,         month = mar,        volume = {141},        number = {3},           eid = {70},         pages = {70},           doi = {10.1088/0004-6256/141/3/70}, archivePrefix = {arXiv},        eprint = {1011.0798},  primaryClass = {astro-ph.SR},        adsurl = {https://ui.adsabs.harvard.edu/abs/2011AJ....141...70B} }

@ARTICLE{2015AJ....149..104B,        author = {{Burgasser}, Adam J. and {Gillon}, Micha{\"e}l and {Melis}, Carl and {Bowler}, Brendan P. and {Michelsen}, Eric L. and {Bardalez Gagliuffi}, Daniella and {Gelino}, Christopher R. and {Jehin}, E. and {Delrez}, L. and {Manfroid}, J. and {Blake}, Cullen H.},         title = "{WISE J072003.20-084651.2: an Old and Active M9.5 + T5 Spectral Binary 6 pc from the Sun}",       journal = {\aj},          year = 2015,         month = mar,        volume = {149},        number = {3},           eid = {104},         pages = {104},           doi = {10.1088/0004-6256/149/3/104}, archivePrefix = {arXiv},        eprint = {1410.4288},  primaryClass = {astro-ph.SR},        adsurl = {https://ui.adsabs.harvard.edu/abs/2015AJ....149..104B} }

@ARTICLE{2011AJ....141....7D,        author = {{Dhital}, Saurav and {Burgasser}, Adam J. and {Looper}, Dagny L. and {Stassun}, Keivan G.},         title = "{Resolved Spectroscopy of M Dwarf/L Dwarf Binaries. IV. Discovery of AN M9 + L6 Binary Separated by Over 100 AU}",       journal = {\aj},          year = 2011,         month = jan,        volume = {141},        number = {1},           eid = {7},         pages = {7},           doi = {10.1088/0004-6256/141/1/7}, archivePrefix = {arXiv},        eprint = {1010.1240},  primaryClass = {astro-ph.SR},        adsurl = {https://ui.adsabs.harvard.edu/abs/2011AJ....141....7D} }

@ARTICLE{2007ApJ...669L..97L,        author = {{Looper}, Dagny L. and {Burgasser}, Adam J. and {Kirkpatrick}, J. Davy and {Swift}, Brandon J.},         title = "{Discovery of an M9.5 Candidate Brown Dwarf in the TW Hydrae Association: DENIS J124514.1-442907}",       journal = {\apjl},          year = 2007,         month = nov,        volume = {669},        number = {2},         pages = {L97-L100},           doi = {10.1086/523812}, archivePrefix = {arXiv},        eprint = {0709.3290},  primaryClass = {astro-ph},        adsurl = {https://ui.adsabs.harvard.edu/abs/2007ApJ...669L..97L} }

@ARTICLE{2011ApJ...726...30M,        author = {{Mainzer}, A. and {Cushing}, Michael C. and {Skrutskie}, M. and {Gelino}, C.~R. and {Kirkpatrick}, J. Davy and {Jarrett}, T. and {Masci}, F. and {Marley}, Mark S. and {Saumon}, D. and {Wright}, E. and {Beaton}, R. and {Dietrich}, M. and {Eisenhardt}, P. and {Garnavich}, P. and {Kuhn}, O. and {Leisawitz}, D. and {Marsh}, K. and {McLean}, I. and {Padgett}, D. and {Rueff}, K.},         title = "{The First Ultra-cool Brown Dwarf Discovered by the Wide-field Infrared Survey Explorer}",       journal = {\apj},          year = 2011,         month = jan,        volume = {726},        number = {1},           eid = {30},         pages = {30},           doi = {10.1088/0004-637X/726/1/30}, archivePrefix = {arXiv},        eprint = {1011.2279},  primaryClass = {astro-ph.GA},        adsurl = {https://ui.adsabs.harvard.edu/abs/2011ApJ...726...30M} }

@ARTICLE{2006ApJ...639.1114R,        author = {{Reid}, I. Neill and {Lewitus}, E. and {Burgasser}, Adam J. and {Cruz}, K.~L.},         title = "{2MASS J22521073-1730134: A Resolved L/T Binary at 14 Parsecs}",       journal = {\apj},          year = 2006,         month = mar,        volume = {639},        number = {2},         pages = {1114-1119},           doi = {10.1086/499484},        adsurl = {https://ui.adsabs.harvard.edu/abs/2006ApJ...639.1114R} }

@ARTICLE{2010AJ....139.1045S,        author = {{Schmidt}, Sarah J. and {West}, Andrew A. and {Burgasser}, Adam J. and {Bochanski}, John J. and {Hawley}, Suzanne L.},         title = "{Discovery of an Unusually Blue L Dwarf Within 10 pc of the Sun}",       journal = {\aj},          year = 2010,         month = mar,        volume = {139},        number = {3},         pages = {1045-1050},           doi = {10.1088/0004-6256/139/3/1045}, archivePrefix = {arXiv},        eprint = {0912.3565},  primaryClass = {astro-ph.SR},        adsurl = {https://ui.adsabs.harvard.edu/abs/2010AJ....139.1045S} }

@ARTICLE{2007AJ....133.2320S,        author = {{Siegler}, Nick and {Close}, Laird M. and {Burgasser}, Adam J. and {Cruz}, Kelle L. and {Marois}, Christian and {Macintosh}, Bruce and {Barman}, Travis},         title = "{Discovery of a 66 mas Ultracool Binary with Laser Guide Star Adaptive Optics}",       journal = {\aj},          year = 2007,         month = may,        volume = {133},        number = {5},         pages = {2320-2326},           doi = {10.1086/513273}, archivePrefix = {arXiv},        eprint = {astro-ph/0702013},  primaryClass = {astro-ph},        adsurl = {https://ui.adsabs.harvard.edu/abs/2007AJ....133.2320S} }

@ARTICLE{2010ApJ...715..561A,        author = {{Allers}, K.~N. and {Liu}, Michael C. and {Dupuy}, Trent J. and {Cushing}, Michael C.},         title = "{Discovery of A Young L Dwarf Binary, SDSS J224953.47+004404.6AB}",       journal = {\apj},          year = 2010,         month = may,        volume = {715},        number = {1},         pages = {561-571},           doi = {10.1088/0004-637X/715/1/561}, archivePrefix = {arXiv},        eprint = {0912.4687},  primaryClass = {astro-ph.SR},        adsurl = {https://ui.adsabs.harvard.edu/abs/2010ApJ...715..561A} }

@ARTICLE{2013ApJ...772...79A,        author = {{Allers}, K.~N. and {Liu}, Michael C.},         title = "{A Near-infrared Spectroscopic Study of Young Field Ultracool Dwarfs}",       journal = {\apj},          year = 2013,         month = aug,        volume = {772},        number = {2},           eid = {79},         pages = {79},           doi = {10.1088/0004-637X/772/2/79}, archivePrefix = {arXiv},        eprint = {1305.4418},  primaryClass = {astro-ph.SR},        adsurl = {https://ui.adsabs.harvard.edu/abs/2013ApJ...772...79A} }

@ARTICLE{2017ApJ...843L...4B,        author = {{Best}, William M.~J. and {Liu}, Michael C. and {Dupuy}, Trent J. and {Magnier}, Eugene A.},         title = "{The Young L Dwarf 2MASS J11193254-1137466 Is a Planetary-mass Binary}",       journal = {\apjl},          year = 2017,         month = jul,        volume = {843},        number = {1},           eid = {L4},         pages = {L4},           doi = {10.3847/2041-8213/aa76df}, archivePrefix = {arXiv},        eprint = {1706.01883},  primaryClass = {astro-ph.SR},        adsurl = {https://ui.adsabs.harvard.edu/abs/2017ApJ...843L...4B} }

@ARTICLE{2009AJ....138.1563B,        author = {{Burgasser}, Adam J. and {Dhital}, Saurav and {West}, Andrew A.},         title = "{Resolved Spectroscopy of M Dwarf/L Dwarf Binaries. III. The ``Wide'' L3.5/L4 Dwarf Binary 2Mass J15500845+1455180AB}",       journal = {\aj},          year = 2009,         month = dec,        volume = {138},        number = {6},         pages = {1563-1569},           doi = {10.1088/0004-6256/138/6/1563}, archivePrefix = {arXiv},        eprint = {0909.3885},  primaryClass = {astro-ph.SR},        adsurl = {https://ui.adsabs.harvard.edu/abs/2009AJ....138.1563B} }

@ARTICLE{2017MNRAS.468.3499D,        author = {{Deacon}, N.~R. and {Magnier}, E.~A. and {Best}, William M.~J. and {Liu}, Michael C. and {Dupuy}, T.~J. and {Chambers}, K.~C. and {Draper}, P.~W. and {Flewelling}, H. and {Metcalfe}, N. and {Tonry}, J.~L. and {Wainscoat}, R.~J. and {Waters}, C.},         title = "{Identification of partially resolved binaries in Pan-STARRS 1 data}",       journal = {\mnras},          year = 2017,         month = jul,        volume = {468},        number = {3},         pages = {3499-3515},           doi = {10.1093/mnras/stx440}, archivePrefix = {arXiv},        eprint = {1702.05491},  primaryClass = {astro-ph.SR},        adsurl = {https://ui.adsabs.harvard.edu/abs/2017MNRAS.468.3499D} }

@ARTICLE{2012ApJS..201...19D,        author = {{Dupuy}, Trent J. and {Liu}, Michael C.},         title = "{The Hawaii Infrared Parallax Program. I. Ultracool Binaries and the L/T Transition}",       journal = {\apjs},          year = 2012,         month = aug,        volume = {201},        number = {2},           eid = {19},         pages = {19},           doi = {10.1088/0067-0049/201/2/19}, archivePrefix = {arXiv},        eprint = {1201.2465},  primaryClass = {astro-ph.SR},        adsurl = {https://ui.adsabs.harvard.edu/abs/2012ApJS..201...19D} }

@ARTICLE{2015ApJ...805...56D,        author = {{Dupuy}, Trent J. and {Liu}, Michael C. and {Leggett}, S.~K. and {Ireland}, Michael J. and {Chiu}, Kuenley and {Golimowski}, David A.},         title = "{The Mass-Luminosity Relation in the L/T Transition: Individual Dynamical Masses for the New J-band Flux Reversal Binary SDSSJ105213.51+442255.7AB}",       journal = {\apj},          year = 2015,         month = may,        volume = {805},        number = {1},           eid = {56},         pages = {56},           doi = {10.1088/0004-637X/805/1/56}, archivePrefix = {arXiv},        eprint = {1503.06212},  primaryClass = {astro-ph.SR},        adsurl = {https://ui.adsabs.harvard.edu/abs/2015ApJ...805...56D} }

@ARTICLE{2014AJ....148....6G,        author = {{Gelino}, Christopher R. and {Smart}, R.~L. and {Marocco}, Federico and {Kirkpatrick}, J. Davy and {Cushing}, Michael C. and {Mace}, Gregory and {Mendez}, Rene A. and {Tinney}, C.~G. and {Jones}, Hugh R.~A.},         title = "{WISEP J061135.13-041024.0 AB: A J-band Flux Reversal Binary at the L/T Transition}",       journal = {\aj},          year = 2014,         month = jul,        volume = {148},        number = {1},           eid = {6},         pages = {6},           doi = {10.1088/0004-6256/148/1/6},        adsurl = {https://ui.adsabs.harvard.edu/abs/2014AJ....148....6G} }

@ARTICLE{2012ApJ...758...57L,        author = {{Liu}, Michael C. and {Dupuy}, Trent J. and {Bowler}, Brendan P. and {Leggett}, S.~K. and {Best}, William M.~J.},         title = "{Two Extraordinary Substellar Binaries at the T/Y Transition and the Y-band Fluxes of the Coolest Brown Dwarfs}",       journal = {\apj},          year = 2012,         month = oct,        volume = {758},        number = {1},           eid = {57},         pages = {57},           doi = {10.1088/0004-637X/758/1/57}, archivePrefix = {arXiv},        eprint = {1206.4044},  primaryClass = {astro-ph.SR},        adsurl = {https://ui.adsabs.harvard.edu/abs/2012ApJ...758...57L} }

@ARTICLE{2006AA...456..253M,        author = {{Mart{\'\i}n}, E.~L. and {Brandner}, W. and {Bouy}, H. and {Basri}, G. and {Davis}, J. and {Deshpande}, R. and {Montgomery}, M.~M.},         title = "{Resolved Hubble space spectroscopy of ultracool binary systems}",       journal = {\aap},          year = 2006,         month = sep,        volume = {456},        number = {1},         pages = {253-259},           doi = {10.1051/0004-6361:20054186}, archivePrefix = {arXiv},        eprint = {physics/0605042},  primaryClass = {physics.space-ph},        adsurl = {https://ui.adsabs.harvard.edu/abs/2006A&A...456..253M} }

@ARTICLE{2015AA...579A..61S,        author = {{Sahlmann}, J. and {Burgasser}, A.~J. and {Mart{\'\i}n}, E.~L. and {Lazorenko}, P.~F. and {Bardalez Gagliuffi}, D.~C. and {Mayor}, M. and {S{\'e}gransan}, D. and {Queloz}, D. and {Udry}, S.},         title = "{DE0823-49 is a juvenile binary brown dwarf at 20.7 pc}",       journal = {\aap},          year = 2015,         month = jul,        volume = {579},           eid = {A61},         pages = {A61},           doi = {10.1051/0004-6361/201425536}, archivePrefix = {arXiv},        eprint = {1505.07972},  primaryClass = {astro-ph.SR},        adsurl = {https://ui.adsabs.harvard.edu/abs/2015A&A...579A..61S} }

@ARTICLE{2010ApJS..190..100K,        author = {{Kirkpatrick}, J. Davy and {Looper}, Dagny L. and {Burgasser}, Adam J. and {Schurr}, Steven D. and {Cutri}, Roc M. and {Cushing}, Michael C. and {Cruz}, Kelle L. and {Sweet}, Anne C. and {Knapp}, Gillian R. and {Barman}, Travis S. and {Bochanski}, John J. and {Roellig}, Thomas L. and {McLean}, Ian S. and {McGovern}, Mark R. and {Rice}, Emily L.},         title = "{Discoveries from a Near-infrared Proper Motion Survey Using Multi-epoch Two Micron All-Sky Survey Data}",       journal = {\apjs},          year = 2010,         month = sep,        volume = {190},        number = {1},         pages = {100-146},           doi = {10.1088/0067-0049/190/1/100}, archivePrefix = {arXiv},        eprint = {1008.3591},  primaryClass = {astro-ph.SR},        adsurl = {https://ui.adsabs.harvard.edu/abs/2010ApJS..190..100K} }

@ARTICLE{spherex,
       author = {{Dor{\'e}}, Olivier and {Bock}, Jamie and {Ashby}, Matthew and {Capak}, Peter and {Cooray}, Asantha and {de Putter}, Roland and {Eifler}, Tim and {Flagey}, Nicolas and {Gong}, Yan and {Habib}, Salman and {Heitmann}, Katrin and {Hirata}, Chris and {Jeong}, Woong-Seob and {Katti}, Raj and {Korngut}, Phil and {Krause}, Elisabeth and {Lee}, Dae-Hee and {Masters}, Daniel and {Mauskopf}, Phil and {Melnick}, Gary and {Mennesson}, Bertrand and {Nguyen}, Hien and {{\"O}berg}, Karin and {Pullen}, Anthony and {Raccanelli}, Alvise and {Smith}, Roger and {Song}, Yong-Seon and {Tolls}, Volker and {Unwin}, Steve and {Venumadhav}, Tejaswi and {Viero}, Marco and {Werner}, Mike and {Zemcov}, Mike},
        title = "{Cosmology with the SPHEREX All-Sky Spectral Survey}",
      journal = {arXiv e-prints},
         year = 2014,
        month = dec,
          eid = {arXiv:1412.4872},
        pages = {arXiv:1412.4872},
          doi = {10.48550/arXiv.1412.4872},
archivePrefix = {arXiv},
       eprint = {1412.4872},
 primaryClass = {astro-ph.CO},
       adsurl = {https://ui.adsabs.harvard.edu/abs/2014arXiv1412.4872D}
}

@ARTICLE{Luminosity_Bardalez,
       author = {{Bardalez Gagliuffi}, Daniella C. and {Burgasser}, Adam J. and {Schmidt}, Sarah J. and {Theissen}, Christopher and {Gagn{\'e}}, Jonathan and {Gillon}, Michael and {Sahlmann}, Johannes and {Faherty}, Jacqueline K. and {Gelino}, Christopher and {Cruz}, Kelle L. and {Skrzypek}, Nathalie and {Looper}, Dagny},
        title = "{The Ultracool SpeXtroscopic Survey. I. Volume-limited Spectroscopic Sample and Luminosity Function of M7-L5 Ultracool Dwarfs}",
      journal = {\apj},
         year = 2019,
        month = oct,
       volume = {883},
       number = {2},
          eid = {205},
        pages = {205},
          doi = {10.3847/1538-4357/ab253d},
archivePrefix = {arXiv},
       eprint = {1906.04166},
 primaryClass = {astro-ph.SR},
       adsurl = {https://ui.adsabs.harvard.edu/abs/2019ApJ...883..205B}
}

@ARTICLE{2008ApJ...689L..53B,
       author = {{Burgasser}, Adam J. and {Tinney}, C.~G. and {Cushing}, Michael C. and {Saumon}, Didier and {Marley}, Mark S. and {Bennett}, Clara S. and {Kirkpatrick}, J. Davy},
        title = "{2MASS J09393548-2448279: The Coldest and Least Luminous Brown Dwarf Binary Known?}",
      journal = {\apjl},
         year = 2008,
        month = dec,
       volume = {689},
       number = {1},
        pages = {L53},
          doi = {10.1086/595747},
       adsurl = {https://ui.adsabs.harvard.edu/abs/2008ApJ...689L..53B}
}

@ARTICLE{2013ApJ...778...36R,
       author = {{Radigan}, Jacqueline and {Jayawardhana}, Ray and {Lafreni{\`e}re}, David and {Dupuy}, Trent J. and {Liu}, Michael C. and {Scholz}, Alexander},
        title = "{Discovery of a Visual T-dwarf Triple System and Binarity at the L/T Transition}",
      journal = {\apj},
         year = 2013,
        month = nov,
       volume = {778},
       number = {1},
          eid = {36},
        pages = {36},
          doi = {10.1088/0004-637X/778/1/36},
archivePrefix = {arXiv},
       eprint = {1308.5702},
 primaryClass = {astro-ph.SR},
       adsurl = {https://ui.adsabs.harvard.edu/abs/2013ApJ...778...36R}
}

@ARTICLE{2021MNRAS.500.5453S,
       author = {{Sahlmann}, J. and {Dupuy}, T.~J. and {Burgasser}, A.~J. and {Filippazzo}, J.~C. and {Mart{\'\i}n}, E.~L. and {Bardalez Gagliuffi}, D.~C. and {Hsu}, C. and {Lazorenko}, P.~F. and {Liu}, Michael C.},
        title = "{Individual dynamical masses of DENIS J063001.4-184014AB reveal a likely young brown dwarf triple}",
      journal = {\mnras},
         year = 2021,
        month = feb,
       volume = {500},
       number = {4},
        pages = {5453-5461},
          doi = {10.1093/mnras/staa3577},
archivePrefix = {arXiv},
       eprint = {2011.08266},
 primaryClass = {astro-ph.SR},
       adsurl = {https://ui.adsabs.harvard.edu/abs/2021MNRAS.500.5453S}
}

@ARTICLE{2010MNRAS.402..620R,
       author = {{Rebassa-Mansergas}, A. and {G{\"a}nsicke}, B.~T. and {Schreiber}, M.~R. and {Koester}, D. and {Rodr{\'\i}guez-Gil}, P.},
        title = "{Post-common envelope binaries from SDSS - VII. A catalogue of white dwarf-main sequence binaries}",
      journal = {\mnras},
         year = 2010,
        month = feb,
       volume = {402},
       number = {1},
        pages = {620-640},
          doi = {10.1111/j.1365-2966.2009.15915.x},
archivePrefix = {arXiv},
       eprint = {0910.4406},
 primaryClass = {astro-ph.SR},
       adsurl = {https://ui.adsabs.harvard.edu/abs/2010MNRAS.402..620R}
}

@ARTICLE{2003AJ....125.2621R,
       author = {{Raymond}, Sean N. and {Szkody}, Paula and {Hawley}, Suzanne L. and {Anderson}, Scott F. and {Brinkmann}, J. and {Covey}, Kevin R. and {McGehee}, P.~M. and {Schneider}, D.~P. and {West}, Andrew A. and {York}, D.~G.},
        title = "{A First Look at White Dwarf-M Dwarf Pairs in the Sloan Digital Sky Survey}",
      journal = {\aj},
         year = 2003,
        month = may,
       volume = {125},
       number = {5},
        pages = {2621-2629},
          doi = {10.1086/374762},
archivePrefix = {arXiv},
       eprint = {astro-ph/0302405},
 primaryClass = {astro-ph},
       adsurl = {https://ui.adsabs.harvard.edu/abs/2003AJ....125.2621R}
}

@ARTICLE{2022MNRAS.513..516F,
       author = {{Feeser}, S. Jean and {Best}, William M.~J.},
        title = "{Using old and new approaches: determining physical properties of brown dwarfs with empirical relations and machine learning models}",
      journal = {\mnras},
         year = 2022,
        month = jun,
       volume = {513},
       number = {1},
        pages = {516-535},
          doi = {10.1093/mnras/stac855},
archivePrefix = {arXiv},
       eprint = {2203.13829},
 primaryClass = {astro-ph.SR},
       adsurl = {https://ui.adsabs.harvard.edu/abs/2022MNRAS.513..516F}
}

@ARTICLE{2024RNAAS...8..102Z,
       author = {{Zhou}, Tianxing and {Theissen}, Christopher A. and {Burgasser}, Adam J. and {Best}, William M.~J. and {Feeser}, S. Jean},
        title = "{Spectral Typing with Artificial Intelligence: Classifying Low-resolution Near-infrared Spectra of Standard M/L/T Dwarfs}",
      journal = {Research Notes of the American Astronomical Society},
         year = 2024,
        month = apr,
       volume = {8},
       number = {4},
          eid = {102},
        pages = {102},
          doi = {10.3847/2515-5172/ad3f16},
       adsurl = {https://ui.adsabs.harvard.edu/abs/2024RNAAS...8..102Z}
}

@ARTICLE{2024ApJS..271...55K,
       author = {{Kirkpatrick}, J. Davy and {Marocco}, Federico and {Gelino}, Christopher R. and {Raghu}, Yadukrishna and {Faherty}, Jacqueline K. and {Bardalez Gagliuffi}, Daniella C. and {Schurr}, Steven D. and {Apps}, Kevin and {Schneider}, Adam C. and {Meisner}, Aaron M. and {Kuchner}, Marc J. and {Caselden}, Dan and {Smart}, R.~L. and {Casewell}, S.~L. and {Raddi}, Roberto and {Kesseli}, Aurora and {Stevnbak Andersen}, Nikolaj and {Antonini}, Edoardo and {Beaulieu}, Paul and {Bickle}, Thomas P. and {Bilsing}, Martin and {Chieng}, Raymond and {Colin}, Guillaume and {Deen}, Sam and {Dereveanco}, Alexandru and {Doll}, Katharina and {Durantini Luca}, Hugo A. and {Frazer}, Anya and {Gantier}, Jean Marc and {Gramaize}, L{\'e}opold and {Grant}, Kristin and {Hamlet}, Leslie K. and {Higashimura}, Hiro and {Hyogo}, Michiharu and {Ja{\l}owiczor}, Peter A. and {Jonkeren}, Alexander and {Kabatnik}, Martin and {Kiwy}, Frank and {Martin}, David W. and {Michaels}, Marianne N. and {Pendrill}, William and {Pessanha Machado}, Celso and {Pumphrey}, Benjamin and {Rothermich}, Austin and {Russwurm}, Rebekah and {Sainio}, Arttu and {Sanchez}, John and {Sapelkin-Tambling}, Fyodor Theo and {Sch{\"u}mann}, J{\"o}rg and {Selg-Mann}, Karl and {Singh}, Harshdeep and {Stenner}, Andres and {Sun}, Guoyou and {Tanner}, Christopher and {Th{\'e}venot}, Melina and {Ventura}, Maurizio and {Voloshin}, Nikita V. and {Walla}, Jim and {W{\k{e}}dracki}, Zbigniew and {Adorno}, Jose I. and {Aganze}, Christian and {Allers}, Katelyn N. and {Brooks}, Hunter and {Burgasser}, Adam J. and {Calamari}, Emily and {Connor}, Thomas and {Costa}, Edgardo and {Eisenhardt}, Peter R. and {Gagn{\'e}}, Jonathan and {Gerasimov}, Roman and {Gonzales}, Eileen C. and {Hsu}, Chih-Chun and {Kiman}, Rocio and {Li}, Guodong and {Low}, Ryan and {Mamajek}, Eric and {Pantoja}, Blake M. and {Popinchalk}, Mark and {Rees}, Jon M. and {Stern}, Daniel and {Su{\'a}rez}, Genaro and {Theissen}, Christopher and {Tsai}, Chao-Wei and {Vos}, Johanna M. and {Zurek}, David and {The Backyard Worlds: Planet 9 Collaboration}},
        title = "{The Initial Mass Function Based on the Full-sky 20 pc Census of {\ensuremath{\sim}}3600 Stars and Brown Dwarfs}",
      journal = {\apjs},
         year = 2024,
        month = apr,
       volume = {271},
       number = {2},
          eid = {55},
        pages = {55},
          doi = {10.3847/1538-4365/ad24e2},
archivePrefix = {arXiv},
       eprint = {2312.03639},
 primaryClass = {astro-ph.SR},
       adsurl = {https://ui.adsabs.harvard.edu/abs/2024ApJS..271...55K}
}

@ARTICLE{2024A&A...690A.198A,
       author = {{Angthopo}, J. and {Granett}, B.~R. and {La Barbera}, F. and {Longhetti}, M. and {Iovino}, A. and {Fossati}, M. and {Ditrani}, F.~R. and {Costantin}, L. and {Zibetti}, S. and {Gallazzi}, A. and {S{\'a}nchez-Bl{\'a}zquez}, P. and {Tortora}, C. and {Spiniello}, C. and {Poggianti}, B. and {Vazdekis}, A. and {Balcells}, M. and {Bardelli}, S. and {Benn}, C.~R. and {Bianconi}, M. and {Bolzonella}, M. and {Busarello}, G. and {Cassar{\`a}}, L.~P. and {Corsini}, E.~M. and {Cucciati}, O. and {Dalton}, G. and {Ferr{\'e}-Mateu}, A. and {Garc{\'\i}a-Benito}, R. and {Gonz{\'a}lez Delgado}, R.~M. and {Gafton}, E. and {Gullieuszik}, M. and {Haines}, C.~P. and {Iodice}, E. and {Ikhsanova}, A. and {Jin}, S. and {Knapen}, J.~H. and {McGee}, S. and {Mercurio}, A. and {Merluzzi}, P. and {Morelli}, L. and {Moretti}, A. and {Murphy}, D.~N.~A. and {Pizzella}, A. and {Pozzetti}, L. and {Ragusa}, R. and {Trager}, S.~C. and {Vergani}, D. and {Vulcani}, B. and {Talia}, M. and {Zucca}, E.},
        title = "{Retrieval of the physical parameters of galaxies from WEAVE-StePS-like data using machine learning}",
      journal = {\aap},
         year = 2024,
        month = oct,
       volume = {690},
          eid = {A198},
        pages = {A198},
          doi = {10.1051/0004-6361/202449979},
archivePrefix = {arXiv},
       eprint = {2406.11748},
 primaryClass = {astro-ph.GA},
       adsurl = {https://ui.adsabs.harvard.edu/abs/2024A&A...690A.198A}
}

@ARTICLE{2022ApJ...930..136L,
       author = {{Lueber}, Anna and {Kitzmann}, Daniel and {Bowler}, Brendan P. and {Burgasser}, Adam J. and {Heng}, Kevin},
        title = "{Retrieval Study of Brown Dwarfs across the L-T Sequence}",
      journal = {\apj},
         year = 2022,
        month = may,
       volume = {930},
       number = {2},
          eid = {136},
        pages = {136},
          doi = {10.3847/1538-4357/ac63b9},
archivePrefix = {arXiv},
       eprint = {2204.01330},
 primaryClass = {astro-ph.EP},
       adsurl = {https://ui.adsabs.harvard.edu/abs/2022ApJ...930..136L}
}

@article{2005ApJ...623.1115C,
	adsurl = {http://ads.ari.uni-heidelberg.de/abs/2005ApJ...623.1115C},
	author = {{Cushing}, M.~C. and {Rayner}, J.~T. and {Vacca}, W.~D.},
	doi = {10.1086/428040},
	eprint = {arXiv:astro-ph/0412313},
	journal = {\apj},
	month = apr,
	pages = {1115-1140},
	title = {{An Infrared Spectroscopic Sequence of M, L, and T Dwarfs}},
	volume = 623,
	year = 2005}

@ARTICLE{2022ApJ...935..167A,
       author = {{Astropy Collaboration} and {Price-Whelan}, Adrian M. and {Lim}, Pey Lian and {Earl}, Nicholas and {Starkman}, Nathaniel and {Bradley}, Larry and {Shupe}, David L. and {Patil}, Aarya A. and {Corrales}, Lia and {Brasseur}, C.~E. and {N{\"o}the}, Maximilian and {Donath}, Axel and {Tollerud}, Erik and {Morris}, Brett M. and {Ginsburg}, Adam and {Vaher}, Eero and {Weaver}, Benjamin A. and {Tocknell}, James and {Jamieson}, William and {van Kerkwijk}, Marten H. and {Robitaille}, Thomas P. and {Merry}, Bruce and {Bachetti}, Matteo and {G{\"u}nther}, H. Moritz and {Aldcroft}, Thomas L. and {Alvarado-Montes}, Jaime A. and {Archibald}, Anne M. and {B{\'o}di}, Attila and {Bapat}, Shreyas and {Barentsen}, Geert and {Baz{\'a}n}, Juanjo and {Biswas}, Manish and {Boquien}, M{\'e}d{\'e}ric and {Burke}, D.~J. and {Cara}, Daria and {Cara}, Mihai and {Conroy}, Kyle E. and {Conseil}, Simon and {Craig}, Matthew W. and {Cross}, Robert M. and {Cruz}, Kelle L. and {D'Eugenio}, Francesco and {Dencheva}, Nadia and {Devillepoix}, Hadrien A.~R. and {Dietrich}, J{\"o}rg P. and {Eigenbrot}, Arthur Davis and {Erben}, Thomas and {Ferreira}, Leonardo and {Foreman-Mackey}, Daniel and {Fox}, Ryan and {Freij}, Nabil and {Garg}, Suyog and {Geda}, Robel and {Glattly}, Lauren and {Gondhalekar}, Yash and {Gordon}, Karl D. and {Grant}, David and {Greenfield}, Perry and {Groener}, Austen M. and {Guest}, Steve and {Gurovich}, Sebastian and {Handberg}, Rasmus and {Hart}, Akeem and {Hatfield-Dodds}, Zac and {Homeier}, Derek and {Hosseinzadeh}, Griffin and {Jenness}, Tim and {Jones}, Craig K. and {Joseph}, Prajwel and {Kalmbach}, J. Bryce and {Karamehmetoglu}, Emir and {Ka{\l}uszy{\'n}ski}, Miko{\l}aj and {Kelley}, Michael S.~P. and {Kern}, Nicholas and {Kerzendorf}, Wolfgang E. and {Koch}, Eric W. and {Kulumani}, Shankar and {Lee}, Antony and {Ly}, Chun and {Ma}, Zhiyuan and {MacBride}, Conor and {Maljaars}, Jakob M. and {Muna}, Demitri and {Murphy}, N.~A. and {Norman}, Henrik and {O'Steen}, Richard and {Oman}, Kyle A. and {Pacifici}, Camilla and {Pascual}, Sergio and {Pascual-Granado}, J. and {Patil}, Rohit R. and {Perren}, Gabriel I. and {Pickering}, Timothy E. and {Rastogi}, Tanuj and {Roulston}, Benjamin R. and {Ryan}, Daniel F. and {Rykoff}, Eli S. and {Sabater}, Jose and {Sakurikar}, Parikshit and {Salgado}, Jes{\'u}s and {Sanghi}, Aniket and {Saunders}, Nicholas and {Savchenko}, Volodymyr and {Schwardt}, Ludwig and {Seifert-Eckert}, Michael and {Shih}, Albert Y. and {Jain}, Anany Shrey and {Shukla}, Gyanendra and {Sick}, Jonathan and {Simpson}, Chris and {Singanamalla}, Sudheesh and {Singer}, Leo P. and {Singhal}, Jaladh and {Sinha}, Manodeep and {Sip{\H{o}}cz}, Brigitta M. and {Spitler}, Lee R. and {Stansby}, David and {Streicher}, Ole and {{\v{S}}umak}, Jani and {Swinbank}, John D. and {Taranu}, Dan S. and {Tewary}, Nikita and {Tremblay}, Grant R. and {de Val-Borro}, Miguel and {Van Kooten}, Samuel J. and {Vasovi{\'c}}, Zlatan and {Verma}, Shresth and {de Miranda Cardoso}, Jos{\'e} Vin{\'\i}cius and {Williams}, Peter K.~G. and {Wilson}, Tom J. and {Winkel}, Benjamin and {Wood-Vasey}, W.~M. and {Xue}, Rui and {Yoachim}, Peter and {Zhang}, Chen and {Zonca}, Andrea and {Astropy Project Contributors}},
        title = "{The Astropy Project: Sustaining and Growing a Community-oriented Open-source Project and the Latest Major Release (v5.0) of the Core Package}",
      journal = {\apj},
         year = 2022,
        month = aug,
       volume = {935},
       number = {2},
          eid = {167},
        pages = {167},
          doi = {10.3847/1538-4357/ac7c74},
archivePrefix = {arXiv},
       eprint = {2206.14220},
 primaryClass = {astro-ph.IM},
       adsurl = {https://ui.adsabs.harvard.edu/abs/2022ApJ...935..167A}
}

@ARTICLE{Ashraf2022,
       author = {{Ashraf}, Afra and {Bardalez Gagliuffi}, Daniella C. and {Manjavacas}, Elena and {Vos}, Johanna M. and {Mechmann}, Claire and {Faherty}, Jacqueline K.},
        title = "{Disentangling the Signatures of Blended-light Atmospheres in L/T Transition Brown Dwarfs}",
      journal = {\apj},
         year = 2022,
        month = aug,
       volume = {934},
       number = {2},
          eid = {178},
        pages = {178},
          doi = {10.3847/1538-4357/ac7aab},
archivePrefix = {arXiv},
       eprint = {2206.09025},
 primaryClass = {astro-ph.SR},
       adsurl = {https://ui.adsabs.harvard.edu/abs/2022ApJ...934..178A}
}

@ARTICLE{2012AJ....144...93M,
       author = {{Morgan}, Dylan P. and {West}, Andrew A. and {Garc{\'e}s}, Ane and {Catal{\'a}n}, Silvia and {Dhital}, Saurav and {Fuchs}, Miriam and {Silvestri}, Nicole M.},
        title = "{The Effects of Close Companions (and Rotation) on the Magnetic Activity of M Dwarfs}",
      journal = {\aj},
         year = 2012,
        month = oct,
       volume = {144},
       number = {4},
          eid = {93},
        pages = {93},
          doi = {10.1088/0004-6256/144/4/93},
archivePrefix = {arXiv},
       eprint = {1205.6806},
 primaryClass = {astro-ph.SR},
       adsurl = {https://ui.adsabs.harvard.edu/abs/2012AJ....144...93M}
}

@ARTICLE{2020JATIS...6d6001M,
       author = {{Mosby}, Gregory and {Rauscher}, Bernard J. and {Bennett}, Chris and {Cheng}, Edward S. and {Cheung}, Stephanie and {Cillis}, Analia and {Content}, David and {Cottingham}, Dave and {Foltz}, Roger and {Gygax}, John and {Hill}, Robert J. and {Kruk}, Jeffrey W. and {Mah}, Jon and {Meier}, Lane and {Merchant}, Chris and {Miko}, Laddawan and {Piquette}, Eric C. and {Waczynski}, Augustyn and {Wen}, Yiting},
        title = "{Properties and characteristics of the Nancy Grace Roman Space Telescope H4RG-10 detectors}",
      journal = {Journal of Astronomical Telescopes, Instruments, and Systems},
         year = 2020,
        month = oct,
       volume = {6},
          eid = {046001},
        pages = {046001},
          doi = {10.1117/1.JATIS.6.4.046001},
archivePrefix = {arXiv},
       eprint = {2005.00505},
 primaryClass = {astro-ph.IM},
       adsurl = {https://ui.adsabs.harvard.edu/abs/2020JATIS...6d6001M}
}

@article{peterson1954theory,
  title={The theory of signal detectability},
  author={Peterson, WWTG and Birdsall, T and Fox, We},
  journal={Transactions of the IRE professional group on information theory},
  volume={4},
  number={4},
  pages={171--212},
  year={1954},
  publisher={IEEE}
}

@ARTICLE{2024ApJ...975..162L,
       author = {{Luhman}, K.~L. and {Alves de Oliveira}, C. and {Baraffe}, I. and {Chabrier}, G. and {Manjavacas}, E. and {Parker}, R.~J. and {Tremblin}, P.},
        title = "{JWST/NIRSpec Observations of Brown Dwarfs in the Orion Nebula Cluster}",
      journal = {\apj},
         year = 2024,
        month = nov,
       volume = {975},
       number = {2},
          eid = {162},
        pages = {162},
          doi = {10.3847/1538-4357/ad7b19},
archivePrefix = {arXiv},
       eprint = {2410.10000},
 primaryClass = {astro-ph.GA},
       adsurl = {https://ui.adsabs.harvard.edu/abs/2024ApJ...975..162L}
}

@article{2004ApJS..155..191B,
	adsurl = {http://ads.ari.uni-heidelberg.de/abs/2004ApJS..155..191B},
	author = {{Burgasser}, A.~J.},
	doi = {10.1086/424386},
	eprint = {arXiv:astro-ph/0407624},
	journal = {\apjs},
	month = nov,
	pages = {191-207},
	title = {{T Dwarfs and the Substellar Mass Function. I. Monte Carlo Simulations}},
	volume = 155,
	year = 2004}

@article{2003A&A...402..701B,
	adsurl = {http://adsabs.harvard.edu/abs/2003A%26A...402..701B},
	author = {{Baraffe}, I. and {Chabrier}, G. and {Barman}, T.~S. and {Allard}, F. and {Hauschildt}, P.~H.},
	doi = {10.1051/0004-6361:20030252},
	eprint = {arXiv:astro-ph/0302293},
	journal = {\aap},
	month = may,
	pages = {701-712},
	title = {{Evolutionary models for cool brown dwarfs and extrasolar giant planets. The case of HD 209458}},
	volume = 402,
	year = 2003}

@article{2013ApJS..208....9P,
	adsurl = {http://adsabs.harvard.edu/abs/2013ApJS..208....9P},
	archiveprefix = {arXiv},
	author = {{Pecaut}, M.~J. and {Mamajek}, E.~E.},
	doi = {10.1088/0067-0049/208/1/9},
	eid = {9},
	eprint = {1307.2657},
	journal = {\apjs},
	month = sep,
	pages = {9},
	primaryclass = {astro-ph.SR},
	title = {{Intrinsic Colors, Temperatures, and Bolometric Corrections of Pre-main-sequence Stars}},
	volume = 208,
	year = 2013}

@ARTICLE{2019ApJ...883..205B,
       author = {{Bardalez Gagliuffi}, Daniella C. and {Burgasser}, Adam J. and {Schmidt}, Sarah J. and {Theissen}, Christopher and {Gagn{\'e}}, Jonathan and {Gillon}, Michael and {Sahlmann}, Johannes and {Faherty}, Jacqueline K. and {Gelino}, Christopher and {Cruz}, Kelle L. and {Skrzypek}, Nathalie and {Looper}, Dagny},
        title = "{The Ultracool SpeXtroscopic Survey. I. Volume-limited Spectroscopic Sample and Luminosity Function of M7-L5 Ultracool Dwarfs}",
      journal = {\apj},
         year = 2019,
        month = oct,
       volume = {883},
       number = {2},
          eid = {205},
        pages = {205},
          doi = {10.3847/1538-4357/ab253d},
archivePrefix = {arXiv},
       eprint = {1906.04166},
 primaryClass = {astro-ph.SR},
       adsurl = {https://ui.adsabs.harvard.edu/abs/2019ApJ...883..205B}
}

@ARTICLE{2024ApJ...973..107B,
       author = {{Beiler}, Samuel A. and {Cushing}, Michael C. and {Kirkpatrick}, J. Davy and {Schneider}, Adam C. and {Mukherjee}, Sagnick and {Marley}, Mark S. and {Marocco}, Federico and {Smart}, Richard L.},
        title = "{Precise Bolometric Luminosities and Effective Temperatures of 23 Late-T and Y Dwarfs Obtained with JWST}",
      journal = {\apj},
         year = 2024,
        month = oct,
       volume = {973},
       number = {2},
          eid = {107},
        pages = {107},
          doi = {10.3847/1538-4357/ad6301},
archivePrefix = {arXiv},
       eprint = {2407.08518},
 primaryClass = {astro-ph.SR},
       adsurl = {https://ui.adsabs.harvard.edu/abs/2024ApJ...973..107B}
}

@ARTICLE{2025arXiv250322497V,
       author = {\v{Z}erjal, M. and {Dominguez-Tagle}, C. and {Sedighi}, N. and {Mart\'{i}n}, E.~L. and {Lodieu}, N. and {Goldman}, B. and {Reyl\'{e}}, C. and {Smart}, R.~L. and {Mohandasan}, A. and {Zapatero Osorio}, M.~R. and {Barrado}, D. and {Mas Buitrago}, P. and {Vitas}, N. and {Cruz}, P. and {B\'{e}jar}, V.~J.~S. and {Bouy}, H. and {Burgasser}, A. and {Mu\~{n}oz Torres}, S. and {Phan-Bao}, N. and {Solano}, E. and {Tata}, R. and {Tsilia}, S. and {Zhang}, J. -Y. and {Aghanim}, N. and {Altieri}, B. and {Amara}, A. and {Andreon}, S. and {Auricchio}, N. and {Baccigalupi}, C. and {Baldi}, M. and {Balestra}, A. and {Bardelli}, S. and {Battaglia}, P. and {Biviano}, A. and {Bonchi}, A. and {Branchini}, E. and {Brescia}, M. and {Brinchmann}, J. and {Camera}, S. and {Ca\~{n}as-Herrera}, G. and {Capobianco}, V. and {Carbone}, C. and {Carretero}, J. and {Casas}, S. and {Castellano}, M. and {Castignani}, G. and {Cavuoti}, S. and {Chambers}, K.~C. and {Cimatti}, A. and {Colodro-Conde}, C. and {Congedo}, G. and {Conselice}, C.~J. and {Conversi}, L. and {Copin}, Y. and {Courbin}, F. and {Courtois}, H.~M. and {Cropper}, M. and {Cuby}, J. -G. and {Da Silva}, A. and {Degaudenzi}, H. and {De Lucia}, G. and {Dolding}, C. and {Dole}, H. and {Douspis}, M. and {Dubath}, F. and {Dupac}, X. and {Dusini}, S. and {Escoffier}, S. and {Farina}, M. and {Faustini}, F. and {Ferriol}, S. and {Fotopoulou}, S. and {Frailis}, M. and {Franceschi}, E. and {Galeotta}, S. and {George}, K. and {Gillis}, B. and {Giocoli}, C. and {G\'{o}mez-Alvarez}, P. and {Gracia-Carpio}, J. and {Granett}, B.~R. and {Grazian}, A. and {Grupp}, F. and {Haugan}, S.~V.~H. and {Hoar}, J. and {Holmes}, W. and {Hormuth}, F. and {Hornstrup}, A. and {Jahnke}, K. and {Jhabvala}, M. and {Keih\"{a}nen}, E. and {Kermiche}, S. and {Kiessling}, A. and {Kubik}, B. and {Kuijken}, K. and {K\"{u}mmel}, M. and {Kunz}, M. and {Kurki-Suonio}, H. and {Le Boulc'h}, Q. and {Le Brun}, A.~M.~C. and {Ligori}, S. and {Lilje}, P.~B. and {Lindholm}, V. and {Lloro}, I. and {Mainetti}, G. and {Maino}, D. and {Maiorano}, E. and {Mansutti}, O. and {Marggraf}, O. and {Martinelli}, M. and {Martinet}, N. and {Marulli}, F. and {Massey}, R. and {Medinaceli}, E. and {Mei}, S. and {Mellier}, Y. and {Meneghetti}, M. and {Merlin}, E. and {Meylan}, G. and {Mora}, A. and {Moresco}, M. and {Moscardini}, L. and {Nakajima}, R. and {Neissner}, C. and {Niemi}, S. -M. and {Padilla}, C. and {Paltani}, S. and {Pasian}, F. and {Pedersen}, K. and {Percival}, W.~J. and {Pettorino}, V. and {Pires}, S. and {Polenta}, G. and {Poncet}, M. and {Popa}, L.~A. and {Pozzetti}, L. and {Raison}, F. and {Rebolo}, R. and {Renzi}, A. and {Rhodes}, J. and {Riccio}, G. and {Romelli}, E. and {Roncarelli}, M. and {Saglia}, R. and {Sakr}, Z. and {Sapone}, D. and {Sartoris}, B. and {Schewtschenko}, J.~A. and {Schirmer}, M. and {Schneider}, P. and {Secroun}, A. and {Seidel}, G. and {Seiffert}, M. and {Serrano}, S. and {Simon}, P. and {Sirignano}, C. and {Sirri}, G. and {Stanco}, L. and {Steinwagner}, J. and {Tallada-Cresp\'{i}}, P. and {Taylor}, A.~N. and {Tereno}, I. and {Toft}, S. and {Toledo-Moreo}, R. and {Torradeflot}, F. and {Tsyganov}, A. and {Tutusaus}, I. and {Valenziano}, L. and {Valiviita}, J. and {Vassallo}, T. and {Verdoes Kleijn}, G. and {Veropalumbo}, A. and {Wang}, Y. and {Weller}, J. and {Zacchei}, A. and {Zamorani}, G. and {Zerbi}, F.~M. and {Zucca}, E. and {Mart\'{i}n-Fleitas}, J. and {Scottez}, V.},
        title = "{Euclid Quick Data Release (Q1): A photometric search for ultracool dwarfs in the Euclid Deep Fields}",
      journal = {arXiv e-prints},
         year = 2025,
        month = mar,
          eid = {arXiv:2503.22497},
        pages = {arXiv:2503.22497},
          doi = {10.48550/arXiv.2503.22497},
archivePrefix = {arXiv},
       eprint = {2503.22497},
 primaryClass = {astro-ph.SR},
       adsurl = {https://ui.adsabs.harvard.edu/abs/2025arXiv250322497V}
}

@ARTICLE{2025arXiv250322559M,
       author = {{Mohandasan}, A. and {Smart}, R.~L. and {Reyl\'{e}}, C. and {Le Brun}, V. and {P\'{e}rez-Garrido}, A. and {Ba\~{n}ados}, E. and {Goldman}, B. and {Casewell}, S.~L. and {Zapatero Osorio}, M.~R. and {Dupuy}, T. and {Rejkuba}, M. and {Mart\'{i}n}, E.~L. and {Dominguez-Tagle}, C. and {\v{Z}erjal}, M. and {Hu\'{e}lamo}, N. and {Lodieu}, N. and {Cruz}, P. and {Rebolo}, R. and {Phillips}, M.~W. and {Zhang}, J. -Y. and {Aghanim}, N. and {Altieri}, B. and {Amara}, A. and {Andreon}, S. and {Auricchio}, N. and {Baccigalupi}, C. and {Baldi}, M. and {Balestra}, A. and {Bardelli}, S. and {Battaglia}, P. and {Biviano}, A. and {Bonchi}, A. and {Branchini}, E. and {Brescia}, M. and {Brinchmann}, J. and {Camera}, S. and {Ca\~{n}as-Herrera}, G. and {Capobianco}, V. and {Carbone}, C. and {Carretero}, J. and {Casas}, S. and {Castellano}, M. and {Castignani}, G. and {Cavuoti}, S. and {Chambers}, K.~C. and {Cimatti}, A. and {Colodro-Conde}, C. and {Congedo}, G. and {Conselice}, C.~J. and {Conversi}, L. and {Copin}, Y. and {Costille}, A. and {Courbin}, F. and {Courtois}, H.~M. and {Cropper}, M. and {Da Silva}, A. and {Degaudenzi}, H. and {De Lucia}, G. and {Dole}, H. and {Douspis}, M. and {Dubath}, F. and {Dupac}, X. and {Dusini}, S. and {Escoffier}, S. and {Farina}, M. and {Faustini}, F. and {Ferriol}, S. and {Fotopoulou}, S. and {Frailis}, M. and {Franceschi}, E. and {Galeotta}, S. and {George}, K. and {Gillard}, W. and {Gillis}, B. and {Giocoli}, C. and {Gracia-Carpio}, J. and {Granett}, B.~R. and {Grazian}, A. and {Grupp}, F. and {Haugan}, S.~V.~H. and {Hoar}, J. and {Holmes}, W. and {Hook}, I.~M. and {Hormuth}, F. and {Hornstrup}, A. and {Jahnke}, K. and {Jhabvala}, M. and {Keih\"{a}nen}, E. and {Kermiche}, S. and {Kiessling}, A. and {Kubik}, B. and {Kuijken}, K. and {K\"{u}mmel}, M. and {Kunz}, M. and {Kurki-Suonio}, H. and {Le Boulc'h}, Q. and {Le Brun}, A.~M.~C. and {Le Mignant}, D. and {Ligori}, S. and {Lilje}, P.~B. and {Lindholm}, V. and {Lloro}, I. and {Mainetti}, G. and {Maino}, D. and {Maiorano}, E. and {Mansutti}, O. and {Marcin}, S. and {Marggraf}, O. and {Martinelli}, M. and {Martinet}, N. and {Marulli}, F. and {Massey}, R. and {Medinaceli}, E. and {Mei}, S. and {Mellier}, Y. and {Meneghetti}, M. and {Merlin}, E. and {Meylan}, G. and {Mora}, A. and {Moresco}, M. and {Moscardini}, L. and {Nakajima}, R. and {Neissner}, C. and {Niemi}, S. -M. and {Padilla}, C. and {Paltani}, S. and {Pasian}, F. and {Pedersen}, K. and {Percival}, W.~J. and {Pettorino}, V. and {Pires}, S. and {Polenta}, G. and {Poncet}, M. and {Popa}, L.~A. and {Pozzetti}, L. and {Raison}, F. and {Renzi}, A. and {Rhodes}, J. and {Riccio}, G. and {Romelli}, E. and {Roncarelli}, M. and {Saglia}, R. and {Sakr}, Z. and {Sapone}, D. and {Sartoris}, B. and {Schewtschenko}, J.~A. and {Schirmer}, M. and {Schneider}, P. and {Schrabback}, T. and {Secroun}, A. and {Seidel}, G. and {Serrano}, S. and {Simon}, P. and {Sirignano}, C. and {Sirri}, G. and {Stanco}, L. and {Steinwagner}, J. and {Surace}, C. and {Tallada-Cresp\'{i}}, P. and {Taylor}, A.~N. and {Tereno}, I. and {Toledo-Moreo}, R. and {Torradeflot}, F. and {Tsyganov}, A. and {Tutusaus}, I. and {Valenziano}, L. and {Valiviita}, J. and {Vassallo}, T. and {Verdoes Kleijn}, G. and {Veropalumbo}, A. and {Vibert}, D. and {Wang}, Y. and {Weller}, J. and {Zacchei}, A. and {Zamorani}, G. and {Zerbi}, F.~M. and {Zucca}, E. and {Mart\'{i}n-Fleitas}, J. and {Scottez}, V.},
        title = "{Euclid Quick Data Release (Q1) Ultracool dwarfs in the Euclid Deep Field North}",
      journal = {arXiv e-prints},
         year = 2025,
        month = mar,
          eid = {arXiv:2503.22559},
        pages = {arXiv:2503.22559},
          doi = {10.48550/arXiv.2503.22559},
archivePrefix = {arXiv},
       eprint = {2503.22559},
 primaryClass = {astro-ph.SR},
       adsurl = {https://ui.adsabs.harvard.edu/abs/2025arXiv250322559M}
}

@ARTICLE{2025ApJ...991...84D,
       author = {{Dominguez-Tagle}, C. and {{\v{Z}}erjal}, M. and {Sedighi}, N. and {Mas-Buitrago}, P. and {Martin}, E.~L. and {Zhang}, J.-Y. and {Vitas}, N. and {B{\'e}jar}, V.~J.~S. and {Tsilia}, S. and {Mu{\~n}oz Torres}, S. and {Lodieu}, N. and {Barrado}, D. and {Solano}, E. and {Cruz}, P. and {Tata}, R. and {Phan-Bao}, N. and {Burgasser}, A.},
        title = "{Euclid Quick Data Release (Q1)--Spectroscopic Search, Classification, and Analysis of Ultracool Dwarfs in the Deep Fields}",
      journal = {\apj},
         year = 2025,
        month = sep,
       volume = {991},
       number = {1},
          eid = {84},
        pages = {84},
          doi = {10.3847/1538-4357/adf72d},
archivePrefix = {arXiv},
       eprint = {2503.22442},
 primaryClass = {astro-ph.SR},
       adsurl = {https://ui.adsabs.harvard.edu/abs/2025ApJ...991...84D}
}

@software{adam_burgasser_2025_17107529,
  author       = {Adam Burgasser and
                  Chris Theissen and
                  ttamiya and
                  Brendan and
                  Caleb Choban and
                  Daniella Bardalez Gagliuffi and
                  Everett Schlawin and
                  jcswong},
  title        = {aburgasser/splat: v1.11},
  month        = sep,
  year         = 2025,
  publisher    = {Zenodo},
  version      = {v1.11},
  doi          = {10.5281/zenodo.17107529},
  url          = {https://doi.org/10.5281/zenodo.17107529},
  swhid        = {swh:1:dir:c48bd6b3b3ee93e801b1efe68890e913f913bde2
                   ;origin=https://doi.org/10.5281/zenodo.17107528;vi
                   sit=swh:1:snp:7ecd992bec214df97ea733944a984d329614
                   fa6d;anchor=swh:1:rel:3fb5f7985ab3a0a575db83ee60bf
                   3ddfd59bf16a;path=aburgasser-splat-80d7575
                  },
}

@article{2003ApJ...586..512B,
	adsurl = {http://ads.ari.uni-heidelberg.de/abs/2003ApJ...586..512B},
	author = {{Burgasser}, A.~J. and {Kirkpatrick}, J.~D. and {Reid}, I.~N. and {Brown}, M.~E. and {Miskey}, C.~L. and {Gizis}, J.~E.},
	doi = {10.1086/346263},
	eprint = {arXiv:astro-ph/0211470},
	journal = {\apj},
	month = mar,
	pages = {512-526},
	title = {{Binarity in Brown Dwarfs: T Dwarf Binaries Discovered with the Hubble Space Telescope Wide Field Planetary Camera 2}},
	volume = 586,
	year = 2003}

@ARTICLE{2013A&A...560A..52M,
       author = {{Manjavacas}, E. and {Goldman}, B. and {Reffert}, S. and {Henning}, T.},
        title = "{Parallax measurements of cool brown dwarfs}",
      journal = {\aap},
         year = 2013,
        month = dec,
       volume = {560},
          eid = {A52},
        pages = {A52},
          doi = {10.1051/0004-6361/201321720},
archivePrefix = {arXiv},
       eprint = {1310.2191},
 primaryClass = {astro-ph.SR},
       adsurl = {https://ui.adsabs.harvard.edu/abs/2013A&A...560A..52M}
}

@ARTICLE{2023AJ....166..226B,
       author = {{Bravo}, Alexia and {Schneider}, Adam C. and {Bardalez Gagliuffi}, Daniella and {Burgasser}, Adam J. and {Meisner}, Aaron M. and {Kirkpatrick}, J. Davy and {Faherty}, Jacqueline K. and {Kuchner}, Marc J. and {Caselden}, Dan and {Sainio}, Arttu and {Hamlet}, Les and {Backyard Worlds: Planet 9 Collaboration}},
        title = "{An Investigation of New Brown Dwarf Spectral Binary Candidates From the Backyard Worlds: Planet 9 Citizen Science Initiative}",
      journal = {\aj},
         year = 2023,
        month = dec,
       volume = {166},
       number = {6},
          eid = {226},
        pages = {226},
          doi = {10.3847/1538-3881/acffc1},
archivePrefix = {arXiv},
       eprint = {2310.06957},
 primaryClass = {astro-ph.SR},
       adsurl = {https://ui.adsabs.harvard.edu/abs/2023AJ....166..226B}
}

@ARTICLE{2019ApJS..240...19K,
       author = {{Kirkpatrick}, J. Davy and {Martin}, Emily C. and {Smart}, Richard L. and {Cayago}, Alfred J. and {Beichman}, Charles A. and {Marocco}, Federico and {Gelino}, Christopher R. and {Faherty}, Jacqueline K. and {Cushing}, Michael C. and {Schneider}, Adam C. and {Mace}, Gregory N. and {Tinney}, Christopher G. and {Wright}, Edward L. and {Lowrance}, Patrick J. and {Ingalls}, James G. and {Vrba}, Frederick J. and {Munn}, Jeffrey A. and {Dahm}, Scott E. and {McLean}, Ian S.},
        title = "{Preliminary Trigonometric Parallaxes of 184 Late-T and Y Dwarfs and an Analysis of the Field Substellar Mass Function into the {\textquotedblleft}Planetary{\textquotedblright} Mass Regime}",
      journal = {\apjs},
         year = 2019,
        month = feb,
       volume = {240},
       number = {2},
          eid = {19},
        pages = {19},
          doi = {10.3847/1538-4365/aaf6af},
archivePrefix = {arXiv},
       eprint = {1812.01208},
 primaryClass = {astro-ph.SR},
       adsurl = {https://ui.adsabs.harvard.edu/abs/2019ApJS..240...19K}
}

@ARTICLE{2020MNRAS.495.1136S,
       author = {{Sahlmann}, J. and {Burgasser}, A.~J. and {Bardalez Gagliuffi}, D.~C. and {Lazorenko}, P.~F. and {S{\'e}gransan}, D. and {Zapatero Osorio}, M.~R. and {Blake}, C.~H. and {Gelino}, C.~R. and {Mart{\'\i}n}, E.~L. and {Bouy}, H.},
        title = "{Astrometric orbits of spectral binary brown dwarfs - I. Massive T dwarf companions to 2M1059-21 and 2M0805+48}",
      journal = {\mnras},
         year = 2020,
        month = jun,
       volume = {495},
       number = {1},
        pages = {1136-1147},
          doi = {10.1093/mnras/staa1235},
archivePrefix = {arXiv},
       eprint = {2004.14889},
 primaryClass = {astro-ph.SR},
       adsurl = {https://ui.adsabs.harvard.edu/abs/2020MNRAS.495.1136S}
}

@ARTICLE{2014A&A...565A..20S,
       author = {{Sahlmann}, J. and {Lazorenko}, P.~F. and {S{\'e}gransan}, D. and {Mart{\'\i}n}, E.~L. and {Mayor}, M. and {Queloz}, D. and {Udry}, S.},
        title = "{Astrometric planet search around southern ultracool dwarfs. I. First results, including parallaxes of 20 M8-L2 dwarfs}",
      journal = {\aap},
         year = 2014,
        month = may,
       volume = {565},
          eid = {A20},
        pages = {A20},
          doi = {10.1051/0004-6361/201323208},
archivePrefix = {arXiv},
       eprint = {1403.1275},
 primaryClass = {astro-ph.SR},
       adsurl = {https://ui.adsabs.harvard.edu/abs/2014A&A...565A..20S}
}

@article{2003PASP..115..362R,
	adsurl = {http://adsabs.harvard.edu/abs/2003PASP..115..362R},
	author = {{Rayner}, J.~T. and {Toomey}, D.~W. and {Onaka}, P.~M. and {Denault}, A.~J. and {Stahlberger}, W.~E. and {Vacca}, W.~D. and {Cushing}, M.~C. and {Wang}, S.},
	journal = {\pasp},
	month = mar,
	pages = {362-382},
	title = {{SpeX: A Medium-Resolution 0.8-5.5 Micron Spectrograph and Imager for the NASA Infrared Telescope Facility}},
	volume = 115,
	year = 2003}

@ARTICLE{2024ApJ...967..115B,
       author = {{Best}, William M.~J. and {Sanghi}, Aniket and {Liu}, Michael C. and {Magnier}, Eugene A. and {Dupuy}, Trent J.},
        title = "{A Volume-limited Sample of Ultracool Dwarfs. II. The Substellar Age and Mass Functions in the Solar Neighborhood}",
      journal = {\apj},
         year = 2024,
        month = jun,
       volume = {967},
       number = {2},
          eid = {115},
        pages = {115},
          doi = {10.3847/1538-4357/ad39ef},
archivePrefix = {arXiv},
       eprint = {2401.09535},
 primaryClass = {astro-ph.SR},
       adsurl = {https://ui.adsabs.harvard.edu/abs/2024ApJ...967..115B}
}

@ARTICLE{2025arXiv251101167M,
       author = {{Morrissey}, Sara J. and {Burgasser}, Adam J. and {de Graaff}, Anna and {McConachie}, Ian and {Brammer}, Gabriel},
        title = "{Discovery of Seven Cold and Distant Brown Dwarfs with JWST RUBIES}",
      journal = {arXiv e-prints},
         year = 2025,
        month = nov,
          eid = {arXiv:2511.01167},
        pages = {arXiv:2511.01167},
          doi = {10.48550/arXiv.2511.01167},
archivePrefix = {arXiv},
       eprint = {2511.01167},
 primaryClass = {astro-ph.SR},
       adsurl = {https://ui.adsabs.harvard.edu/abs/2025arXiv251101167M}
}

@ARTICLE{2025arXiv251002026T,
       author = {{Tu}, Zhijun and {Wang}, Shu and {Chen}, Xiaodian and {Liu}, Jifeng},
        title = "{A Large Sample of JWST/NIRSpec Brown Dwarfs: New Distant Discoveries}",
      journal = {arXiv e-prints},
         year = 2025,
        month = oct,
          eid = {arXiv:2510.02026},
        pages = {arXiv:2510.02026},
          doi = {10.48550/arXiv.2510.02026},
archivePrefix = {arXiv},
       eprint = {2510.02026},
 primaryClass = {astro-ph.SR},
       adsurl = {https://ui.adsabs.harvard.edu/abs/2025arXiv251002026T}
}
\bibliographystyle{aasjournal}

\end{document}